\documentclass[twoside,11pt]{article}

\usepackage{amsmath,amssymb,booktabs, algorithm, algorithmic, natbib}
\usepackage[english]{babel}
\usepackage[utf8]{inputenc}
\usepackage{xcolor, bbm, subcaption,multirow}
\usepackage[title]{appendix}

\newtheorem{assumption}{Assumption}

\def\E{\mathbb E}
\def\Hbb{\mathbb H}
\def\Ibb{\mathbb I}
\def\P{\mathbb P}
\def\Rbb{\mathbb R}
\def\R{\mathbb R}

\def\cbf{c} 
\def\xbf{x}
\def\zbf{z} 
\def\epsilonbf{\epsilon} 
\def\thetabf{\theta} 
\def\Sigmabf{\Sigma} 

\def\Ccal{\mathcal C}
\def\Fcal{\mathcal F}
\def\Hcal{\mathcal H}
\def\Pcal{\mathcal P}
\def\Rcal{\mathcal R}
\def\Tcal{\mathcal T}
\def\TT{\mathcal T}
\def\Xcal{\mathcal X}
\def\HH{\mathcal H}

\usepackage{jmlr2e}
\usepackage{lastpage}
\jmlrheading{25}{2024}{1-\pageref{LastPage}}{5/22; Revised
1/24}{2/24}{22-0487}{Xiaowu Dai and Lexin Li}
\ShortHeadings{Post-Regularization Confidence Bands for ODE}{Dai and Li}

\firstpageno{1}

\begin{document}

\title{Post-Regularization Confidence Bands \\
for Ordinary Differential Equations}

\author{\name Xiaowu Dai \email dai@stat.ucla.edu\\
       \addr 
       Department of Statistics and Data Science and Department of Biostatistics\\
       University of California, Los Angeles, CA 90095-1554, USA
       \AND
       \name Lexin Li \email lexinli@berkeley.edu \\
       \addr 
       Department of Biostatistics and Epidemiology\\
       University of California, Berkeley, CA 94720-1776, USA
       }
\editor{Mladen Kolar}

\maketitle

\begin{abstract}%
Ordinary differential equation (ODE) is an important tool to study a system of biological and physical processes. A central question in ODE modeling is to infer the significance of individual regulatory effect of one signal variable on another. However, building confidence band for ODE with unknown regulatory relations is challenging, and it remains largely an open question. In this article, we construct the post-regularization confidence band for the individual regulatory function in ODE with unknown functionals and noisy data observations. Our proposal is the first of its kind, and is built on two novel ingredients. The first is a new localized kernel learning approach that combines reproducing kernel learning with local Taylor approximation, and the second is a new de-biasing method that tackles infinite-dimensional functionals and additional measurement errors. We show that the constructed confidence band has the desired asymptotic coverage probability, and the recovered regulatory network approaches the truth with probability tending to one. We establish the theoretical properties when the number of variables in the system can be either smaller or larger than the number of sampling time points, and we study the regime-switching phenomenon. We demonstrate the efficacy of the proposed method through both simulations and illustrations with two data applications. 
\end{abstract}


%%%%%%%%%%%%%%%%%%%%%%%%%%%%%%%%%%%%%%%%%%%%%%%%%%%
\section{Introduction}
\label{sec:introduction}

\noindent
Characterizing the dynamics of biological and physical processes is of fundamental interest in a large variety of scientific fields, and ordinary differential equation (ODE) is a frequently used tool to address such type of questions. Examples include infectious disease \citep{Liang2008}, genomics \citep{CaoZhao2008, Ma2009, Wu2014}, neuroscience \citep{Izhikevich2007, Zhang2015, Zhang2017, CaoLuo2019}, economics \citep{dai2023continuous}, among many others. An ODE system models the changes of a set of variables, quantified by their derivatives with respect to time, as functions of all other variables in the system. Typically, the system is observed on discrete time points with some additive measurement errors. In recent years, there have been an increased number of proposals for ODE modeling. One family of ODE models adopt linear function forms in the ODE system; for instance, \cite{LuLiang2011} proposed a set of linear ODEs, \citet{Zhang2015} extended to include two-way interactions, and \citet{Dattner2015} further extended to a generalized linear form using a finite set of known basis functions. Another family of ODE models consider additive functionals; for instance, \citet{HendersonMichailidis2014, Wu2014, Chen2017} proposed the generalized additive models with a set of common basis functions plus an unknown residual function. The third family studies the ODE system when the functional forms are completely known \citep{gonzalez2014reproducing, li2015regularized, zhang2015selection, mikkelsen2017learning}. Recently, \citet{dai2021kernel} proposed reproducing kernel-based ODEs to flexibly model the unknown functionals of both main effects and two-way interactions. 

A central question in ODE modeling is about \emph{inference} of significance of individual regulatory relations among the variables in the system \citep{Ma2009}. The majority of existing ODE solutions, however, have been focusing on regularized sparse \emph{estimation} of the ODEs. {Even though sparse estimation can in effect identify such relations, it does not produce a quantification of statistical significance, nor explicitly controls the false discovery rate (FDR). By contrast, inference provides both an explicit uncertainty quantification and an explicit FDR control. As such, inference is generally a different problem from sparse estimation.}  When the functional forms are completely \emph{known} in an ODE system, confidence intervals for the ODE parameters have been studied and been mostly built upon asymptotic normality of finite-dimensional parameters \citep{Ramsay2007, qi2010asymptotic, xue2010sieve, miao2014generalized, zhang2015selection, wu2019parameter}. When the functional forms are \emph{unknown}, however, there is \emph{no} existing solution for individual ODE parameter inference. \citet{dai2021kernel} derived the confidence interval of the entire signal trajectory in kernel ODEs with unknown functionals. Nevertheless, their method could \emph{not} infer the \emph{individual} regulatory effect of one variable on another, but instead only the \emph{sum} of all individual effects. Building confidence intervals for individual ODE parameters with unknown regulatory relations is particularly challenging, as it involves infinite-dimensional functionals. There is a clear gap in the current literature on ODE inference. 

In this article, we tackle the ODE inference problem with unknown functionals and noisy data observations. Our goal is to directly construct the confidence band for any individual regulatory functional that measures the effect of one signal variable on another. The constructed confidence band provides both an uncertainty quantification for the individual regulatory relation, and also a sparse recovery of the entire regulatory system when coupled with a proper false discovery rate (FDR) control. We establish the confidence band for both low-dimensional and high-dimensional settings, where the number of variables in the system can be smaller or larger than the number of sampling time points, and we study the regime-switching phenomenon. We show that the constructed confidence band has the desired asymptotic coverage probability, and the recovered regulatory network approaches the truth with probability tending to one. Toward our goal, we propose and develop two novel methodological ingredients: a new \emph{localized kernel learning} approach that combines reproducing kernel learning with local Taylor approximation, and a new de-biasing method that tackles infinite-dimensional functionals and additional measurement errors. Consequently, our proposal makes useful contributions on multiple fronts, including ODE inference, nonparametric modeling, as well as high-dimensional inference with measurement errors. 

The first component is a new localized kernel learning approach that in effect fuses two widely used nonparametric modeling techniques, reproducing kernel learning methods \citep{wahba1990}, and local polynomial methods \citep{fan1996local}. More specifically, we adopt the kernel ODE model of \citet{dai2021kernel}, which is built upon the learning framework of reproducing kernel Hilbert space \citep[RKHS,][]{Aronszajn1950, wahba1990} and smoothing spline analysis of variance \citep[SS-ANOVA,][]{Wahba1995, huang1998projection, LinZhang2006}. This model allows highly flexible and unknown forms for the functions in the ODE system as well as interactions. Meanwhile, we employ the Taylor expansion and local approximation idea, which is frequently employed in local polynomial nonparametric regressions \citep{fan1996local, Opsomer1997}. Such a fused method essentially characterizes the regulatory effect of one variable on another in the ODE system through a scalar quantity, which in turn allows us to derive the corresponding confidence band. Moreover, this localized kernel learning approach is potentially useful for other nonparametric modeling problems beyond ODEs. 

We remark that our method is related to but also substantially different from the kernel ODE method of \citet{dai2021kernel}, and the kernel-sieve hybrid method of \citet{lu2020kernel}. Compared to \citet{dai2021kernel}, although we adopt the same ODE model framework, our estimator is utterly different after introducing the local approximation. Actually, our localized kernel estimator has a slower convergence rate compared to the minimax rate of the estimator of \citet{dai2021kernel} in a low-dimensional setting; see Section \ref{sec:estimation-rate}. On the other hand, this slower rate is sufficient for constructing an asymptotically valid confidence band. {More importantly, the method of \citet{dai2021kernel} can only obtain the confidence interval for the entire signal trajectory, which is the \emph{sum} of \emph{all} individual effects, but not for a \emph{single} individual effect of one variable on another. Directly applying the estimator of \citet{dai2021kernel} \emph{cannot} obtain the individual confidence band we target. Later, we also numerically demonstrate that the proposed method outperforms a naive modification of \citet{dai2021kernel} that aggregates the point-wise confidence intervals at grid points with Bonferroni correction.}  Compared to \citet{lu2020kernel} who tackled the inference problem of a nonparametric additive model, our ODE inference method differs in multiple ways, including that the signals are not directly observed but need to be estimated from the data with error, there are pairwise interaction terms, and the ODE estimation involves integrals. These differences have introduced a whole new set of challenges than \citet{lu2020kernel} and the classical sieve method of \citet{shen1994convergence}, and thus a different solution. 

We also remark that, by choosing a proper linear kernel or additive kernel, the kernel ODE model includes the linear ODE \citep{Zhang2015} and the additive ODE \citep{Chen2017} as special cases. Consequently, our inference solution is applicable to a range of different ODE models. Our work thus addresses a scientific question that is crucial but currently still remains open, and makes a useful addition to the ODE toolbox. 

The second component is a new de-biasing method for the localized kernel ODE estimator. We observe that the individual regulatory effects in ODE models are nonparametric functionals, which are estimated through proper regularization in terms of the RKHS norms. However, the regularization introduces bias, and essentially offers a trade-off between bias and overfitting. Consequently, it is crucial to perform de-biasing in post-regularization high-dimensional statistical inference \citep{Zhang2014CI, Vandegeer2014, Ning2017, ZhangCheng2017}.  On the other hand, there are some extra layers of complications for de-biasing in ODEs. First of all, the signal variables themselves are not directly observed, but only their noisy counterparts, and they need to be estimated given the noisy data. Besides, the objects of inference are infinite-dimensional functionals, and their estimation involves integrals.  To overcome those difficulties, we introduce a new bias correction score with integral of the estimated functional. We also generalize \citet{chernozhukov2014anti} and perform a new approximation analysis for the Gaussian multiplier bootstrap within the RKHS framework. Our method is the first de-biasing solution for ODE models, and thus also contributes to the literature on de-biasing. In addition, it is potentially useful for high-dimensional inference of other statistical models that involve latent variables and measurement errors \citep{wansbeek2001measurement}.

The rest of the article is organized as follows. Section \ref{sec:method} introduces the kernel ODE model, then develops the localized kernel learning approach. Section \ref{sec:estimation} presents the parameter estimation, and Section \ref{sec:inference} derives the confidence band formula. Section \ref{sec:theory} establishes the convergence rate and coverage property of the proposed method. Section \ref{sec:simulation} investigates the finite-sample performance, and Section \ref{sec:application} illustrates with two real data examples. {Section \ref{sec:discussion} concludes the paper with a discussion,} and the Appendix collects all technical proofs.

%%%%%%%%%%%%%%%%%%%%%%%%%%%%%%%%%%%%%%%%%%%%%%%%%%%
\section{Localized Kernel Learning for ODE}
\label{sec:method}

\noindent
In this section, {we first present our kernel ODE model system, which consists of models \eqref{eqn:intermodel}, \eqref{eqn:obserdata} and \eqref{eqn:nonadditivemodel}.} We then propose the localized kernel learning method, which fuses reproducing kernel learning and local polynomial approximation, and is crucial for ODE inference.

%%%%%%%%%%%%%%%%%%%%%%%%%%%%%%%%%%%%%%%%%%%%%%%%%%%
\subsection{Kernel ODE Model}
\label{sec:model}

\noindent
Let $x(t) = (x_1(t), \ldots, x_p(t))^\top \in \Rbb^p$ denote the set of $p$ variables of interest, and $t$ index the time in a standardized interval $\Tcal = [0,1]$. We consider the ODE system, 
\begin{equation} \label{eqn:intermodel}
\frac{d x(t)}{dt} = \left(
\begin{array}{c}
\dfrac{d x_1(t)}{d t} \\
\vdots\\
\dfrac{d x_p(t)}{d t} \\
\end{array}
\right) = \left(
\begin{array}{c}
F_1(x(t))  \\
\vdots\\
F_p(x(t)) \\
\end{array}
\right) 
= F(x(t)), 
\end{equation}
where $F = \{F_1, \ldots,$ $F_p\}$ denotes the set of unknown functionals that characterize the regulatory relations among $x(t)$. Typically, the system \eqref{eqn:intermodel} is observed on a set of $n$ discrete time points $\{t_1, \ldots, t_n\}$, with additional measurement errors, 
\begin{equation} \label{eqn:obserdata}
y_i = x(t_i) + \epsilon_i, \quad i=1,\ldots,n,
\end{equation}
where $y_i  = (y_{i1},\ldots,y_{ip})^\top \in \Rbb^p$ denotes the observed data, and $\epsilon_i = (\epsilon_{i1}, \ldots, \epsilon_{ip})^\top \in \Rbb^p$ denotes the vector of measurement errors that are usually assumed to follow an independent normal distribution with mean $0$ and variance $\sigma_j^2$,  $j=1,\ldots,p$. Besides, the system \eqref{eqn:intermodel} usually starts with an initial condition $x(0) \in \Rbb^p$. 

In a biological and physical system, given the observed noisy time-course data $\{ y_i \}_{i=1}^{n}$, a central question of interest is to uncover the structure of the system of ODEs in terms of which variables regulate which other variables. We say that $x_{k}(t)$ regulates $x_{j}(t)$, if $F_{j}$ is a non-zero functional of $x_{k}(t)$. That is, $x_{k}(t)$ affects $x_{j}(t)$ through the functional $F_{j}$ on its derivative $d x_{j}(t) / d t$, for $j, k = 1, \ldots, p$. We consider the following model for $F_j$,  
\vspace{-0.01in}
\begin{equation} \label{eqn:nonadditivemodel}
F_{j}(x(t)) = \theta_{j0} + \sum_{k=1}^p F_{jk}(x_k(t)) + \sum_{k=1,k \neq l}^p \sum_{l=1}^p F_{jkl}(x_k(t),x_l(t)), \quad j = 1, \ldots, p, 
\end{equation}
where $\theta_{j0} \in \Rbb$ denotes the global intercept, $F_{jk}$ and $F_{jkl}$ denote the main effect and two-way interaction, respectively. Higher-order interactions are possible, but two-way interactions are most common in ODEs \citep{Ma2009, Zhang2015}.

We next build model \eqref{eqn:nonadditivemodel} within the smoothing spline ANOVA framework \citep{Wahba1995, gu2013, LinZhang2006}. Specifically, let $\Hcal_k$ denote a space of functions of $x_k(t)$ with zero marginal integral, which is specified through the averaging operator, $\int_\Tcal F_{jk}(x_k(t))dt = 0$ for any $k,j=1,\ldots,p$. 
The zero marginal integrals of functions in $\Hcal_{k}$ ensure that the decomposition in \eqref{eqn:nonadditivemodel} is well defined over its domain, and that the terms $\theta_{j0}, F_{jk},$ and $F_{jkl}$ in \eqref{eqn:nonadditivemodel}  are identifiable and can be estimated uniquely. More discussion of the averaging operator is given in Section \ref{sec:ave-operator} of the Appendix.  Let $x_k(t) \in \Xcal$, where $\Xcal$ is a compact set in $\Rbb$. Let $\{1\}$ denote the space of constant functions. Construct the tensor product space as
\begin{equation} \label{eqn:spaceH}
\Hcal = \{1\} \; \oplus \; \sum_{k=1}^p\Hcal_k \; \oplus \sum_{k=1,k \neq l}^p \sum_{l=1}^p \left( \Hcal_k\otimes\Hcal_l \right),
\end{equation}
where $\oplus$ and $\otimes$ denote the direct sum operator and the tensor product operator, respectively. The space $\Hcal_k\otimes\Hcal_l$ includes the tensor products of the functions in $\Hcal_k$ and $\Hcal_l $.  We assume the functionals $F_j$, $j=1, \ldots, p$, in the ODE model \eqref{eqn:nonadditivemodel} are located in the space of $\Hcal$.

We note that our kernel ODE model is the same as the model used in \citet{dai2021kernel}. Nevertheless, one  \emph{cannot} achieve the goal of inferring individual regulatory functional using the approach of \citet{dai2021kernel}, which estimates the collective functional $F_j$, rather than the individual functional $F_{jk}$. Instead, we propose a completely new localized kernel learning approach for inference, which is a key novelty of this article.

%%%%%%%%%%%%%%%%%%%%%%%%%%%%%%%%%%%%%%%%%%%%%%%%%%%
\subsection{Localized Kernel Learning}
\label{sec:functionalestimation}

\noindent
We next introduce the Taylor expansion and local approximation idea into our kernel ODE model framework. In effect, we fuse two popular nonparametric modeling techniques, reproducing kernel learning \citep{wahba1990} and local polynomial learning \citep{fan1996local}. Our primary goal is to infer the individual regulatory effect of $x_k(t)$ on $x_j(t)$, for any given pair of $j, k = 1, \ldots, p$, in the ODE system \eqref{eqn:intermodel}. Toward that goal, we observe that such a regulatory effect is encoded in two parts: the main effect term, $F_{jk}(x_k(t))$, $j, k = 1, \ldots, p$, and the two-way interaction terms,  $F_{jkl}(x_k(t), x_l(t))$, $l = 1, \ldots, p, l \neq k$. We next study the two parts separately.

First, for the main effect term $F_{jk}(x_k(t))$, {we consider the Taylor expansion with Lagrange remainder at a fixed time point $t = t_0$ \citep[Section 7.7,][]{apostol1991calculus}.} Specifically, letting $\tilde{t}$ be a point locating between $t_0$ and $t$, by the chain rule, we have 
\begin{equation} \label{eqn:taylorexpansionfjk} 
F_{jk}(x_k(t)) = F_{jk}(x_k(t_0)) + \frac{dF_{jk}(x_k(\tilde{t}))}{d x_k} \frac{d x_k(\tilde{t})}{dt}(t-t_0). 
\end{equation}
Since $F_{jk}(x_k(t_0))$ is a constant at the fixed $t_0$ and does not vary with $t$, we denote $F_{jk}(x_k(t_0))$ $\equiv \alpha_{jk, t_0}$. Besides, let $\Fcal$ be the function space where $x_k(t)$ reside in, and $\Fcal$ does not have to be the same as $\Hcal_k$. Suppose the functions in $\Hcal_k$ and $\Fcal$ have continuous first derivatives, which is true for most of commonly used RKHS, including those generated by the Gaussian, Laplace, or Mat\'{e}rn kernel \citep{scholkopf2018learning}. As such, both ${dF_{jk}(x_k(t))} / {dx_k}$ and ${dx_k(t)}/{dt}$ are continuous functions in $t$, and when $t \to t_0$, the second term in \eqref{eqn:taylorexpansionfjk} goes to zero. Henceforth, we can approximate $F_{jk}(x_k(t))$ by $\alpha_{jk, t_0}$. We remark that, we consider the first-order Taylor expansion, instead of the zeroth-order, in \eqref{eqn:taylorexpansionfjk}. Nevertheless, due to the smoothness property of RKHS, we  approximate $F_{jk}(x_k(t))$ by $\alpha_{jk, t_0}$, which is a constant at a fixed time point $t_0$ while it changes as $t_0$ varies. Moreover, when inferring the effect of $x_k(t)$ on $x_j(t)$, we focus on the main effect term of interest $F_{jk}(x_k(t))$, while treating the rest of the main effect terms $F_{jl}(x_{l}(t))$, $l=1,\ldots,p, l \neq k$, as nuisance parameters.

Next, for the interaction terms $F_{jkl}(x_k(t), x_l(t))$, $l = 1, \ldots, p, l \neq k$, since $\{1\}\otimes \Hcal_l = \Hcal_l$ \citep{wahba1990}, $F_{jkl}(x_k(t), x_l(t))$ is absorbed into the main effect term $F_{jl}(x_l(t))$, when $F_{jk}(x_k(t))$ is approximated by the constant $\alpha_{jk, t_0}$. 
In other words, when the effect of $x_k(t)$ on $x_j(t)$ holds constant, the change of the joint effect of $(x_k(t), x_l(t))$ on $x_j(t)$ only depends on the change of the effect of $x_l(t)$ on $x_j(t)$. Another way to see this is that, since $F_{jkl} \in \Hcal_k \otimes \Hcal_l$ in \eqref{eqn:spaceH}, there exists a finite integer $s$ and functions $F_{jk_\nu} \in\Hcal_k, F_{jl_\nu} \in\Hcal_l$ for $\nu=1,\ldots,s$, such that
\begin{align} \label{eqn:interaction-part}
F_{jkl} = \sum_{\nu=1}^s F_{jk_\nu} F_{jl_\nu} = \sum_{\nu=1}^s\alpha_{jk_\nu, t_0} F_{jl_\nu}   \in \Hcal_l, 
\end{align}
where the second equality is due to \eqref{eqn:taylorexpansionfjk} with $F_{jk_\nu}(x_k(t)) = \alpha_{jk_\nu,t_0}$ as $t  \to t_0$, and the last step is due to the fact that the linear combination of RKHS functions is still in the RKHS \citep{wahba1990}.

Combining the above results, and following the Taylor approximation that $F_{jk}(x_k(t)) = \alpha_{jk, t_0}$ as $t \to t_0$, the regulatory effect of $x_k(t)$ on $x_j(t)$ is now captured by the scalar $\alpha_{jk, t_0}$. Consequently, we estimate $\alpha_{jk, t_0}$ at any given time point $t_0 \in \Tcal$, along with other nuisance terms in $F_j$, then build a confidence band based on $\alpha_{jk, t_0}$ to infer the effect of $x_k(t)$ on $x_j(t)$. {Specifically, write $F_j = \theta_{j0} + \tilde{F}_j$, where $\theta_{j0}$ is the global mean, and $\tilde{F}_j$ is the centralized functional, $j = 1, \ldots, p$. For any given $k = 1, \ldots, p$, if $t$ is within a local neighborhood of $t_0$ in that $|t - t_0| < \varepsilon$ for some small $\varepsilon > 0$, we write 
\begin{align} \label{eqn:local}
\begin{split}
\tilde{F}_j(x(t)) & = \alpha_{jk, t_0} + \sum_{k'=1, k' \neq k}^p F_{j k'}(x_{k'}(t)) + \sum_{k'=1, k' \neq k,l}^p \sum_{l=1, l\neq k}^p  F_{j k'l}(x_{k'}(t), x_l(t)) \\ 
& \equiv \alpha_{jk, t_0} + H_{jk}(x(t)), 
\end{split}
\end{align}
where $H_{jk}$ collects all the nuisance terms when evaluating the effect of $x_k(t)$ on $x_j(t)$.} We next develop a procedure to estimate the global mean $\theta_{j0}$ and the functional $\tilde{F}_j$ in \eqref{eqn:local}.

%%%%%%%%%%%%%%%%%%%%%%%%%%%%%%%%%%%%%%%%%%%%%%%%%%%
\section{ODE Estimation}
\label{sec:estimation}

\noindent
In this section, we develop an estimation procedure, where we adopt a two-step collocation approach \citep{Varah1982}. We first estimate model \eqref{eqn:nonadditivemodel} under the constraint that $F_j \in \Hcal$ in \eqref{eqn:spaceH}, then incorporate the local approximation in \eqref{eqn:local}. Next, we derive the corresponding optimization algorithm to estimate the unknown parameters.

%%%%%%%%%%%%%%%%%%%%%%%%%%%%%%%%%%%%%%%%%%%%%%%%%%%
\subsection{Two-Step Collocation Estimation}
\label{sec:twostep}

\noindent
We adopt the two-step collocation method that is commonly used in ODE estimation. 

The first step is to obtain a smoothing estimate $\widehat{x}(t) = (\widehat{x}_1(t), \ldots, \widehat{x}_p(t))^\top$, 
\vspace{-0.01in}
\begin{equation} \label{eqn:sshatxj}
\widehat{x}_j(t) = \underset{z_j \in \Fcal}{\arg\min} \left\{\frac{1}{n}\sum_{i=1}^n \left[ y_{ij} - z_{j}(t_i) \right]^2 + \lambda_{nj}\|z_j(t)\|_{\Fcal}^2\right\}, \quad j=1,\ldots,p, 
\end{equation}
where $\|\cdot\|_{\Fcal}$ is the norm of the RKHS $\Fcal$, $z_j$ is a function in $\Fcal$ we minimize over, and $\lambda_{nj}\geq 0$ is the smoothness parameter often tuned by generalized cross-validation \citep{wahba1990}.

The second step is to estimate $F_j$ in \eqref{eqn:nonadditivemodel}. We first follow \citet{dai2021kernel} and obtain an estimator of the global mean $\theta_{j0}$ in $F_j$ as,
\begin{align} \label{eqn:theta0}
\widehat{\theta}_{j0} = \bar{y}_j - \int_\Tcal\bar{T}(t) \tilde{F}_j(\widehat{x}(t)) dt, 
\end{align}
where $\bar{y}_j =  n^{-1}\sum_{i=1}^ny_{ij}$, $\bar{T}(t) =n^{-1}\sum_{i=1}^nT_i(t)$, $T_i(t) = \Ibb\{0\leq t\leq t_i\}$, and $\Ibb(\cdot)$ is the indicator function. We then plug in this global mean estimator and estimate the centralized component $\tilde{F}_j$ in $F_j$ by solving the penalized optimization,
\begin{equation} \label{eqn:kode2step-Ftilde}
\begin{aligned}
\underset{\tilde{F}_j\in\Hcal}{\min} \left\{\frac{1}{n}\sum_{i=1}^n\left[(y_{ij} - \bar{y}_j)-\int_{\Tcal}\left( T_i(t)-\bar{T}(t)\right) \tilde{F}_j(\widehat{x}(t))dt\right]^2 \right. \\ 
\left. + \; \tau_{nj} \left( \sum_{k=1}^p \|\Pcal^k \tilde{F}_j\|_{\Hcal} + \sum_{k=1, k \neq l}^p \sum_{l=1}^p \|\Pcal^{kl} \tilde{F}_j\|_\Hcal \right) \right\},
\end{aligned}
\end{equation}
where $\|\cdot\|_{\Hcal}$ is the norm of $\Hcal$, $\Pcal^{k}\tilde{F}_j$ and $\Pcal^{kl}\tilde{F}_j$ are the orthogonal projections of $\tilde{F}_j$, or equivalently $F_j$, onto $\Hcal_{k}$ and  $\Hcal_{k}\otimes \Hcal_{l}$, respectively, and $\tau_{nj}$ is the penalty parameter.  Note that the optimization problem \eqref{eqn:kode2step-Ftilde} deals with the integral $ \int_0^{t_i}\tilde{F}_j(\widehat{x}(u))du$, rather than the derivative $d \widehat{x}_j(t) / d t$. This follows a similar spirit as \citet{Dattner2015}, and is to produce a more robust estimate. Moreover, the penalty function in \eqref{eqn:kode2step-Ftilde} is a sum of RKHS norms on the main effects and pairwise interactions, which is similar in spirit as the component selection and smoothing operator penalty of \citet{LinZhang2006}.

Next, we introduce the localization as specified in \eqref{eqn:local}. Let $R : \Tcal \mapsto \Rbb$ be a symmetric density function with bounded support. Denote $R_h(\cdot)=h^{-1}R(\cdot/h)$, where $h>0$ is the bandwidth. Following \eqref{eqn:local} and \eqref{eqn:kode2step-Ftilde}, we estimate $\alpha_{jk, t_0} \in \Rbb$ and $H_{jk} \in \Hcal$ through the localized and penalized optimization, 
\begin{equation} \label{eqn:DKL}
\begin{aligned}
 \underset{\alpha_{jk, t_0}, H_{jk}}{\min} \left\{ \frac{1}{n}\sum_{i=1}^nR_h\left(t_i-t_0\right)\left[(y_{ij} - \bar{y}_j) - \alpha_{jk, t_0} \bar{t}_i -\int_{\Tcal}\left( T_i(t)-\bar{T}(t) \right) H_{jk}(\widehat{x}(t))dt\right]^2\right. \\
\left. + \; \tau_{nj} \left( \sum_{k'=1,  k'\neq k}^p\|\Pcal^{k'} H_{jk}\|_{\Hcal} + \sum_{k'=1, k' \neq k,l}^p \sum_{l=1, l\neq k}^p \|\Pcal^{k'l} H_{jk}\|_\Hcal \right)\right\}, 
\end{aligned}
\end{equation}
where $\bar{t}_i = t_i - n^{-1}\sum_{i=1}^nt_i$. The estimate $\widehat{\alpha}_{jk, t_0}$ obtained from \eqref{eqn:DKL} captures the individual regulatory effect of $x_k(t)$ on $x_j(t)$ in a local neighborhood of $t_0 \in \Tcal$. {The local weight $R_h$ introduced in \eqref{eqn:DKL} is to facilitate both the estimation and subsequent inference. It places more weight to the data observations close to $t_0$ and less weight to those far away from $t_0$, in the same spirit as the local polynomial method \citep{fan1996local}. Besides, it allows us to later construct  confidence bands using tools such as the extreme value theory as well as the Gaussian multiplier bootstrap procedure.}

Now, given the estimates $\widehat{x}_j(t)$ from \eqref{eqn:sshatxj}, $\widehat{\theta}_{j0} $ from \eqref{eqn:theta0}, and $\widehat{\alpha}_{jk, t_0}, \widehat{H}_{jk}$ from \eqref{eqn:DKL}, we estimate the $p$-dimensional functional $F_j(x(t_0)) = \theta_{j0} +F_{jk}(x_k(t_0)) + \sum_{k'=1,k'\neq k}^p F_{jk'}(x_k'(t_0)) + \sum_{k'=1,k' \neq l}^p \sum_{l=1}^p F_{jk'l}(x_k'(t_0),x_l(t_0))$ at any time point $t_0 \in \Tcal$ by
\begin{equation} \label{eqn:estnonadditivemodel}
\widehat{F}_{j}(\widehat{x}(t_0)) = \widehat{\theta}_{j0} + \widehat{\alpha}_{jk, t_0}+ \widehat{H}_{jk}(\widehat{x}(t_0)), \quad j = 1, \ldots, p.
\end{equation}
We comment that, due to the Taylor approximation and localization, our localized kernel ODE estimator in \eqref{eqn:estnonadditivemodel} is different from the kernel ODE estimator of \citet{dai2021kernel} obtained from \eqref{eqn:kode2step-Ftilde}. In Section \ref{sec:estimation-rate}, we study its convergence rate, and compare it to the minimax optimal rate of the kernel ODE estimator of \citet{dai2021kernel}.

%%%%%%%%%%%%%%%%%%%%%%%%%%%%%%%%%%%%%%%%%%%%%%%%%%%
 \subsection{Optimization Algorithm}
 \label{sec:compalgo}

\noindent
We next develop an optimization algorithm to solve \eqref{eqn:DKL}. {Toward that end, we first propose an optimization problem that is equivalent to \eqref{eqn:DKL} but is computationally easier to tackle. We then develop an iterative algorithm to solve this equivalent optimization problem. We also remark that, the new algorithm differs from that of \citet{dai2021kernel}, in that a local weight $R_h$ is introduced, and the estimations of the parameter of interest $\alpha_{jk,t_0}$ and the rest of nuisance parameters are separated.} 

Specifically, we consider the following optimization problem that is equivalent to \eqref{eqn:DKL}, 
\begin{align} \label{eqn:kodeequiv}
\begin{split}
\underset{\theta_j,\alpha_{jk, t_0},H_{jk}}{\min}\left\{\frac{1}{n}\sum_{i=1}^n  R_h\left(t_i-t_0\right)\left[(y_{ij} - \bar{y}_j) - \alpha_{jk, t_0} \bar{t}_i -\int_{\Tcal}\left( T_i(t)-\bar{T}(t) \right) H_{jk}(\widehat{x}(t))dt\right]^2 \right. \\
+ \; \eta_{nj}\left( \sum_{k'=1, k'\neq k}^p \theta_{jk'}^{-1} \|\Pcal^{k'} H_{jk}\|^2_{\Hcal} + \sum_{k'=1, k'\neq k}^p \sum_{l=1,  l\neq k', k}^p \theta_{jk'l}^{-1}  \|\Pcal^{k'l} H_{jk}\|^2_\Hcal \right) \\
\left. + \kappa_{nj} \left( \sum_{k'=1, k'\neq k}^p\theta_{jk'} +  \sum_{k'=1, k'\neq k}^p \sum_{l=1,  l\neq k', k}^p \theta_{jk'l} \right)\right\}, 
\end{split}
\end{align}
subject to $\theta_{j}\geq 0$, where $\theta_j = \{\theta_{jk'}\}_{k' \neq k}\cup\{\theta_{jk'l}\}_{k'\neq k; l\neq k',k} \in \Rbb^{(p-1)^2}$ collects all the parameters to estimate, and $\eta_{nj}, \kappa_{nj} \geq 0$ are the tuning parameter, $j = 1, \ldots, p$.  The optimization problem \eqref{eqn:kodeequiv} utilizes the parameter $\theta_{jk'}$ to control the sparsity of each component $\Pcal^{k'} H_{jk}$, and $\theta_{jk'l}$ to control the sparsity of $\Pcal^{k'l} H_{jk}$, $k'\neq k$ and $l\neq k',k$. The shrinkage of $\theta_j$ gives rise to zero function components in the final estimate.   The two optimizations \eqref{eqn:DKL} and \eqref{eqn:kodeequiv} are equivalent in the sense that, if $( \widehat{\alpha}_{jk,t_0},\widehat{H}_{jk})$ minimizes \eqref{eqn:DKL}, then $(\widehat{\theta}_{j},\widehat{\alpha}_{jk,t_0}, \widehat{H}_{jk})$ minimizes \eqref{eqn:kodeequiv}, with $\widehat{\theta}_{jk'} = \eta_{nj}^{1/2}\kappa_{nj}^{-1/2}\|\Pcal^{k'}\widehat{H}_{jk}\|_\Hcal$, and $\widehat{\theta}_{jk'l}= \eta_{nj}^{1/2}\kappa_{nj}^{-1/2}\|\Pcal^{k'l}\widehat{H}_{jk}\|_\Hcal$, $k',l = 1, \ldots, p, k',l \neq k, k' \neq l$. Meanwhile, if $(\widehat{\theta}_{j},\widehat{\alpha}_{jk,t_0}, \widehat{H}_{jk})$  minimizes \eqref{eqn:kodeequiv}, then $( \widehat{\alpha}_{jk,t_0},\widehat{H}_{jk})$ minimizes \eqref{eqn:DKL}. {The reason of introducing $\theta_j$ is to benefit the computation. The optimization in \eqref{eqn:DKL} is challenging. By contrast, we develop an iterative procedure to solve the equivalent optimization problem \eqref{eqn:kodeequiv}, where each iteration either has a closed-form solution, or becomes a standard Lasso regression. As such, the computation is greatly simplified. That is, we employ the representer theorem to obtain a closed-form estimate for $(\alpha_{jk,t_0}, H_{jk})$ given $\theta_j$. We then employ the Lasso method to obtain a sparse estimate of $\theta_j$ given $(\alpha_{jk,t_0}, H_{jk})$.} 

Specifically, for a given estimate $\widehat\theta_j$, the optimization problem \eqref{eqn:kodeequiv} becomes, 
\begin{equation} \label{eqn:solveF}
\begin{aligned}
\underset{\alpha_{jk, t_0},H_{jk}}{\min} \left\{\frac{1}{n}\sum_{i=1}^nR_h\left(t_i-t_0\right)\left[(y_{ij} - \bar{y}_j) - \alpha_{jk, t_0} \bar{t}_i -\int_{\Tcal}\left( T_i(t)-\bar{T}(t) \right) H_{jk}(\widehat{x}(t))dt\right]^2\right. \\
\left. + \; \eta_{nj} \left(\sum_{k'=1, k'\neq k}^p \widehat\theta_{jk'}^{-1} \|\Pcal^{k'} H_{jk}\|^2_{\Hcal} +\sum_{k'=1, k'\neq k}^p \sum_{l=1, l \neq k',k}^p \widehat\theta_{jk'l}^{-1}  \|\Pcal^{k'l} H_{jk}\|^2_\Hcal  \right) \right\}.
\end{aligned}
\end{equation}
Let $K_k'(\cdot,\cdot):\Xcal\times\Xcal\mapsto\Rbb$ denote the kernel generating the RKHS $\Hcal_k'$, $k'\neq k$. Then $K_{k'l} = K_k'K_l$ is the reproducing kernel of the RKHS $\Hcal_k' \otimes \Hcal_l$ \citep{Aronszajn1950}. {Let $K_{\theta_j} = \sum_{k'=1, k'\neq k}^p \widehat\theta_{jk'} K_k' + \sum_{k'=1, k'\neq k}^p\sum_{l=1, l \neq k',k}^p \widehat\theta_{jk'l} K_{k'l}$; see also \citet[Section 2]{Wahba1995} for a discussion on the weighted kernel}. {Employing the representer theorem of \citet[Theorem 1.3.1]{wahba1990}, the solution $\widehat{H}_{jk}$ to \eqref{eqn:solveF} is of the form,
\begin{equation} \label{eqn:hatfjxt}
\widehat{H}_{jk}(\widehat{x}(t)) = \sum_{i=1}^nc_{ij}\int_\Tcal K_{\theta_j}\left( \widehat{x}(t),\widehat{x}(s) \right) \left( T_i(s)-\bar{T}(s) \right) ds
\end{equation}
for some $c_j = (c_{1j}, \ldots, c_{nj})^\top \in \Rbb^n$.} Define two $n \times n$ matrices, 
\begin{align} \label{eqn:2matrices}
\begin{split}
\Sigma & =  (\Sigma_{ii'}) \in \Rbb^{n\times n}, \Sigma_{ii'} = \int_\Tcal\int_\Tcal \{T_i(s)-\bar{T}(s)\} K_{\theta_j}(\widehat{x}(t),\widehat{x}(s)) \{T_{i'}(t)-\bar{T}(t)\} ds dt, \\
R_{t_0} & = \text{diag}\{R_h(t_1-t_0),\ldots,R_h(t_n-t_0)\} \in \Rbb^{n\times n}.
\end{split} 
\end{align}
Write $y_j = (y_{1j},\ldots,y_{nj})^\top \in \Rbb^n$. Plugging \eqref{eqn:hatfjxt} into \eqref{eqn:solveF}, we reach the following weighted quadratic minimization problem in terms of $\{\alpha_{jk,t_0}, c_j\}$, 
\begin{equation} \label{eqn:weightedlsforh}
\underset{\alpha_{jk,t_0}, c_j}{\min}
\left\{\frac{1}{n}[(y_j - \bar{y}_j) -\alpha_{jk,t_0}\bar{t} -  \Sigma c_j]^\top R_{t_0} [(y_j - \bar{y}_j) -\alpha_j\bar{t}-  \Sigma c_j]+ \eta_{nj} c_j^\top \Sigma c_j\right\},
\end{equation}
where $\bar{t} = (\bar{t}_1,\ldots,\bar{t}_n)^\top\in\Rbb^n$. The optimization problem \eqref{eqn:weightedlsforh} has a closed-form solution; see also \citet{dai2021kernel}. We tune the parameter $\eta_{nj}\geq 0$ using generalized cross-validation, following \citet{wahba1990}.

Next, for a given estimate $(\widehat\alpha_{jk,t_0}, \widehat{H}_{jk})$, the optimization problem \eqref{eqn:kodeequiv} becomes, 
\begin{equation} \label{eqn:regfortheta}
\begin{aligned}
\min_{\theta_j}\left\{\frac{1}{n}(z_j - G\theta_j)^\top R_{t_0} (z_j - G\theta_j) + \kappa_{nj} \left( \sum_{k'=1, k'\neq k}^p\theta_{jk'}+\sum_{k'=1, k'\neq k}^p\sum_{l=1,  l\neq k', k}^{p}\theta_{jk'l} \right)\right\},
\end{aligned}
\end{equation}
subject to $\theta_{jk'}\geq 0,\theta_{jk'l} \geq 0$, where the ``response" is $z_j = (y_j - \bar{y}_j) - \widehat{\alpha}_{jk,t_0}\bar{t} - (1/2)n\eta_{nj} c_j$, the ``predictor" is $G \in \Rbb^{n \times (p-1)^2}$, whose first $(p-1)$ columns are $\Sigma^{k'} c_j$ with $k'=1,\ldots,k-1,k+1,\ldots,p$, and the last $(p-1)(p-2)$ columns are $\Sigma^{lr} c_j$ with $k',l=1,\ldots,k-1,k+1,\ldots,p, k' \neq l$, and $\Sigma^{k'} = (\Sigma^{k'}_{ii'}), \Sigma^{k'l} = (\Sigma^{k'l}_{ii'})$ are both $n\times n$ matrices whose $(i,i')$th entries are $\Sigma^{k'}_{ii'} = \int_\Tcal\int_\Tcal\{ T_i(s)-\bar{T}(s) \} K_{k'}(\widehat{x}(t),\widehat{x}(s)) \{ T_{i'}(t)-\bar{T}(t) \} dsdt$, and $\Sigma^{k'l}_{ii'} = \int_\Tcal\int_\Tcal\{ T_i(s)-\bar{T}(s) \} K_{k'l}(\widehat{x}(t),\widehat{x}(s)) \{ T_{i'}(t)-\bar{T}(t) \} dsdt$, respectively, where $i, i'=1,\ldots,n, j=1,\ldots,p$. We employ the standard Lasso for \eqref{eqn:regfortheta} in our implementation, and tune the parameter $\kappa_{nj}$ using tenfold cross-validation, following the usual Lasso literature. 

We repeat the above optimization steps iteratively until some stopping criterion is met, e.g., when the estimates in two consecutive iterations are close enough, or when the number of iterations reaches some maximum number. We summarize the above iterative procedure, along with the confidence band derived in the next section, in  Algorithm \ref{alg:trainofkode}.

\begin{algorithm}[t!]
\caption{Estimation and inference procedure for a given pair $(j,k) \in \{1, \ldots, p\}$.} 
\begin{algorithmic}[1]
\STATE Initialization with the values for $\theta_{jk'}=\theta_{jk'l}=1$, for $k',l=1,\ldots,p, k' \neq k, l\neq k',k$.
\STATE Obtain the smoothing spline estimate $\widehat{x}_j(t)$ from \eqref{eqn:sshatxj}.
\REPEAT  
\STATE Solve $(\widehat{\alpha}_{jk,t_0},\widehat{H}_{jk})$ in \eqref{eqn:solveF} given $\widehat\theta_j$ through \eqref{eqn:hatfjxt} and \eqref{eqn:weightedlsforh}.
\STATE Solve $\widehat{\theta}_j$ in \eqref{eqn:regfortheta} given $(\widehat{\alpha}_{jk,t_0}, \widehat{H}_{jk})$ through the Lasso penalized regression \eqref{eqn:regfortheta}. 
\UNTIL{the stopping criterion is met.}
\STATE Construct the confidence band by Gaussian multiplier bootstrap from \eqref{eqn:debiasedcb}.
\end{algorithmic} 
\label{alg:trainofkode}
\end{algorithm}

We remark on the computational complexity of the proposed estimation method. Specifically, the computational cost is $O(n^3+np^4)$, where $O(n^3)$ is due to the reproducing kernel-type regression in \eqref{eqn:solveF}, and $O(np^4)$ is due to the Lasso-type regression in \eqref{eqn:regfortheta} with $O(p^2)$ parameters. We note that this computational complexity is comparable to the ones of existing ODE estimation methods such as \citet{dai2021kernel} and \citet{Chen2017}. Moreover, the computational complexity of the inference method we develop in \eqref{eqn:debiasedcb} is $O(n^2)$, which is substantially smaller than that for the estimation method when $n\to\infty$.

%%%%%%%%%%%%%%%%%%%%%%%%%%%%%%%%%%%%%%%%%%%%%%%%%%%
\section{ODE Inference}
\label{sec:inference}

\noindent 
In this section, we construct the confidence band for the regulatory effect $F_{jk}(x_k(t))$  of $x_k(t)$ on $x_j(t)$ for any given pair $(j, k) \in \{1, \ldots, p\}$. {Toward that end, we first construct a de-biased estimator for $\widehat{\alpha}_{jk,t}$, since $\widehat{\alpha}_{jk,t}$ is obtained through the regularization in \eqref{eqn:DKL}, which has an $\ell_1$-type penalty with the sum of RKHS norms and inevitably introduces bias  \citep{Zhang2014CI, Javanmard2014, Vandegeer2014}. 
Then based on our de-biased estimator, we employ the Gaussian multiplier process to construct a valid confidence band.}  We also discuss hypothesis testing for a single pair of variables, then multiple testing for all pairs of variables to reconstruct the entire regulatory system. Finally, we briefly discuss the extension from a single experiment to multiple experiments.

%%%%%%%%%%%%%%%%%%%%%%%%%%%%%%%%%%%%%%%%%%%%%%%%%%%
\subsection{Confidence Band}
\label{sec:band}

\noindent
As the first step, we propose the following de-biased estimator, given that the functional $F_{jk}$  is approximated by $\widehat{\alpha}_{jk, t_0}$ in our localized kernel learning, to reduce the bias in the estimate $\widehat{\alpha}_{jk, t_0}$,
\begin{equation} \label{eqn:debiasedestimator}
\widehat{F}_{jk}(x_k(t_0)) = \widehat{\alpha}_{jk,t_0} + \frac{1}{n} \sum_{i=1}^n \Sigma_{i\cdot} R_{t_0} \left[(y_j - \bar{y}_j) -  \int_{\Tcal}\left( T_i(t)-\bar{T}(t) \right) \widehat{H}_{jk}(\widehat{x}(t))dt\right],
\end{equation}
where $\Sigma_{i\cdot}, i=1,\ldots,n$, is the $k$th row of the $n\times n$ matrix $\Sigma$ defined in \eqref{eqn:2matrices}, and $\widehat{H}_{jk}(\widehat{x}(t))$ is obtained from \eqref{eqn:hatfjxt}. We make a few remarks. First of all, we employ the integral of the infinite-dimensional functional $\widehat{H}_{jk}(\widehat{x}(t))$ in \eqref{eqn:debiasedestimator}  to correct the bias in $\widehat{\alpha}_{jk,t_0}$. As a result, the inference of the de-biased estimator $\widehat{F}_{jk}(x_k(t_0))$ relies on analyzing the distribution of the integral, $\int_{\Tcal}\{ T_i(t)-\bar{T}(t) \} \widehat{H}_{jk}(\widehat{x}(t))dt$, and the measurement error introduced by the estimated trajectory $\widehat{x}(t)$. These features clearly differentiate our de-biasing solution from the existing ones. 
Second, we note that a similar approach to constructing a de-biased solution involving an infinite dimensional object  has been studied by \citet{lu2020kernel} in a different context.
Third, we briefly examine an alternative de-biased estimator that uses the derivative instead of the integral in \eqref{eqn:debiasedestimator}. The convergence of this alternative de-biased estimator hinges on the estimation error of the derivative term, $\mathbb E\int_{\Tcal}\{d\widehat{x}_j(t)/dt-dx_j(t)/dt\}^2dt$, which has a slower convergence rate than its integral counterpart \citep{Chen2017, dai2021kernel}. As such, it is to have inferior inference properties, and we choose to use the integral instead of the derivative in our de-biased estimator  \eqref{eqn:debiasedestimator}. 

Next, we obtain the critical value for the confidence band based on the de-biased estimator \eqref{eqn:debiasedestimator} using Gaussian multiplier bootstrap. Specifically, we consider the distribution of the supremum of the empirical process, $\sup_{t_0\in\Tcal}\Hbb_n(t_0)$, where 
\begin{equation*}
\Hbb_n(t_0)  \equiv \sqrt{nh}\left[ \widehat{F}_{jk}(x_k(t_0))- F_{jk}(x_k(t_0)) \right], \;\; \textrm{ for any } t_0\in\Tcal.
\end{equation*} 
Recognizing that the finite sample distribution of $\Hbb_n(t_0)$ is unknown, we approximate the distribution of $\Hbb_n(t_0)$ by the Gaussian multiplier process \citep{chernozhukov2014anti},
\begin{equation*}
\widehat{\Hbb}_n(t_0) \equiv \frac{1}{\sqrt{nh^{-1}}}\sum_{i=1}^n\xi_i\cdot\frac{\widehat{\sigma}_jR_h(t_{i}-t_0) R_{t_0,i\cdot}^\top \Sigma_{k\cdot}}{\widehat{\sigma}_n(t_0)}, \;\; \textrm{ for any } t_0\in\Tcal,
\end{equation*}
where $\xi_1,\ldots,\xi_n$ are independent standard normal random variables, the error variance estimator $\widehat{\sigma}_j^2 = \| A_{j}(y_j - \bar{y}_j) - (y_j - \bar{y}_j) \|^2 / \textrm{trace}(I_{n \times n} - A_{j})$, with $A_{j} = I_{n\times n} -n\eta_{nj}M_{t_0}^{-1}\big[ I_{n\times n}-\bar{t}\left(\bar{t}^\top M_{t_0}^{-1}\bar{t}\right)^{-1}\bar{t}^\top M_{t_0}^{-1} \big] \in\Rbb^{n\times n}$ being the smoothing matrix, and $M_{t_0} = \Sigma R_{t_0}^{-1}+n\eta_{nj}R_{t_0}^{-2}\in\Rbb^{n\times n}$, and $\widehat{\sigma}^2_n(t_0) = n^{-1}\Sigma_{k\cdot} R^2_{t_0}\Sigma_{k\cdot}$,  and $R_{t_0, i\cdot}, i=1,\ldots,n$, being the $i$th row of the $n\times n$ matrix $R_{t_0}$ defined in \eqref{eqn:2matrices}. We then compute the critical value $\widehat{c}_n(\alpha)$ as the $(1-\alpha)$-quantile of the supremum of the empirical process $\sup_{t_0\in\Tcal} \widehat{\Hbb}_n(t_0)$  given the observed data \citep{gine1990bootstrapping, chernozhukov2014anti}. 

Finally, we construct the $100\times(1-\alpha)\%$ confidence band for $F_{jk}(x_k(t))$ based on the de-biased estimator $\widehat{F}_{jk}(x_k(t_0))$ in \eqref{eqn:debiasedestimator}. The de-biasing step essentially removes the impact of regularization bias on the estimation of the individual regulatory effect, which in turn guarantees the validity of the confidence band. We also remark that our proposed de-biasing is built upon but also considerably extends the existing de-biasing methods in high-dimensional inference such as \citet{Zhang2014CI, Vandegeer2014}, because our setting is more challenging, and it involves the dynamic ODE system with an infinite-dimensional functional object as well as additional measurement error. Specifically, we construct the $100\times(1-\alpha)\%$ confidence band as,
\begin{equation} \label{eqn:debiasedcb}
\Ccal_{n,\alpha} = \left.\left\{  \left[\widehat{F}_{jk}(x_k(t_0)) - \frac{\widehat{c}_n(\alpha)\widehat{\sigma}_n(t_0)}{\sqrt{nh}}, \;\; \widehat{F}_{jk}(x_k(t_0)) + \frac{\widehat{c}_n(\alpha)\widehat{\sigma}_n(t_0)}{\sqrt{nh}}\right] \ \right| \ t_0 \in \Tcal \right\},
\end{equation} 

Given the confidence band in \eqref{eqn:debiasedcb}, we can perform hypothesis testing for any given pair $(j,k) \in \{1, \ldots, p\}$ and any $t_0 \in \Tcal$ that, 
\begin{align*}
& H_{0,jk} : x_k(t_0) \text{ has no regulatory effect on }x_j(t_0) \\ 
& H_{1,jk} : x_k(t_0) \text{ has nonzero regulatory effect on }x_j(t_0). 
\end{align*}
We use the standardized regulatory effect as the test statistic, 
\begin{align*}
z_{jk}(t_0) = \frac{\widehat{F}_{jk}(x_k(t_0))\sqrt{nh}}{\widehat{\sigma}_n(t_0)}, \;\; \textrm{ for any }  t_0\in\Tcal, 
\end{align*}
which follows the asymptotic distribution $\Hbb_n(t_0)$ under the null hypothesis. In practice, we apply this test to a set of grid points $t_0 \in \Tcal$, and we reject the null that $x_k(t)$ has no regulatory effect on $x_j(t)$, if any single test rejects the null at a given $t_0$.

Moreover, the confidence band \eqref{eqn:debiasedcb} is for the inference of the regulatory effect of $x_k(t)$ on $x_j(t)$ for a given pair $(j, k)$. We can easily couple it with existing multiple testing procedure for \emph{all} pairs of $(j, k)$ to recover the entire regulatory system, e.g., the Benjamini–Hochberg procedure, while controlling the FDR \citep{benjamini1995controlling}.

%%%%%%%%%%%%%%%%%%%%%%%%%%%%%%%%%%%%%%%%%%%%%%%%%%%
\subsection{Extension to Multiple Experiments}
\label{sec:multiexperiment}

\noindent
The localized kernel learning method we have developed so far focuses on a single experiment. Meanwhile, it can be easily generalized to incorporate multiple experiments. Specifically, let $\big\{ y_{ij}^{(s)};i=1,\ldots,n, j=1,\ldots,p, s=1,\ldots,S \big\}$ denote the observed data from $n$ subjects for $p$ variables under $S$ experiments, with unknown initial conditions $x^{(s)}(0) \in \Rbb^p, s=1,\ldots,S$. Then we modify the localized kernel learning in \eqref{eqn:DKL}, by seeking $\widehat{H}_{jk}\in\Hcal$ and $\widehat{\alpha}_{jk, t_0}\in\Rbb$ that minimize 
\begin{eqnarray*} \label{eqn:multiplelocal}
 \frac{1}{Sn}\sum_{s=1}^S\sum_{i=1}^nR_h\left(t_i-t_0\right)\left[\left(y^{(s)}_{ij} - \bar{y}^{(s)}_j\right) - \alpha_{jk, t_0} \bar{t}_i -\int_{\Tcal}\left( T_i(t)-\bar{T}(t) \right) H_{jk}(\widehat{x}^{(s)}(t))dt\right]^2 \\
+ \; \tau_{nj} \left( \sum_{k'=1, k'\neq k}^p\|\Pcal^l H_{jk}\|_{\Hcal} + \sum_{k'=1, k'\neq k,l}^p \sum_{l=1, l \neq k}^p \|\Pcal^{k'l} H_{jk}\|_\Hcal \right), 
\end{eqnarray*}
where $\widehat{x}^{(s)}(t) = (\widehat{x}^{(s)}_1(t), \ldots, \widehat{x}^{(s)}_p(t))^\top$ is the smoothing spline estimate obtained by, 
\begin{equation*}
\widehat{x}^{(s)}_j(t) = \underset{z_j\in\mathcal F}{\arg\min} \left\{\frac{1}{n} \sum_{i=1}^n (y^{(s)}_{ij} - z_{j}(t_i))^2 + \lambda_{nj}\|z_j(t)\|_{\mathcal F}^2\right\}, \quad j=1,\ldots,p,\ s=1,\ldots,S.
\end{equation*}
The de-biased estimator in \eqref{eqn:debiasedestimator} becomes, 
\begin{equation*}
\widehat{F}_{jk}(x_k(t_0)) = \widehat{\alpha}_{jk,t_0} + \frac{1}{Sn}\sum_{s=1}^S\sum_{i=1}^n \Sigma^{(s)}_{i\cdot} R_{t_0}  \left[\left(y^{(s)}_j - \bar{y}^{(s)}_j\right) -  \int_{\Tcal}\left\{ T_i(t)-\bar{T}(t) \right\} \widehat{H}_{jk}(\widehat{x}^{(s)}(t))dt\right].
\end{equation*}
The rest of Algorithm \ref{alg:trainofkode} for estimation and inference remains largely the same.

%%%%%%%%%%%%%%%%%%%%%%%%%%%%%%%%%%%%%%%%%%%%%%%%%%%
\section{Theoretical Properties}
\label{sec:theory}

\noindent
In this section, {we first establish the convergence rate for the localized kernel ODE estimator, which characterizes its estimation accuracy and is needed for establishing the properties of the subsequent inference. We then establish the asymptotic validity in terms of the coverage property for both the constructed confidence band and the recovered regulatory system.} Our theoretical results hold for both the low-dimensional setting and the high-dimensional setting, where the number of variables $p$ can be smaller or larger than the sample size $n$. We also study the regime-switching phenomenon.

%%%%%%%%%%%%%%%%%%%%%%%%%%%%%%%%%%%%%%%%%%%%%%%%%%%
\subsection{Statistical Convergence Rate}
\label{sec:estimation-rate}

\noindent 
We begin with three regularity conditions. 

\begin{assumption}
\label{assump:kerneldensity}
The kernel density $R(t)$ is a continuous function that has a bounded support, and satisfies that $\int_\Tcal R(t)dt=1$ and $\int_\Tcal tR(t)dt=0$. 
\end{assumption}

\begin{assumption}
\label{assump:complexity}
The number of nonzero functional components, $ \text{card}\big( \{k': F_{jk'}\neq 0\}\cup\{1\le k'\neq l\le p: F_{jk'l}\neq 0\} \big)$, is bounded, for any $j=1,\ldots,p$.
\end{assumption}

\begin{assumption}
\label{assump:fluctuation}
For any $F_j\in\Hcal$, there exists a random variable $B$, with $\mathbb E(B)<\infty$, and $\left|{\partial F_j(x)} / {\partial x_k}\right| \leq B \|F_j\|_{L_2}$ almost surely.
\end{assumption}

\noindent 
Assumption \ref{assump:kerneldensity} is standard in the local polynomial regression literature \citep{fan1996local}. Assumption \ref{assump:complexity} regards the complexity of the functionals. Similar assumptions have been adopted in the sparse additive model over RKHS without interactions \citep{Koltchinskii2010, Raskutti2011}. Assumption \ref{assump:fluctuation} is an inverse Poincar\'e inequality type condition, which places a regularization on the fluctuation in $F_j$ relative to the $L_2$-norm. The same assumption has also been used in additive models in RKHS \citep{Zhu2014,dai2021kernel}.

{Recall $F_j(x(t))$ represents the true functional in \eqref{eqn:nonadditivemodel}, and $\widehat{F}_{j}(\widehat{x}(t))$ the proposed localized kernel estimator in \eqref{eqn:estnonadditivemodel}.} The next theorem obtains the rate of convergence of $\widehat{F}_{j}(\widehat{x}(t))$. 

\begin{theorem}
\label{thm:optimalestoffunctional}
Suppose Assumptions \ref{assump:kerneldensity} to \ref{assump:fluctuation} hold. Suppose $x_j(t) \in \Fcal$, and the RKHS $\Fcal$ is embedded to a $\beta_1$th-order Sobolev space with $\beta_1 > 1/2$. Suppose $F_j \in \Hcal$, where $\Hcal$ satisfies \eqref{eqn:spaceH}, and the RKHS $\Hcal_k$ in \eqref{eqn:spaceH} is embedded to a $\beta_2$th-order Sobolev space with $\beta_2>1$, $j,k = 1, \ldots, p$. Then, the localized kernel ODE estimator $\widehat{F}_{j}(\widehat{x}(t))$ in \eqref{eqn:estnonadditivemodel} satisfies that, 
\begin{equation}
\label{eqn:termsoferror}
\begin{aligned}
\min_{\lambda_{nj},\tau_{nj}\geq 0}\int_\Tcal \left[ \widehat{F}_{j}(\widehat{x}(t))-F_{j}(x(t)) \right]^2dt  
= O_p\left(  \left( \frac{nh}{\log n} \right)^{-\frac{2\beta_2}{2\beta2+1}} + h^{2\beta_2}+ \frac{\log p}{n} + n^{-\frac{2\beta_1}{2\beta_1+1}} \right).
\end{aligned}
\end{equation}
Furthermore, if $h=O\left((n/\log n)^{-1/(2\beta_2+2)}\right)$, then the localized kernel ODE estimator in \eqref{eqn:estnonadditivemodel} satisfies that,
\begin{equation}
\label{eqn:optimalrateest}
\begin{aligned}
\min_{\lambda_{nj},\tau_{nj}\geq 0}\int_\Tcal \left[ \widehat{F}_{j}(\widehat{x}(t))-F_{j}(x(t)) \right]^2dt = O_p\left( \left(\frac{n}{\log n}\right)^{-\frac{\beta_2}{\beta_2+1}}+\frac{\log p}{n} + n^{-\frac{2\beta_1}{2\beta_1+1}} \right).
\end{aligned}
\end{equation}
\end{theorem}

{We first note that the convergence rate in \eqref{eqn:termsoferror} is established for the estimator $\widehat{F}_{j}(\widehat{x}(t))$ of the entire regulatory effect $F_j(x(t))$. We can further obtain the convergence rate for the estimators of the individual effect and nuisance parameter $(\alpha_{jk,t}, H_{jk})$, which turns to be the same as the rate in \eqref{eqn:termsoferror}. Specifically, for the estimators $(\widehat{\alpha}_{jk, t}, \widehat{H}_{jk})$ from \eqref{eqn:DKL}, we have, 
\begin{equation*}
\begin{aligned}
\min_{\lambda_{nj},\tau_{nj}\geq 0}\int_\Tcal \left[ \widehat{\alpha}_{jk,t} - F_{jk}(x(t)) \right]^2dt  
& = O_p\left(  \left( \frac{nh}{\log n} \right)^{-\frac{2\beta_2}{2\beta2+1}} + h^{2\beta_2}+ \frac{\log p}{n} + n^{-\frac{2\beta_1}{2\beta_1+1}} \right),\\
\min_{\lambda_{nj},\tau_{nj}\geq 0}\int_\Tcal \left[ \widehat{H}_{jk}(\widehat{x}(t))-H_{jk}(x(t)) \right]^2dt  
& = O_p\left(  \left( \frac{nh}{\log n} \right)^{-\frac{2\beta_2}{2\beta2+1}}  + h^{2\beta_2}+ \frac{\log p}{n} + n^{-\frac{2\beta_1}{2\beta_1+1}} \right).
\end{aligned}
\end{equation*}
}

We next examine the sources of the estimation error in \eqref{eqn:termsoferror}. There are totally four sources. Specifically, the first source of error $O_p((nh/\log n)^{-2\beta_2/(2\beta_2+1)})$ comes from the error in estimating the interaction terms in the true functional $F_j$. The second term $O_p(h^{2\beta_2})$ comes from the localized estimation. The third term $O_p(\log p/n)$ comes from the bias introduced by the Lasso estimation. The last term $O_p(n^{-2\beta_1/(2\beta_1+1)})$ comes from the measurement errors in $x(t)$.   

We also observe some interesting regime-switching phenomenon in \eqref{eqn:optimalrateest}. That is, when $p$ is ultrahigh-dimensional, in that $p>\exp[\{n(\log n)^{\beta_2}\}^{1/(\beta_2+1)}]$, then the convergence rate in  \eqref{eqn:optimalrateest}  becomes $O_p(\log p/n+n^{-2\beta_1/(2\beta_1+1)})$. In this case, it matches with the minimax optimal rate for estimating the functional $F_j$ in \eqref{eqn:nonadditivemodel} obtained in \citet{dai2021kernel}. Henceforth, we pay no extra price in terms of the rate of convergence for adopting the localized estimation in this ultrahigh-dimensional setting. On the other hand, when $p$ is low-dimensional, in that $p<\exp[\{n(\log n)^{\beta_2}\}^{1/(\beta_2+1)}]$, then the convergence rate in  \eqref{eqn:optimalrateest} becomes $O_p((n/\log n)^{-\beta_2/(\beta_2+1)}+n^{-2\beta_1/(2\beta_1+1)})$. Here, the first term $O_p((n/\log n)^{-\beta_2/(\beta_2+1)})$ matches, up to some logarithmic factors, with the rate as if we knew a priori that $F_j$ is an additive model with pairwise interactions. Moreover, the rate  $O_p((n/\log n)^{-\beta_2/(\beta_2+1)})$ in our proposed method is slower than the optimal rate $O_p((n/\log n)^{-2\beta_2/(2\beta_2+1)})$ of the kernel ODE estimator of \citet{dai2021kernel}. Such a slower rate is due to the weight matrix $R_h$ utilized in the localized estimation in \eqref{eqn:DKL}, which increases the variance of estimating $F_j$ by $O_p(h^{-1})$, when compared to the non-localized estimation method in \citet{dai2021kernel}. Nevertheless, as we show next, this slower rate is sufficient to establish an asymptotically valid confidence band.

%%%%%%%%%%%%%%%%%%%%%%%%%%%%%%%%%%%%%%%%%%%%%%%%%%%
\subsection{Coverage Property}
\label{sec:theoryconfidence}

\noindent 
A confidence band $\Ccal_{n,\alpha}$ is said to be asymptotically valid with level $100\times(1-\alpha)\%$ for $F_{jk}$, if it satisfies that, for some constants $c,C>0$, 
\vspace{-0.01in}
\begin{equation} \label{eqn:defofah}
\mathbb P\left(F_{jk}(x_k(t_0)) \in \Ccal_{n,\alpha}, \ \forall t_0\in\Tcal\right)\geq 1-\alpha-Cn^{-c}. 
\end{equation}
The condition \eqref{eqn:defofah} implies that the confidence band $ \Ccal_{n,\alpha}$ has an asymptotic coverage probability of at least $1-\alpha$ for a given data generating process.

The next theorem establishes the theoretical property of our proposed confidence band $\Ccal_{n,\alpha}$ in \eqref{eqn:debiasedcb}. We introduce another regularity condition. For any $t\in\Tcal$, let $p_j$ denote the marginal density of $x_j(t)$, $p_{jk}$ denote the bivariate density of $(x_j(t),x_k(t))$, and $p_{jkl}$ denote the joint density of $(x_j(t),x_k(t),x_l(t))$, for $j, k, l = 1, \ldots, p$.  

\begin{assumption} 
\label{assump:weakdependency}
The density function of $x(t)$ satisfies that, 
\begin{equation*}
\sum_{j=1,j\neq k}^p\|p_{jk} - p_j p_k\|_2\leq \frac{\rho_{\min}}{2B}, \ \text{ and } \
\sup_{l\neq k} \sum_{j,k=1,j\neq k}^p\|p_{j,k,l}-p_jp_kp_l\|_2\leq \frac{\rho_{\min}}{2B}, 
\end{equation*}
for some constant $\rho_{\min}>0$, and $B$ is as defined in Assumption \ref{assump:fluctuation}.
\end{assumption}
\noindent 
Assumption \ref{assump:weakdependency}  quantifies how weak the dependency between the signal variables can be. Similar conditions have also been used in the inference of the high-dimensional linear regression \citep{Zhang2014CI, Vandegeer2014}, and the nonparametric additive regression \citep{lu2020kernel}. We show this condition is also sufficient for establishing the validity of the confidence band in \eqref{eqn:debiasedcb} in a complex ODE system.

\begin{theorem} 
\label{thm:postselection}
Suppose Assumptions \ref{assump:kerneldensity} to \ref{assump:weakdependency} hold. If $h=O(n^{-r})$,  for $r\in(1/5,3/13)$, then there exist constants $c,C_1>0$, such that, for any $\alpha\in(0,1)$, the confidence band $\Ccal_{n,\alpha}$ in \eqref{eqn:debiasedcb} is asymptotically valid.
\end{theorem}

Putting all the functionals $\{F_1, \ldots,$ $F_p\}$ together forms a network of regulatory relations among the $p$ variables $\{x_1(t),$ $\ldots, x_p(t)\}$. The next theorem shows that the estimated system based on our localized kernel ODE approaches the truth with probability tending to one. Denote the set of the true and the estimated regulators of $x_j(t)$ by
\begin{align*}
S^*_j & = \big\{1 \leq k \leq p : F_{jk}\neq 0, \textrm{ or } F_{jkl}\neq 0 \textrm{ for some } 1 \le l \neq k \le p \big\}, \\
\widehat{S}_j & = \big\{ 1 \leq k \leq p: \widehat{\alpha}_{jk,t_0}\neq 0\textrm{ for any }  t_0\in\Tcal \big\}.
\end{align*}

\noindent
We also need two additional regularity conditions. Let $s_j = \text{card}(S^*_j)$. Recall the definitions of $R_{t_0} \in \Rbb^{n\times n}$ in \eqref{eqn:2matrices}, $G \in \Rbb^{n \times (p-1)^2}$ in \eqref{eqn:regfortheta}, and the tuning parameter $\kappa_{nj}$ in \eqref{eqn:kodeequiv}. Let $G_{S^*_j} \in \Rbb^{n \times s_j}$ denote the sub-matrix of $G$ with the column indices in the set $S^*_j$.

\begin{assumption} 
\label{assump:dependency-incoherence}
Suppose there exists a constant $C_{\min}>0$, such that the minimal eigenvalue of the matrix $G_{S^*_j}^\top R_{t_0}G_{S^*_j}$ is no smaller than $C_{\min}/2$ as $n\to\infty$. In addition, suppose there exists a constant $0 \leq \xi_G < 1$, such that $\max_{(k,l)\not\in S^*_j}\left\|G_{kl}^\top R_{t_0}G_{S^*_j}(G_{S^*_j}^\top R_{t_0}G_{S^*_j})^{-1}\right\|_{\ell_2} \leq \xi_G$.
\end{assumption}

\begin{assumption} 
\label{assump:minregeffect}
Let $\theta_{\min} = \min_{(k,l)\in S_j}\|\theta_{jkl}\|_{L_2}$, and $\eta_{\mathcal R}$ is given in Lemma \ref{lem:conds3} in Section \ref{sec:auxlemnew} of the Appendix. Suppose the following inequalities hold: 
\begin{equation*}
\frac{\eta_{\mathcal R}\sqrt{s_j}}{C_{\min}}+n\kappa_{nj}\frac{\sqrt{s_j}}{C_{\min}}\leq \frac{2}{3}\theta_{\min}, \quad \textrm{ and } \quad 
\frac{(\xi_G+1)\sqrt{s_j}}{n\kappa_{nj}}\eta_{\mathcal R}+\xi_G\sqrt{s_j}<1.
\end{equation*}  
\end{assumption}

\noindent 
Assumption \ref{assump:dependency-incoherence} ensures the sub-matrix $G_{S^*_j}$ is not degenerated, and the irrelevant variables would not exert too strong an effect on the relevant variables. {It is similar to Assumptions 3 and 4 in \citet{dai2021kernel} except for the additional term $R_{t_0}$. It is interesting to note that the bandwidth $h$ in the localization and $R_{t_0}$ does not affect the validity of this assumption, as long as $h\to 0$ when $n\to\infty$. More discussion on this assumption is given in Section \ref{sec:assumption5} of the Appendix.}  Assumption \ref{assump:minregeffect} imposes some regularity on the minimum effect $\theta_{\min}$, and also characterizes the relationship among $\xi_G$, $\kappa_{nj}$, and $s_j$. Both assumptions are mild, and similar conditions as Assumptions \ref{assump:dependency-incoherence} and \ref{assump:minregeffect} have been commonly imposed in the literature \citep[see, e.g.,][]{Zhao2006, meinshausen2006high, Ravikumar2010, Chen2017, dai2021kernel}.

\begin{theorem}
\label{thm:optimalrecovery}
Suppose Assumptions \ref{assump:kerneldensity}, \ref{assump:complexity}, \ref{assump:dependency-incoherence} and \ref{assump:minregeffect} hold. Then, the localized kernel ODE estimator correctly recovers the true regulatory system, in that, 
\begin{eqnarray*}
\mathbb P\left( \widehat{S}_j = S^*_j \right) \to 1, \; \textrm{ as } \; n \to \infty, \; \textrm{ for all } j = 1, \ldots, p. 
\end{eqnarray*}
\end{theorem}

%%%%%%%%%%%%%%%%%%%%%%%%%%%%%%%%%%%%%%%%%%%%%%%%%%%
\section{Simulation Studies}
\label{sec:simulation}

\noindent
In this section, we study the finite-sample performance of the confidence band as well as the localized kernel ODE estimator using two well-known ODE systems. In the first example, we focus on the coverage of the confidence band for some \emph{given pairs}, while in the second example, we study the performance of recovery of the \emph{entire} regulatory system.

%%%%%%%%%%%%%%%%%%%%%%%%%%%%%%%%%%%%%%%%%%%%%%%%%%%
\subsection{Enzymatic Regulation Equations}
\label{sec:enzymatic}

\noindent 
The first example is a three-node enzyme regulatory system of a negative feedback loop with a buffering node \citep[NFBLB]{Ma2009}. The ODE system is given by, 
\begin{eqnarray} \label{eqn:NFBLB}
\begin{split}
\frac{dx_1(t)}{dt} & = & 10\frac{x_0 \{1-x_1(t)\}}{\{1-x_1(t)\}+0.1} - 10 \frac{x_1(t)}{x_1(t)+0.1},  \\
\frac{dx_2(t)}{dt} & = & 10\frac{\{1-x_2(t)\}  x_3(t)}{\{1-x_2(t)\}+0.1} - 0.2  \frac{x_2(t)}{x_2(t)+0.1},\\
\frac{dx_3(t)}{dt} & = & 10\frac{x_1(t) \{1-x_3(t)\}}{\{1-x_3(t)\}+0.1} - 10 \frac{x_2(t) x_3(t)}{x_3(t)+0.1}. 
\end{split}
\end{eqnarray}
The coefficient $x_0 \in \Rbb$ is drawn uniformly from $[0.5,1.5]$. The initial values are chosen as $(x_1(0),x_2(0),x_3(0)) = (0,0,0)$. The errors are drawn independently from Normal$(0,\sigma_j^2)$, with three noise levels, $\sigma_j \in \{0.1,0.3,0.5\}$. The time points are evenly distributed,  $t_i=(i-1)/20, i=1,\ldots,n$, with the sample size fixed at $n=40$. In this example, $p=3$, and there are $p^2 = 9$ functions in \eqref{eqn:NFBLB} to estimate for each $j=1,2,3$, and in total there are $27$ unknown functions. 

In this example, we focus on the performance of the confidence band for some given node pairs. Specifically, we examine a nonzero effect $F_{23}(x_3(t))$ that captures the regulatory effect of $x_3(t)$ on $x_2(t)$, and a zero effect $F_{12}(x_2(t))$ of $x_2(t)$ on $x_1(t)$. Correspondingly, we construct the confidence band for $F_{23}(x_3(t))$ and $F_{12}(x_2(t))$, with the $95\%$ significance level, on 500 evenly distributed grid points on $[0,1]$. In our implementation, we use a first-order Mat\'{e}rn kernel for both steps in \eqref{eqn:sshatxj} and \eqref{eqn:kode2step-Ftilde} of the collocation method, where $K_{\Hcal_1}(x,x') = (1+\sqrt{3}\|x-x'\|/\nu)\exp(-\sqrt{3}\|x-x'\|/\nu)$, and $\nu$ is chosen by tenfold cross-validation. We have found the inference results are not overly sensitive to the choice of kernel functions here. {Moreover, we use the quadratic density $R_h(t) = (15/16)\cdot(1-t^2/h^2)^2\mathbf 1(|t|<h)$ for the local weight function, where the bandwidth $h$ is chosen by tenfold cross-validation. We have carried out a sensitivity analysis in Section \ref{sec:sensitivity} of the Appendix, and show that the inference results are again not sensitive to the choice of the weight function or the bandwidth.}  We compute the quantile $\widehat{c}_n(\alpha)$ in \eqref{eqn:debiasedcb} by bootstrap with $500$ repetitions.  

We note that there is \emph{no} direct competitor to our confidence band solution in the literature. Alternatively, we compare to three commonly used ODE solutions, the linear ODE with interactions \citep{Zhang2015}, the additive ODE  \citep{Chen2017}, and the kernel ODE  \citep{dai2021kernel}, and couple them with a confidence band that  aggregates the point-wise confidence intervals at 500 grid points on $[0,1]$. For a fair comparison, we adjust the significance level at each of these 500 time points with the Bonferroni correction \citep{holm1979simple}, i.e., $(1 - \alpha / 500)\%$, where $\alpha = 0.05$.  

We consider two evaluation criteria. One is the empirical coverage probability of the confidence band, and the other is the area of the confidence band, defined as 
\vspace{-0.01in}
\begin{equation*}
\int_{t_0\in\Tcal}2\widehat{c}_n(\alpha)(nh)^{-1/2}\widehat{\sigma}_n(t_0)dt_0, 
\end{equation*}
where the integration is computed by discretizing the interval into $1000$ grids. A larger coverage probability and a smaller area indicates a better performance.

\begin{table}[t!]
\centering
\resizebox{\textwidth}{!}{
\begin{tabular}{cclcccccc}
\toprule
& &&& \multicolumn{2}{c}{Nonzero functional $F_{23}(x_3(t))$} & & \multicolumn{2}{c}{Zero functional $F_{12}(x_2(t))$} \\ \cline{5-6} \cline{8-9}
  & & && Coverage & Confidence  & & Coverage & Confidence  \\ 
 Noise level & & Method & & probability & band area & & probability & band area \\ [0.2ex]
 \midrule
$\sigma_j=0.1$ & & Linear ODE & &$0.212$ & $1.550$ & & $0.274$ & $2.145$ \\ [0.2ex]
 & & & &$(0.205,0.219)$ & $(1.514,1.586)$ & & $(0.266,0.282)$ & $(2.113,2.177)$ \\ [0.2ex]
 & &  Additive ODE && $0.224$ & $1.332$ & & $0.292$ & $2.008$  \\ [0.2ex]
 & &  && $(0.216,0.232)$ & $(1.290,1.374)$ & & $(0.285,0.299)$ & $(1.964,2.052)$  \\ [0.2ex]
 & &  Kernel ODE && $0.939$ & $1.270$ & & $0.928$ & $1.848$  \\ [0.2ex]
 & &   && $(0.926,0.952)$ & $(1.250,1.290)$ & & $(0.916,0.940)$ & $(1.824,1.872)$  \\ [0.2ex]
 & &  Localized kernel ODE && $\textbf{0.974}$ & $\textbf{0.102}$  & & $\textbf{0.957}$ & $\textbf{0.217}$   \\ [0.2ex]
 & &   && $(\textbf{0.968},\textbf{0.980})$ & $(\textbf{0.092},\textbf{0.112})$  & & $(\textbf{0.952},\textbf{0.962})$ & $(\textbf{0.210},\textbf{0.224})$   \\ [0.2ex]
$\sigma_j=0.3$ & & Linear ODE & &$0.178$ & $1.827$ & & $0.224$ & $2.441$ \\ [0.2ex]
 & &  & &$(0.167,0.189)$ & $(1.776,1.878)$ & & $(0.212,0.236)$ & $(2.382,2.500)$ \\ [0.2ex]
 & &  Additive ODE && $0.194$ & $1.663$ & & $0.262$ & $2.158$  \\ [0.2ex]
 & &   && $(0.184,0.204)$ & $(1.596,1.730)$ & & $(0.252,0.272)$ & $(2.089,2.227)$  \\ [0.2ex]
 & &  Kernel ODE && $0.914$ & $ 1.381$ & & $0.911$ & $1.856$  \\ [0.2ex]
 & &  && $(0.891,0.937)$ & $ (1.350,1.412)$ & & $(0.891,0.931)$ & $(1.821,1.891)$  \\ [0.2ex]
 & &  Localized kernel ODE && $\textbf{0.962}$ & $\textbf{0.163}$  & & $\textbf{0.957}$ & $\textbf{0.290}$   \\ [0.2ex]
 & &  && $(\textbf{0.952},\textbf{0.972})$ & $(\textbf{0.150},\textbf{0.176})$  & & $(\textbf{0.949},\textbf{0.965})$ & $(\textbf{0.281},\textbf{0.299})$   \\ [0.2ex]
$\sigma_j=0.5$ & & Linear ODE & &$0.123$ & $3.541$ & & $0.191$ & $2.538$ \\ [0.2ex]
 & & & &$(0.103,0.143)$ & $(3.456,3.626)$ & & $(0.172,0.210)$ & $(2.447,2.629)$ \\ [0.2ex]
 & &  Additive ODE && $0.141$ & $2.679$ & & $0.225$ & $2.350$  \\ [0.2ex]
 & &   && $(0.122,0.160)$ & $(2.586,2.772)$ & & $(0.205,0.245)$ & $(2.251,2.449)$  \\ [0.2ex]
 & &  Kernel ODE && $0.876$ & $1.637$ & & $0.902$ & $2.031$  \\ [0.2ex]
 & &   && $(0.843,0.909)$ & $(1.595,1.679)$ & & $(0.873,0.931)$ & $(1.980,2.082)$  \\ [0.2ex]
 & &  Localized kernel ODE && $\textbf{0.956}$ & $\textbf{0.231}$  & & $\textbf{0.948}$ & $\textbf{0.401}$   \\ [0.2ex]
 & & && $(\textbf{0.943},\textbf{0.969})$ & $(\textbf{0.216},\textbf{0.246})$  & & $(\textbf{0.937},\textbf{0.959})$ & $(\textbf{0.389},\textbf{0.413})$   \\ [0.2ex]
\bottomrule
\end{tabular}}
\caption{The NFBLB example: the empirical coverage probability and area of the confidence band, and their $95\%$ confidence intervals, for the varying noise level $\sigma_j$. The results are based on $500$ data replications.}
\label{table:enzymatic}
\end{table}

Table \ref{table:enzymatic} reports the results based on 500 data replications. We see that the proposed confidence band achieves the desired coverage. By contrast, the confidence bands of the additive and linear ODEs mostly fail to include the truth. This is because there is a discrepancy between the additive and linear ODE model specifications and the true ODE model in \eqref{eqn:NFBLB}, and this discrepancy accumulates as the course of the ODE evolves. Meanwhile, the kernel ODE has a much larger confidence band compared to our method. {This is because the Bonferroni correction makes the confidence band of kernel ODE overly conservative.} We report some additional results graphically in Section \ref{sec:add-ex1} of the Appendix.

%%%%%%%%%%%%%%%%%%%%%%%%%%%%%%%%%%%%%%%%%%%%%%%%%%%
\subsection{Lotka-Volterra Equations}
\label{sec:lotkavolterra}

\noindent
The second example is the classical Lotka-Volterra system, which consists of pairs of first-order nonlinear differential equations describing the dynamics of biological system in which predators and prey interact \citep{Volterra1928}. The ODE is given by,  
\begin{eqnarray} \label{eqn:lotkavolterra}
\begin{split}
\frac{dx_{2j-1}(t)}{dt} & = &  0.1(2j+11)x_{2j-1}(t)- 0.2(j+1) x_{2j-1}(t)x_{2j}(t),  \\
\frac{dx_{2j}(t)}{dt} & = &  0.1(2j-1) x_{2j-1}(t)x_{2j}(t)- 0.2(j+1)x_{2j}(t),
\end{split}
\end{eqnarray}
for $j=1,\ldots,5$. Here $dx_{2j-1}(t)/dt$ and $dx_{2j}(t)/dt$ are nonadditive functions of $x_{2j-1}$ and $x_{2j}$, where $x_{2j-1}$ is the prey and $x_{2j}$ is the predator. The initial values are set as $x_{2j-1}(0) = x_{2j}(0)$. The measurement errors are independent Normal$(0,\sigma^2_j)$, where the noise level $\sigma_j \in \{1,2,\ldots,10\}$. The time points are evenly distributed in $[0,100]$, with $n=200$. In this example, $p=10$, and there are $p^2 = 100$ functions in \eqref{eqn:lotkavolterra} to estimate for each ODE of $x_{2j-1}$ and $x_{2j}$, $j=1,\ldots,5$, and in total there are $1000$ unknown functions. 
 
In this example, we focus on the performance of recovery of the entire regulatory system through the proposed confidence band coupled with the Benjamini–Hochberg (BH) procedure for multiple testing correction \citep{benjamini1995controlling}. Since the ODE equations in \eqref{eqn:lotkavolterra} only involve the linear and interaction terms, we use the first-order Mat\'{e}rn kernel for the step in \eqref{eqn:sshatxj}, and use the linear kernel in \eqref{eqn:kode2step-Ftilde}. As such, the linear and kernel ODEs yield the same estimates. {We continue to use the quadratic density for the local weight function $R_h(t)$.} We control the FDR at the level of $20\%$. 

We consider three evaluation criteria, the false discovery proportion, the empirical power, and the trajectory prediction accuracy. The false discovery proportion is defined as the proportion of falsely selected edges in the system out of the total number of edges, and the empirical power is defined as the proportion of selected true edges in the system. We also evaluate the prediction accuracy of the entire regulatory effect $\widehat{F}_{j}(\widehat{x}(t))$ as given in \eqref{eqn:estnonadditivemodel}, by the squared root of the sum of predictive mean squared errors for $F_j(x_j(t)),j=1,\ldots,10$, at the unseen “future” time point $t\in[100,200]$, i.e., $\left\{\sum_{j=1}^{10}\int_{100}^{200}[\widehat{F}_j(\widehat{x}_j(t))-F_j(x_j(t))]^2dt\right\}^{1/2}$, where the integral is evaluated at $10000$ evenly distributed time points in $[100,200]$.  

\begin{figure}[t!]
\centering
\includegraphics[width=\textwidth, height=2.25in]{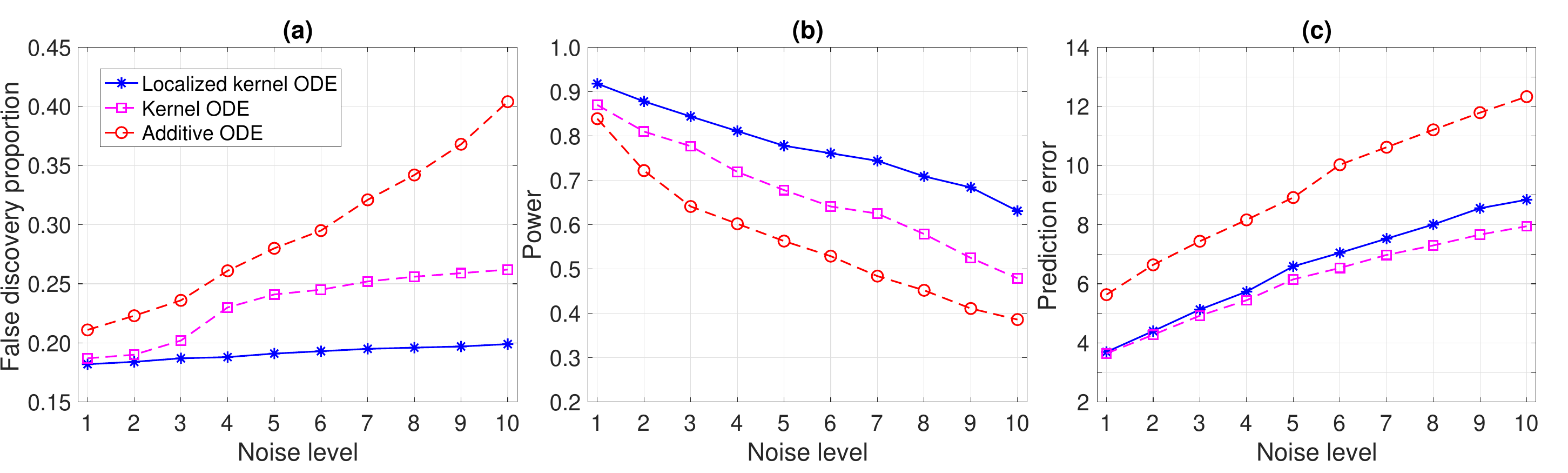}
\caption{The Lotka-Volterra example: the empirical FDR, power, and trajectory prediction error for the varying noise level $\sigma_j$. The results are averaged over $500$ data replications.}
\label{fig:lotkavolterra}
\vspace{-0.2in}
\end{figure}

Figure \ref{fig:lotkavolterra} reports the results averaged over 500 data replications. We see that our method successfully controls the FDR under the nominal level across all noise levels, whereas the additive and kernel ODEs both suffer some inflations, especially when the noise level is high. Meanwhile, our method achieves the best empirical power. In addition, we see that the prediction error of our localized kernel ODE estimator is slightly worse than that of kernel ODE, which agrees with Theorem \ref{thm:optimalestoffunctional}. Nevertheless, this does not affect the inference performance of our method.

%%%%%%%%%%%%%%%%%%%%%%%%%%%%%%%%%%%%%%%%%%%%%%%%%%%
\section{Data Applications}
\label{sec:application}

\noindent
In this section, we illustrate our method with two data applications, a gene regulatory network analysis given time-course gene expression data, and a brain effective connectivity analysis given electrocorticography (ECoG) data.

%%%%%%%%%%%%%%%%%%%%%%%%%%%%%%%%%%%%%%%%%%%%%%%%%%%
\subsection{Gene Regulatory Network}
\label{sec:generegulatory}

\noindent 
Gene regulation plays a central role in biological activities such as cell growth, development, and response to environmental stimulus \citep{peng2009partial, gonzalez2013inferring}. Thanks to the advancement of high-throughput DNA microarray technologies, it becomes feasible to measure the dynamic features of gene expression profiles on a genome scale. Such time-course gene expression data allow investigators to study gene regulatory networks, and ODE modeling is frequently employed for such a purpose \citep{LuLiang2011}. The data we analyze is the in silico benchmark gene expression data generated by GeneNetWeaver (GNW) using dynamical models of gene regulations and nonlinear ODEs \citep{Schaffter2011}. GNW extracts two regulatory networks of \emph{E.coli}, \emph{E.coli1}, \emph{E.coli2}, and three regulatory networks of yeast, \emph{yeast1}, \emph{yeast2}, \emph{yeast3}, each of which has two values of dimension, $p=10$ nodes and $p=100$ nodes. The system of ODEs for each extracted network is based on a thermodynamic approach, and the resulting ODE system is non-additive and nonlinear \citep{Marbach2010}. For the $10$-node network, GNW provides $S=4$ perturbation experiments, and for the 100-node network, GNW provides $S=46$ experiments. In each perturbation experiment, GNW generates the time-course data with different initial conditions of the ODE system to emulate the diversity of gene expression trajectories \citep{Marbach2009}. All the trajectories are measured at $n=21$ evenly spaced time points in $[0,1]$. We add independent measurement errors from Normal$(0, 0.025^2)$, which is the same as the DREAM3 competition and the data analysis in \citet{HendersonMichailidis2014}. 

\begin{table}[t!]
\centering
\resizebox{\textwidth}{!}{
\begin{tabular}{ccccccccccc}
\toprule
& & \multicolumn{4}{c}{$p=10$} & & \multicolumn{4}{c}{$p=100$} \\ \cline{3-6} \cline{8-11} 
& & Localized & Kernel & Additive & Linear & & Localized & Kernel & Additive & Linear \\ 
& & kernel ODE & ODE & ODE & ODE & & kernel ODE & ODE & ODE & ODE \\ [0.2ex]
 \midrule
\emph{E.coli1} &FDR& $\textbf{0.182}$ & $0.212$ & $0.241$ & $0.206$ && $\textbf{0.191}$ & $0.194$ & $0.231$ & $0.232$  \\
& & $(0.178, 0.186)$ & $(0.207, 0.217)$ & $(0.235, 0.247)$ & $(0.199, 0.213)$ && $(0.187,0.195)$ & $(0.191,0.197)$ & $(0.226,0.236)$ & $(0.229,0.235)$  \\
& Power & $\textbf{0.793}$ & $0.587$ & $0.547$ & $0.467$ && $\textbf{0.823}$ & $0.611$ & $0.512$ & $0.481$ \\ 
& & $(0.789, 0.797)$ & $(0.582, 0.592)$ & $(0.541, 0.553)$ & $(0.460, 0.474)$ && $(0.818,0.828)$ & $(0.608,0.614)$ & $(0.507,0.517)$ & $(0.478,0.484)$ \\ [0.2ex]
\emph{E.coli2} &FDR&  $\textbf{0.193}$ &  $0.214$ & $0.332$ & $0.362$ && $\textbf{0.186}$ &  $0.201$ & $0.312$ & $0.278$  \\
& &  $(0.190,0.196)$ &  $(0.210,0.218)$ & $(0.325,0.339)$ & $(0.355,0.369)$ && $(0.181,0.191)$ &  $(0.197,0.205)$ & $(0.305,0.319)$ & $(0.278,)$  \\
& Power  & $\textbf{0.736}$ & $0.666$ & $ 0.639$ & $0.569$  &&  $\textbf{0.787}$ & $0.684$ & $ 0.649$ & $0.575$ \\ 
& & $(0.733,0.739)$ & $(0.662,0.670)$ & $(0.632,0.646)$ & $(0.562,0.576)$  &&  $(0.782,0.792)$ & $(0.680,0.688)$ & $(0.642,0.656)$ & $(0.569,0.681)$ \\ [0.2ex]
\emph{Yeast1} &FDR& $\textbf{0.181}$ & $0.203$ & $0.214$ & $0.336$ && $0.195$ & $\textbf{0.193}$ & $0.224$ & $0.288$ \\
& & $(0.177,0.185)$ & $(0.199,0.207)$ & $(0.209,0.219)$ & $(0.330,0.342)$ && $(0.192,0.198)$ & $(0.190,0.196)$ & $(0.216,0.232)$ & $(0.281,0.295)$ \\
& Power  & $\textbf{0.887}$ & $0.607$ & $ 0.546$ & $0.442$ && $\textbf{0.917}$ & $0.642$ & $ 0.522$ & $0.498$ \\
& & $(0.883,0.891)$ & $(0.603,0.611)$ & $(0.541,0.551)$ & $(0.436,0.448)$ && $(0.914,0.920)$ & $(0.639,0.645)$ & $(0.514,0.530)$ & $(0.49810.505)$ \\ [0.2ex]
\emph{Yeast2} &FDR& $\textbf{0.174}$ & $0.199$ & $0.262$ & $0.236$ && $\textbf{0.189}$ & $0.211$ & $0.241$ & $0.241$ \\
& & $(0.169,0.179)$ & $(0.195,0.203)$ & $(0.254,0.270)$ & $(0.230,0.242)$ && $(0.185,0.193)$ & $(0.208,0.214)$ & $(0.235,0.247)$ & $(0.236,0.246)$ \\
& Power & $\textbf{0.744}$ & $0.603$ & $0.570$ & $0.542$ && $\textbf{0.729}$ & $0.582$ & $0.535$ & $0.611$ \\
& & $(0.739,0.749)$ & $(0.599,0.607)$ & $(0.562,0.578)$ & $(0.536,0.548)$ && $(0.725,0.733)$ & $(0.579,0.585)$ & $(0.529,0.541)$ & $(0.606,0.616)$ \\ [0.2ex]
\emph{Yeast3} &FDR& $0.184$ & $\textbf{0.181}$ & $0.189$ & $0.248$ && $\textbf{0.190}$ & $0.208$ & $0.196$ & $0.223$ \\
& & $(0.180,0.188)$ & $(0.177,0.185)$ & $(0.184,0.194)$ & $(0.242,0.254)$ && $(0.185,0.195)$ & $(0.204,0.212)$ & $(0.191,0.201)$ & $(0.219,0.227)$ \\
& Power  & $\textbf{0.845}$  & $0.616$ & $0.573$ & $0.493$ &&  $\textbf{0.812}$  & $0.577$ & $0.539$ & $0.472$ \\
& & $(0.841,0.849)$  & $(0.612,0.620)$ & $(0.568,0.578)$ & $(0.487,0.499)$ &&  $(0.807,0.817)$  & $(0.573,0.581)$ & $(0.534,0.544)$ & $(0.468,0.476)$ \\ [0.2ex]
\bottomrule
\end{tabular}}
\caption{The gene regulatory network example: the empirical FDR and power, and their $95\%$ confidence intervals, for 10 combinations of network structures from GNW. The results are based on $500$ data replications.}
\label{table:gnw}
\vspace{-0.2in}
\end{table}

We apply the proposed confidence band approach, coupled with the BH procedure at the $20\%$ FDR level, to this data.  Table \ref{table:gnw} reports the empirical FDR and power for the recovered regulatory network for all ten combinations of network structures, and the results are based on 500 data replications. We compare with the alternative methods of linear, additive, and kernel ODEs, similarly as in Section \ref{sec:enzymatic}. We see clearly that the proposed method performs competitively in all cases. This example also shows that the proposed method can scale up and work with reasonably large networks. For instance, for the network with $p = 100$ nodes, there are $p^2 = 10,000$ functions to estimate.

%%%%%%%%%%%%%%%%%%%%%%%%%%%%%%%%%%%%%%%%%%%%%%%%%%%
\subsection{Brain Effective Connectivity Analysis}

\noindent 
Brain effective connectivity refers explicitly to the directional influence that one neural system exerts over another \citep{Friston2011}, and is of central interest in neuroscience research.  Effective connectivity analysis uncovers such directional influences among different brain regions through imaging techniques such as electrocorticographic (ECoG), and modeling techniques such as ODE \citep{Zhang2015}. The data we analyze is an ECoG study of the brain during decision making \citep{saez2018encoding}. It consists of the ECoG recordings of $p=61$ electrodes placed in the orbitofrontal cortex (OFC) region of an epilepsy patient when performing gambling tasks with different levels of winning risk. The patient performed 72 rounds of gambling games in total, half of which are low-risk games, and half are high-risk games. We analyze the low-risk and high-risk games separately, with $S=36$. The length of the ECoG signals for each round of game is $n=3001$. See \citet{saez2018encoding} for more details about the data collection and processing.

\begin{figure}[t!]
\centering
\begin{tabular}{ll}
low-risk & \\ 
\includegraphics[width=1.5in,height=1.75in]{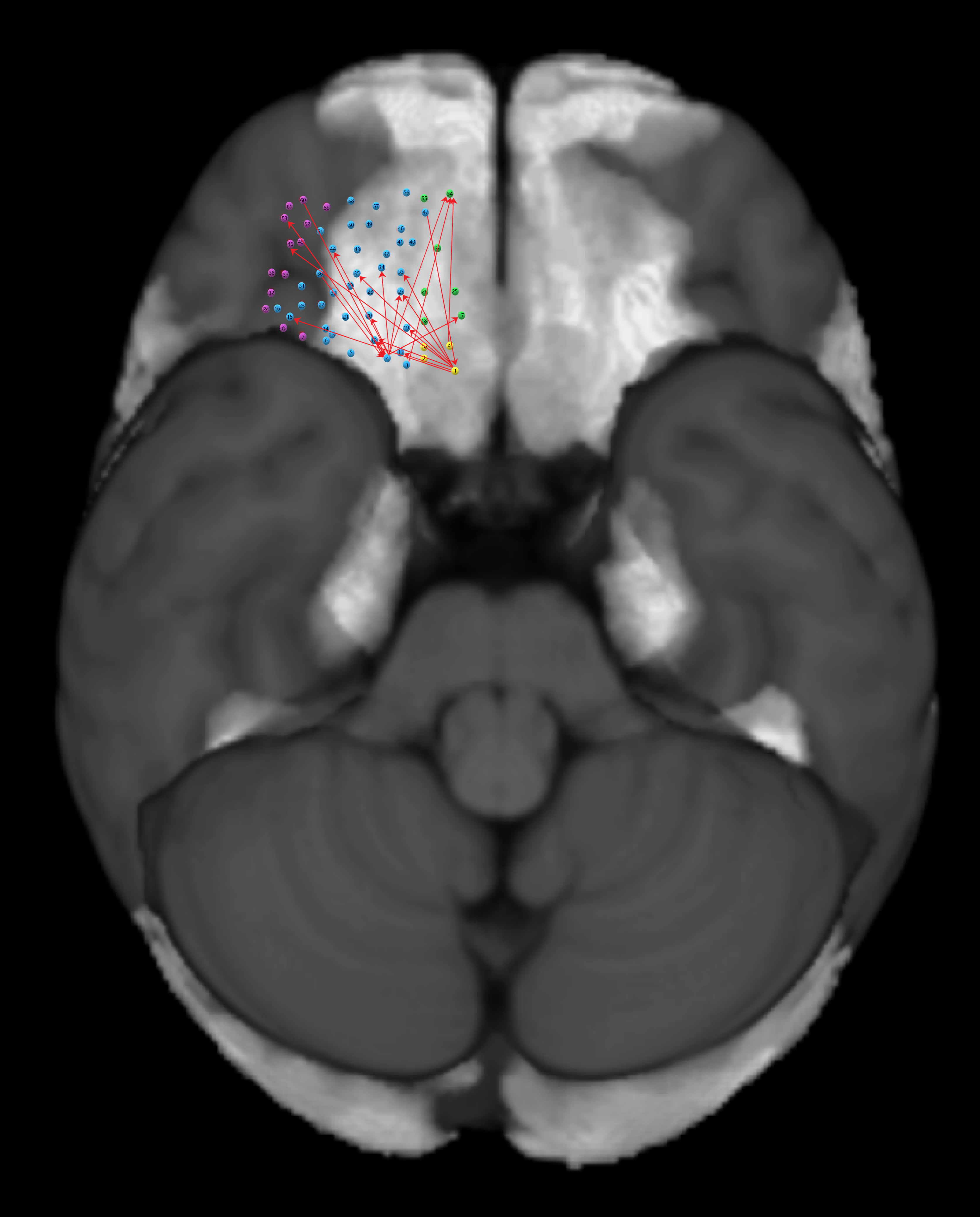} & 
\includegraphics[width=3.1in,height=1.75in]{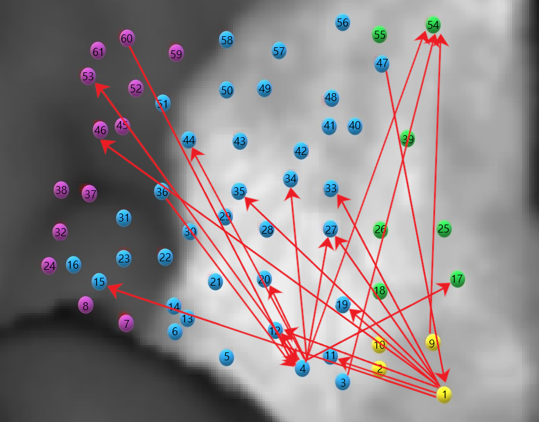} \\
high-risk & \\
\includegraphics[width=1.5in,height=1.75in]{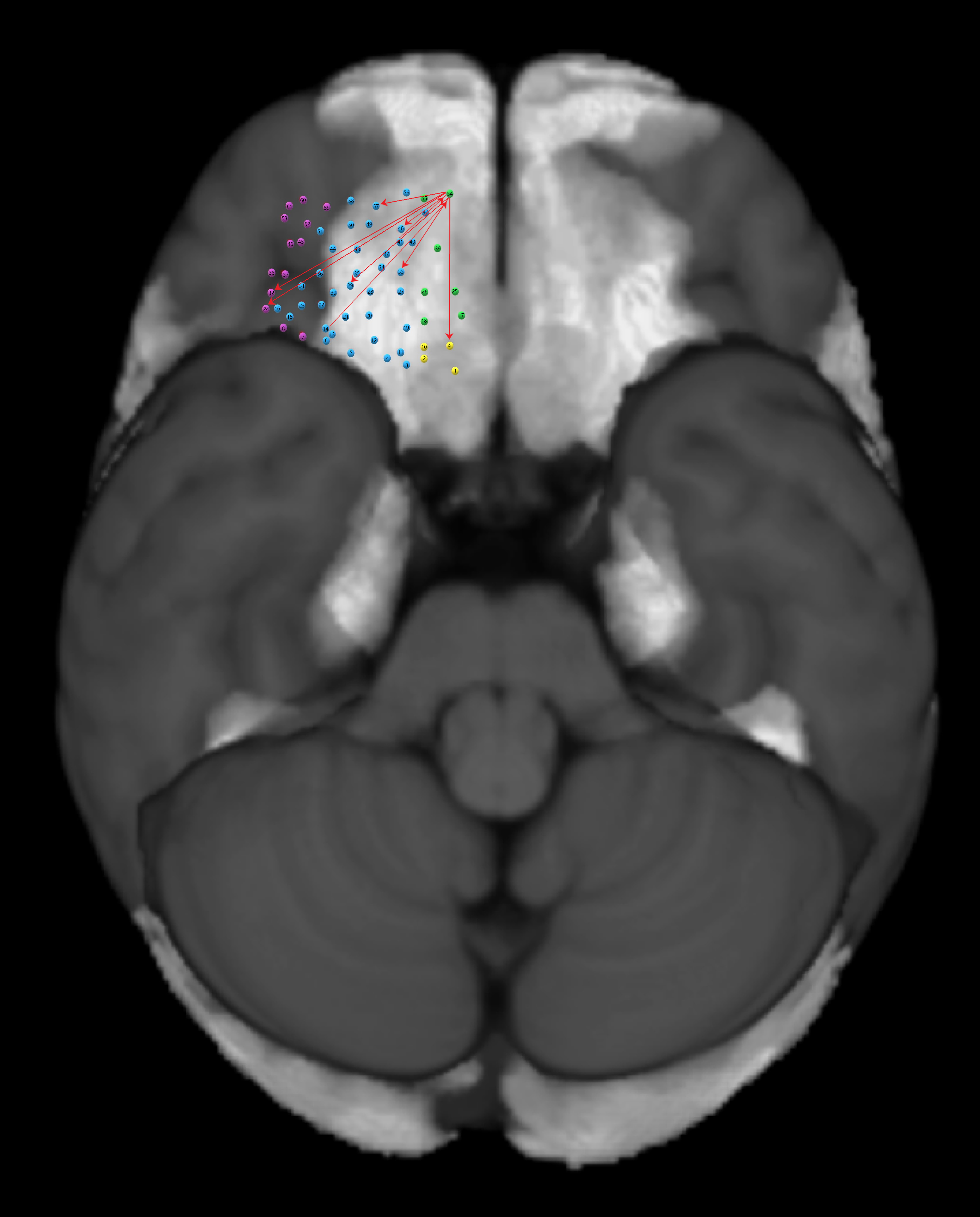} & 
\includegraphics[width=3.1in,height=1.75in]{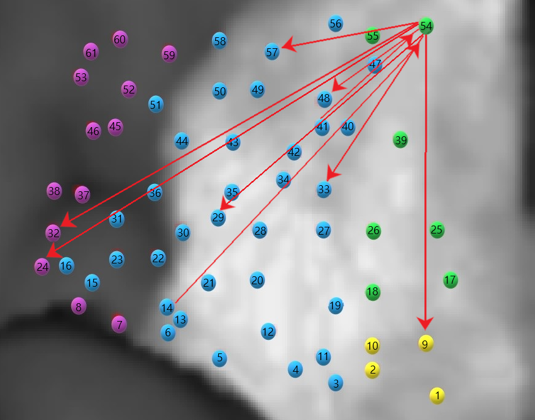} \\
\end{tabular}
\caption{The brain effective connectivity example: the connectivity patterns during the low-risk and high-risk games. The colored nodes correspond to different cytoarchitectural regions of orbitofrontal cortex. Green: Fo1; yellow: Fo2; blue: Fo3; purple: other regions. The left panels are for the entire brain, and the right panels for the enlarged areas.}  
\label{fig:ecog}
\vspace{-0.2in}
\end{figure}

We again apply the proposed confidence band approach, coupled with the BH procedure at the $20\%$ FDR level, to this data. To identify the nodes that show different connectivity patterns under two risk groups, we focus on those nodes whose total number of inward and outward edges is no more than 2 in one risk type, but no fewer than 9 in the other risk type. This results in node 1 that is located in the cytoarchitectural region of OFC called Fo2, node 4 located in Fo3, and node 54 located in Fo1 \citep{Henssen2016}. Figure \ref{fig:ecog} plots the estimated connectivity patterns of these three nodes, first on the entire brain, then in an amplified area. We see that, for nodes 1 and 4, there are many more outward edges during the low-risk games than the high-risk games, whereas for node 54, there are many more outward edges during the high-risk games than the low-risk games. Note that both Fo2 and Fo3 belong to the posterior OFC, which is more involved in simple reward type decision making, whereas Fo1 belongs to the anterior OFC, which is involved in abstract reward \citep{Kringelbach2004}. Our results suggest that the posterior OFC is more active during the low-risk games, in which the reward is relatively simple and clear. Meanwhile, the anterior OFC tends to more actively influence other nodes during the high-risk games, which involve more calculations and harder decisions. {We briefly comment that, the arrow direction in the plot indicates if the estimated effect is from node $x_k(t)$ on $x_j(t)$ or from $x_j(t)$ on $x_k(t)$, $j, k = 1, \ldots, p$. In our analysis, we primarily focus on the total numbers of inward and outward edges. }

%%%%%%%%%%%%%%%%%%%%%%%%%%%%%%%%%%%%%%%%%%%%%%%%%%%
\section{Discussion}
\label{sec:discussion}

In this article, we aim at a central question in ODE modeling; that is, to infer the significance of individual regulatory effect of one signal variable on another. This question is challenging for ODE with unknown regulatory relations and noisy data observations, and remains largely untapped in the literature. We propose a new post-regularization confidence band method, which provides both an uncertainty quantification for the individual regulatory relation, and also a sparse recovery of the entire regulatory system when coupled with a proper FDR control. Our proposal involves two key ingredients: a new localized kernel learning approach that combines reproducing kernel learning with local polynomial learning, and a new de-biasing method that tackles infinite-dimensional functionals and additional measurement errors. We establish the theoretical guarantees, and demonstrate the efficacy of the proposed method through numerical analyses. 

An interesting extension of the proposed method is to tackle the scenario of multiple experiments or subjects. In this article, we have primarily focused on the scenario of a single experiment or a single subject, and we propose to sum together the objective functions for multiple experiments in Section \ref{sec:multiexperiment}. However, in numerous applications, e.g., neuroscience, there may be considerable experiment-to-experiment, or subject-to-subject variability. How to effectively account for such variability is important, and warrants future research.

%%%%%%%%%%%%%%%%%%%%%%%%%%%%%%%%%%%%%%%%%%%%%%%%%%%
 \acks{The authors thank two anonymous reviewers and the Action Editor for their
invaluable feedback.
Xiaowu Dai  acknowledges support of the California Center for Population Research as a part of the
Eunice Kennedy Shriver National Institute of Child Health and Human Development (NICHD) population research infrastructure grant P2C-HD041022. Lexin Li acknowledges support of NSF grant CIF-2102227, and NIH grants R01AG061303 and R01AG062542.}

\newpage

%%%%%%%%%%%%%%%%%%%%%%%%%%%%%%%%%%%%%%%%%%%%%%%%%%%
 \appendix

\renewcommand{\theequation}{S\arabic{equation}}
\renewcommand{\thefigure}{S\arabic{figure}}
\renewcommand{\thetable}{S\arabic{table}}
\setcounter{equation}{0}
\setcounter{figure}{0}
\setcounter{table}{0}

\section*{Appendices}

%%%%%%%%%%%%%%%%%%%%%%%%%%%%%%%%%%%%%%%%%%%%%%%%%%%
\section{Parameters}
\label{sec:notations}
We present a list of main parameters in Table \ref{table:notation}, along with their meanings and dimensions.
\begin{table}[ht!]
\centering
\resizebox{\textwidth}{!}{
\begin{tabular}{lclcl}
\toprule
parameter & & definition & & dimension \\ [0.2ex]
 \midrule
$x(t)= (x_1(t), \ldots, x_p(t))^\top$ & & $p$ variables of interest & & $\Rbb^p$  \\ [0.2ex]
$F = \{F_1, \ldots,$ $F_p\}$ & & $p$ functionals of regulatory relations  & & $\Rbb^p$  \\ [0.2ex]
$\theta_{j0} $ & &  global intercept  & & $\Rbb^1$   \\ [0.2ex]
$\alpha_{jk, t_0}$ & &  regulatory effect of $x_k(t)$ on $x_j(t)$ at $t=t_0$  & & $\Rbb^1$   \\ [0.2ex]
$H_{jk}$ & & nuisance terms of evaluating effect of $x_k(t)$ on $x_j(t)$ && $\Rbb^1$      \\ [0.2ex]
$c_j = (c_{1j}, \ldots, c_{nj})^\top $ & & parameters in the optimization  \eqref{eqn:weightedlsforh}  && $\Rbb^n$      \\ [0.2ex]
$\theta_j = \{\theta_{jk'}\}_{k' \neq k}\cup\{\theta_{jk'l}\}_{k'\neq k; l\neq k',k}$ & & parameters in the optimization  \eqref{eqn:regfortheta}  && $\Rbb^{(p-1)^2}$      \\ [0.2ex]
\bottomrule
\end{tabular}}
\caption{List of parameters of the proposed ODE modeling.}
\label{table:notation}
\end{table}

%%%%%%%%%%%%%%%%%%%%%%%%%%%%%%%%%%%%%%%%%%%%%%%%%%%
\section{Proofs}

%%%%%%%%%%%%%%%%%%%%%%%%%%%%%%%%%%%%%%%%%%%%%%%%%%%
\subsection{Proof of Theorem \ref{thm:optimalestoffunctional}}

\noindent 
We divide the proof of this theorem into two parts. We first present the main proof in Section \ref{sec:uppbdfunctional}, then give two auxiliary lemmas useful for the proof of this theorem in Section \ref{sec:auxlem}.

%%%%%%%%%%%%%%%%%%%%%%%%%%%%%%%%%%%%%%%%%%%%%%%%%%%
\subsubsection{Main proof}
\label{sec:uppbdfunctional}

\begin{proof}
For $j=1,\ldots,p$, write $\widehat{F}_j(\widehat{x}(t_0)) = \widehat{\theta}_{j0} + \widehat{\alpha}_{jk, t_0}+ \widehat{H}_{jk}(\widehat{x}(t_0))$ for any $t_0\in\Tcal$. Write $F_j(x(t_0)) = \theta_{j0} + \alpha_{jk, t_0} + H_{jk}(x(t_0))$. Considering $\widehat{\theta}_{j0}$ that is given by \eqref{eqn:theta0}, where $\widehat{\theta}_{j0} = \bar{y}_j - \int_\Tcal\bar{T}(t) \tilde{F}_j(\widehat{x}(t)) dt$, its convergence rate is the same as that of $\widehat{\alpha}_{jk, t_0} + \widehat{H}_{jk}(\widehat{x}(t_0))$. Therefore, we focus our attention on $\widehat{\alpha}_{jk, t_0}$ and $\widehat{H}_{jk}(\widehat{x}(t_0))$ in the subsequent proof. 

Recall that $\widehat{\alpha}_{jk, t_0}$ and $\widehat{H}_{jk}$ are obtained from 
\begin{equation*}
\begin{aligned}
 \underset{\alpha_{jk, t_0}, H_{jk}}{\min} \left\{ \frac{1}{n}\sum_{i=1}^nR_h\left(t_i-t_0\right)\left[y_{ij}  - \alpha_{jk, t_0} \bar{t}_i -\int_{0}^{t_i} H_{jk}(\widehat{x}(t))dt\right]^2 + \tau_{nj}J(H_{jk}) \right\}, 
\end{aligned}
\end{equation*}
where $J(H_{jk})\equiv  \sum_{k' =1, k' \neq k}^p\|\Pcal^l H_{jk}\|_{\Hcal} + \sum_{k' =1, k' \neq k,l}^p \sum_{l=1, l \neq k}^p \|\Pcal^{k' l} H_{jk}\|_\Hcal$, and $\bar{t}_i = t_i - n^{-1}\sum_{i=1}^nt_i$. Then we have that, 
\begin{equation*}
\begin{aligned}
& \frac{1}{n}\sum_{i=1}^nR_h\left(t_i-t_0\right)\left\{  \alpha_{jk, t_0} \bar{t}_i + \int_0^{t_i}H_{jk}(\xbf(t))dt+\epsilonbf_{ij} -  \widehat{\alpha}_{jk, t_0} \bar{t}_i - \int_0^{t_i}\widehat{H}_{jk}(\widehat{\xbf}(t))dt\right\}^2\\
&\quad\quad\quad\quad\quad\quad\quad\quad\quad\quad\quad\quad\quad\quad\quad\quad\quad\quad\quad\quad\quad\quad\quad\quad\quad\quad\quad\quad\quad\quad+ \tau_{nj} J(\widehat{H}_{jk})\\
\leq \;\; &   
\frac{1}{n}\sum_{i=1}^nR_h\left(t_i-t_0\right)\left\{\alpha_{jk, t_0} \bar{t}_i  + \int_0^{t_i}H_{jk}(\xbf(t))dt+\epsilonbf_{ij} -  \alpha_{jk, t_0} \bar{t}_i  - \int_0^{t_i}H_{jk}(\widehat{\xbf}(t))dt\right\}^2\\
&\quad\quad\quad\quad\quad\quad\quad\quad\quad\quad\quad\quad\quad\quad\quad\quad\quad\quad\quad\quad\quad\quad\quad\quad\quad\quad\quad\quad\quad\quad+ \tau_{nj} J(H_{jk}).
\end{aligned}
\end{equation*}
With the rearrangement of the terms, we have that,
\begin{equation}
\label{eqn:totaldiff}
\begin{aligned}
& \frac{1}{n}\sum_{i=1}^nR_h\left(t_i-t_0\right)\left[  (\alpha_{jk, t_0} -\widehat{\alpha}_{jk, t_0})\bar{t}_i + \int_0^{t_i}\left\{ H_{jk}(\xbf(t)) - \widehat{H}_{jk}(\widehat{\xbf}(t)) \right\} dt \right]^2\\
&\quad\quad\quad\quad\quad\quad\quad\quad\quad\quad\quad\quad\quad\quad\quad\quad\quad\quad\quad\quad\quad\quad\quad\quad\quad\quad\quad+ \tau_{nj} J(\widehat{H}_{jk})\\
\leq \;\; & \frac{2}{n}\sum_{i=1}^nR_h\left(t_i-t_0\right)\epsilonbf_{ij}\left[ (\widehat{\alpha}_{jk, t_0}-\alpha_{jk, t_0})\bar{t}_i + \int_0^{t_i} \left\{ \widehat{H}_{jk}(\widehat{\xbf}(t))-H_{jk}(\widehat{\xbf}(t)) \right\}dt \right] \\
&\quad\quad\quad +
\frac{1}{n}\sum_{i=1}^nR_h\left(t_i-t_0\right) \left[\int_0^{t_i}\left\{ H_{jk}(\xbf(t))-H_{jk}(\widehat{\xbf}(t)) \right\} dt \right]^2+ \tau_{nj} J(H_{jk}).
\end{aligned}
\end{equation}
By Assumption \ref{assump:complexity} and the Taylor expansion, 
\begin{equation*}
(\widehat{H}_{jk} - H_{jk})(\widehat{x}) =(\widehat{H}_{jk}-H_{jk})(x) + \frac{\partial}{\partial t}(\widehat{H}_{jk}-H_{jk})(x)(\widehat{x}-x) + o_p\left(\max_{l=1,\ldots,p}\|\widehat{x}_l-x_l\|_{L_2}\right), 
\end{equation*}
where the Fr\'echet derivative of any  $H_{jk}(\cdot)\in\HH$ is defined as,
\begin{equation*}
\frac{\partial}{\partial t} H_{jk}(x)(\widehat{x}-x) = \sum_{k=1}^p\frac{\partial H_{jk}(x)}{\partial x_k}(\widehat{x}_k-x_k).
\end{equation*}
Then the first term on the right-hand-side of \eqref{eqn:totaldiff} can be written as,
\begin{equation*}
\begin{aligned}
&\frac{2}{n}\sum_{i=1}^nR_h\left(t_i-t_0\right)\epsilonbf_{ij}  \left[(\widehat{\alpha}_{jk, t_0}-\alpha_{jk, t_0})\bar{t}_i+\int_0^{t_i}\left\{ \widehat{H}_{jk}(\widehat{\xbf}(t))-H_{jk}(\widehat{\xbf}(t)) \right\}dt \right] \\
&= \frac{2}{n}\sum_{i=1}^nR_h\left(t_i-t_0\right)\epsilonbf_{ij}\left[(\widehat{\alpha}_{jk, t_0}-\alpha_{jk, t_0})\bar{t}_i+\int_0^{t_i}(\widehat{H}_{jk}-H_{jk})(x(t))dt\right] \\
& + \frac{2}{n}\sum_{i=1}^nR_h\left(t_i-t_0\right)\epsilon_{ij}\left[ \int_0^{t_i} \frac{\partial}{\partial t}(\widehat{H}_{jk}-H_{jk})(x(t))\{\widehat{x}(t)-x(t)\}dt + o_p\left(\max_{l=1,\ldots,p}\|\widehat{x}_l-x_l\|_{L_2}\right) \right]\\
&\equiv \Delta_1 + \Delta_2.
\end{aligned}
\end{equation*}
Meanwhile, by the Taylor expansion, the first term on the left-hand-side of \eqref{eqn:totaldiff} can be written as,
\begin{equation*}
\begin{aligned}
& \frac{1}{n}\sum_{i=1}^nR_h\left(t_i-t_0\right)\left[  (\alpha_{jk, t_0} -\widehat{\alpha}_{jk, t_0})\bar{t}_i + \int_0^{t_i}\left\{ H_{jk}(\xbf(t))-\widehat{H}_{jk}(\widehat{\xbf}(t)) \right\} dt \right]^2 \\
= \;\; & \frac{1}{n}\sum_{i=1}^nR_h\left(t_i-t_0\right)\left[  (\alpha_{jk, t_0} -\widehat{\alpha}_{jk, t_0})\bar{t}_i + \int_0^{t_i}\left\{ H_{jk}(\xbf(t))-\widehat{H}_{jk}(\xbf(t)) \right\} dt  \right.\\
&\quad\quad\quad\quad\quad\quad\quad\quad\quad\left. + \int_0^{t_1} \frac{\partial}{\partial t}\widehat{H}_{jk}(x(t))\{x(t)-\widehat{x}(t)\}dt + o_p\left( \max_{l=1,\ldots,p}\|\widehat{x}_l-x_l\|_{L_2} \right)\right]^2.
\end{aligned}
\end{equation*}
The right-hand-side of the above equation can be rewritten as
\vspace{-0.05in}
\begin{equation*}
\begin{aligned}
& \frac{1}{n}\sum_{i=1}^nR_h\left(t_i-t_0\right)\left[ (\alpha_{jk, t_0} -\widehat{\alpha}_{jk, t_0})\bar{t}_i +\int_0^{t_i}\left\{ H_{jk}(\xbf(t))-\widehat{H}_{jk}(\xbf(t)) \right\}dt\right]^2 \\
&+ \frac{1}{n}\sum_{i=1}^nR_h\left(t_i-t_0\right)\left[ \int_0^{t_i} \frac{\partial}{\partial t} \widehat{H}_{jk}(x(t))\{x(t)-\widehat{x}(t)\}dt \right]^2\\
&+\frac{2}{n}\sum_{i=1}^nR_h\left(t_i-t_0\right)\left[ (\alpha_{jk, t_0} -\widehat{\alpha}_{jk, t_0})\bar{t}_i +\int_0^{t_i}\left\{ H_{jk}(\xbf(t))-\widehat{H}_{jk}(\xbf(t)) \right\}dt\right] \\
&\quad\quad\quad \times \int_0^{t_i} \frac{\partial}{\partial t}\widehat{H}_{jk}(x(t))\{ \widehat{x}(t)-x(t) \}dt + \Rcal_1\\
&\equiv \widetilde\Delta_1 + \widetilde\Delta_2 + \widetilde\Delta_3 + \Rcal_1,
\end{aligned}
\end{equation*}
where the remainder term $\Rcal_1$ is of the form, 
\begin{equation*}
\begin{aligned}
\Rcal_1 & = \frac{1}{n}\sum_{i=1}^nR_h\left(t_i-t_0\right)\left( o_p\left( \max_{l=1,\ldots,p}\|\widehat{x}_l-x_l\|^2_{L_2} \right) -  o_p\left( \max_{l=1,\ldots,p}\|\widehat{x}_l-x_l\|_{L_2} \right) \right. \\
&\quad \left. \times\left[  (\alpha_{jk, t_0} -\widehat{\alpha}_{jk, t_0})\bar{t}_i +\int_0^{t_i}\left\{ H_{jk}(x(t))-\widehat{H}_{jk}(x(t)) - \frac{\partial}{\partial t} \widehat{H}_{jk}(x(t))\{ \widehat{x}(t)-x(t) \} \right\} dt \right]\right).
\end{aligned}
\end{equation*}
Denote $\Delta_3 \equiv  n^{-1} \sum_{i=1}^n R_h\left(t_i-t_0\right) \left[\int_0^{t_i}\left\{ H_{jk}(\xbf(t))-H_{jk}(\widehat{\xbf}(t)) \right\} dt \right]^2$. Then \eqref{eqn:totaldiff} becomes,
\begin{equation} \label{eqn:newtotaldiff}
\begin{aligned}
 \widetilde\Delta_1 + \widetilde\Delta_2 + \widetilde\Delta_3 + \Rcal_1 + \tau_{nj} J(\widehat{H}_{jk})  
\leq  \Delta_1 + \Delta_2 + \Delta_3 + \tau_{nj} J(H_{jk})  \\
\end{aligned}
\end{equation}
Our proof strategy is to derive the upper and lower bounds for the left and right-hand sides of \eqref{eqn:newtotaldiff}, respectively, then put them together. We first study the convergence rate of the estimator of the nuisance parameter $H_{jk}$. We then apply a similar proof procedure to obtain the rate of the estimator of the individual effect $\alpha_{jk,t_0}$. Together, we obtain the rate for the estimator of the entire regulatory effect $F_{j}(x(t))$.

\bigskip
\noindent
\textbf{Step 1: Bounding the right-hand-side of (\ref{eqn:newtotaldiff})}. 
We first bound the three terms $\Delta_1, \Delta_2, \Delta_3$ on the right-hand-side of \eqref{eqn:newtotaldiff}. 

For $\Delta_1$, by Lemma \ref{lem:bdonradamacher1} and the Minkowski inequality, we have that, as $h=o(1)$, 
\begin{equation*}
\begin{aligned}
\Delta_1 \leq \; & O_p \bigg\{ \left\| \widehat{H}_{jk}-H_{jk} \right\|_{L_2}^2 \log^{-2}\left\| \widehat{H}_{jk}-H_{jk} \right\|_{L_2}  +h^{2\beta_2}\\
&\quad\quad\quad\quad+ \left(\frac{nh}{\log n}\right)^{-\frac{2\beta_2}{2\beta_2+1}} + \frac{\log p}{n} + \sqrt{\frac{\log p}{n}} \left\| \widehat{H}_{jk} - H_{jk} \right\|_{L_2} \bigg\}.
\end{aligned}
\end{equation*}

For $\Delta_2$, since $\beta_2>1$, $\partial K(x,\cdot)/\partial x_k\in\HH$, and by the reproducing property, we have, 
\begin{equation*}
\frac{\partial (\widehat{H}_{jk}-H_{jk})(x)}{\partial x_k} = \left\langle\widehat{H}_{jk}-H_{jk},\frac{\partial K(x,\cdot)}{\partial x_k}\right\rangle_\HH\leq \|\widehat{H}_{jk}-H_{jk}\|_\HH^{1/2} \left\|\frac{\partial K(x,\cdot)}{\partial x_k}\right\|^{1/2}_\HH<\infty.
\end{equation*}
Henceforth, $\partial (\widehat{H}_{jk}-H_{jk})(x)/\partial x_k\in\HH$ for $k=1,\ldots,p$. By Assumption \ref{assump:fluctuation}, we have, 
\begin{equation*}
\max_{k=1,\ldots,p}\left\{|\partial (\widehat{H}_{jk}-H_{jk})(x)/\partial x_k|\right\} \leq B\|\widehat{H}_{jk}-H_{jk}\|_{L_2},\quad  \text{almost surely}.
\end{equation*} 
By the Cauchy-Schwarz inequality, we have, 
\begin{equation*}
\begin{aligned}
\Delta_2 \leq \;\; & \frac{2c}{n}\sum_{i=1}^n|\epsilon_{ij}|R_h\left(t_i-t_0\right)\int_0^{t_i} B\|\widehat{H}_{jk}(x(t))-H_{jk}(x(t))\|_{L_2} \max_{k=1,\ldots,p}|\widehat{x}_k(t)-x_k(t)|dt \\
&\quad\quad\quad\quad\quad\quad\quad\quad\quad\quad\quad\quad\quad\quad\quad\quad + o_p\left( n^{-1/2}\max_{k=1,\ldots,p}\|x_k-\widehat{x}_k\|^2_{L_2} \right) \\
\leq \;\; & 2c\max_{k=1,\ldots,p}\|\widehat{x}_k-x_k\|_{L_2} \left\| \widehat{H}_{jk}(x(t))-H_{jk}(x(t)) \right\|_{L_2} \frac{1}{n}\sum_{i=1}^n|\epsilon_{ij} B|R_h\left(t_i-t_0\right) \\
&\quad\quad\quad\quad\quad\quad\quad\quad\quad\quad\quad\quad\quad\quad\quad\quad+ o_p\left(n^{-1/2}\max_{k=1,\ldots,p}\|x_k-\widehat{x}_k\|^2_{L_2}\right)\\
= \;\; & O_p\left( n^{\frac{-\beta_1}{2\beta_1+1}}\|\widehat{H}_{jk}(x(t))-H_{jk}(x(t))\|_{L_2} \right),
\end{aligned}
\end{equation*}
for some constant $c$, where the last step is due to the strong law of large numbers and  Theorem 3 of  \citet{dai2021kernel}.

For $\Delta_3$, by the Taylor expansion and Assumption \ref{assump:complexity}, we have, 
\begin{equation}
\label{eqn:bdonerrorinx}
\begin{aligned}
\Delta_3 \leq \; & \frac{c}{n}\sum_{i=1}^n R_h\left(t_i-t_0\right) \left[ \int_0^{t_i} \frac{\partial}{\partial t} H_{jk}(\xbf(t))\{ \xbf(t)-\widehat{\xbf}(t) \} + o_p\left( \max_{k=1,\ldots,p}\|x_k-\widehat{x}_k\|^2_{L_2} \right)dt \right]^2\\
\leq \; &c' \|H_{jk}\|_\HH^2\max_{k=1,\ldots,p}\|x_k-\widehat{x}_k\|_{L_2}^2 + c' o_p\left(\max_{k=1,\ldots,p}\|x_k-\widehat{x}_k\|^2_{L_2}\right) 
= O_p\left( n^{\frac{-2\beta_1}{2\beta_1+1}} \right).
\end{aligned}
\end{equation}
for some constant $c,c'$, where the second step is by the Jensen's inequality.

\medskip
\noindent
\textbf{Step 2: Bounding the left-hand-side of \eqref{eqn:newtotaldiff}}. 
We next bound the terms $\widetilde\Delta_1, \widetilde\Delta_2, \widetilde\Delta_3$ and $\Rcal_1$ on the left-hand-side of \eqref{eqn:newtotaldiff}. 

For $\widetilde\Delta_1$, by Lemma \ref{lem:bdonradamacher}, with probability at least $1-2p^{-c_1}$, for some constant $C>0$, 
\begin{equation}
\label{eqn:step2term1}
\begin{aligned}
\widetilde\Delta_1 \geq \ \left\| H_{jk}-\widehat{H}_{jk} \right\|_{L_2}^2 - & C \bigg\{ \left\| H_{jk} - \widehat{H}_{jk} \right\|^2_{L_2}\log^{-2} \left\|H_{jk}-\widehat{H}_{jk} \right\|_{L_2} +h^{2\beta_2}\\
 &\quad\quad+ \left(\frac{nh}{\log n}\right)^{-\frac{2\beta_2}{2\beta_2+1}} + \frac{\log p}{n} + n^{-1/2}e^{-p}  \bigg\}.
\end{aligned}
\end{equation}

For $\widetilde\Delta_2$, we can drop this term, because $\widetilde\Delta_2 \geq 0$. 

For $\widetilde\Delta_3$, by the Cauchy-Schwarz inequality,
\begin{equation*}
\begin{aligned}
\widetilde\Delta_3 \geq \; & -2\left( \frac{1}{n}\sum_{i=1}^n\left[ \int_0^{t_i}\left\{ R_h(t_i-t_0)\left[\widehat{H}_{jk}(\xbf(t))-H_{jk}(\xbf(t)) \right]\right\} dt \right]^2 \right)^{1/2} \\
& \times \left( \frac{1}{n}\sum_{i=1}^n\left[ \int_0^{t_i} \frac{\partial}{\partial t} \widehat{H}_{jk}(x(t))\{ \widehat{x}(t)-x(t) \}dt \right]^2 \right)^{1/2}.
\end{aligned}
\end{equation*}
Henceforth,
\begin{equation*}
\begin{aligned}
\widetilde\Delta_3 
\geq \; & -2 \left\| \widehat{F}_j(x(t))-F_j(x(t)) \right\|_{L_2} \|F_j\|_\HH\max_{k=1,\ldots,p}\|x_k-\widehat{x}_k\|_{L_2}\\
= \; & O_p\left( n^{\frac{-\beta_1}{2\beta_1+1}} \left\| \widehat{F}_j(x(t))-F_j(x(t)) \right\|_{L_2}  \right),
\end{aligned}
\end{equation*}
where the second step is due to the Minkowski inequality.

For the remainder term $\Rcal_1$ on the left-hand-side of \eqref{eqn:newtotaldiff}, by Assumption \ref{assump:complexity} and the Cauchy-Schwarz inequality, we have, 
\begin{equation*}
\begin{aligned}
\Rcal_1 
& = o_p\left(\max_{k=1,\ldots,p}\|x_k-\widehat{x}_k\|_{L_2} \, \left\| \widehat{H}_{jk}-H_{jk} \right\|_{L_2} + \max_{k=1,\ldots,p}\|x_k-\widehat{x}_k\|^2_{L_2} \, \|H_{jk}\|_\HH\right)\\
& = o_p\left( n^{\frac{-\beta_1}{2\beta_1+1}} \left\| \widehat{H}_{jk} - H_{jk} \right\|_{L_2} \right) + o_p\left( n^{\frac{-2\beta_1}{2\beta_1+1}} \right),
\end{aligned}
\end{equation*}
where the second step is again due to the Minkowski inequality.

\medskip
\noindent
\textbf{Step 3: Putting the two bounds together}. 
Combining the bounds for each term in \eqref{eqn:newtotaldiff}, we obtain that, for any $c_1>0$ and $c_2>1$, with probability at least $1-4p^{-c_1}$, there exists a constant $C>0$, such that 
\begin{equation*}
\begin{aligned}
 & \left\| H_{jk}-\widehat{H}_{jk} \right\|_{L_2}^2 \log^{-2} \left\| H_{jk}-\widehat{H}_{jk} \right\|_{L_2}\\
\leq \; & C \left[ c_2^{-\frac{4\beta_2}{2\beta_2-1}} \left\| H_{jk}-\widehat{H}_{jk} \right\|^2_{L_2} \log^{-2} \left\| H_{jk} -\widehat{H}_{jk} \right\|_{L_2} + c_2^{\frac{4\beta_2}{4\beta_2+1}} \left(\frac{nh}{\log n}\right)^{-\frac{2\beta_2}{2\beta2+1}} \right. \\
& \quad\quad + (c_1+1) \frac{\log p}{n} + \sqrt{(c_1+1) \frac{\log p}{n}} \left\| H_{jk}-\widehat{H}_{jk} \right\|_{L_2} + h^{2\beta_2}\\
& \quad\quad \left. + n^{-1/2}e^{-p}+ n^{\frac{-\beta_1}{2\beta_1+1}} \left\| \widehat{H}_{jk} - H_{jk} \right\|_{L_2} + n^{\frac{-2\beta_1}{2\beta_1+1}}+ \tau_{nj} \left\{ J(H_{jk})-J(\widehat{H}_{jk}) \right\} \right].
\end{aligned}
\end{equation*}
Taking $c_2$ large enough such that $C c_2^{-4\beta_2/(2\beta_2-1)} \leq 1/2$, then we obtain that, 
\begin{equation*}
\begin{aligned}
& \left\|H_{jk} - \widehat{H}_{jk} \right\|_{L_2}^2 \log^{-2} \left\| H_{jk} - \widehat{H}_{jk} \right\|_{L_2} \leq 2C \left[ c_2^{\frac{4\beta_2}{4\beta_2+1}}\left(\frac{nh}{\log n}\right)^{-\frac{2\beta_2}{2\beta2+1}} \right. \\
& \quad\quad\quad\quad\; +(c_1+1)\frac{\log p}{n} + \sqrt{(c_1+1)\frac{\log p}{n}} \left\| H_{jk}-\widehat{H}_{jk} \right\|_{L_2} + h^{2\beta_2}\\
& \left. \quad\quad\quad\quad\; + n^{-1/2}e^{-p}+  n^{\frac{-\beta_1}{2\beta_1+1}} \left\| \widehat{H}_{jk} - H_{jk} \right\|_{L_2} + n^{\frac{-2\beta_1}{2\beta_1+1}}+\tau_{nj} \left\{ J(H_{jk})-J(\widehat{H}_{jk}) \right\} \right].
\end{aligned}
\end{equation*}
Therefore, 
\begin{equation} \label{eqn:bdonl2}
\begin{aligned}
\left\| H_{jk}(x(t))-\widehat{H}_{jk}(x(t)) \right\|_{L_2}^2 = O_p\left\{ \left( \frac{nh}{\log n} \right)^{-\frac{2\beta_2}{2\beta2+1}} + h^{2\beta_2} + \frac{\log p}{n}  + n^{-\frac{2\beta_1}{2\beta_1+1}} \right\}.
\end{aligned}
\end{equation}

\medskip
\noindent
\textbf{Step 4: Estimation of $\alpha_{jk,t_0}$}. 
We now apply a similar proof procedure in the above Steps 1-3 to estimate $\alpha_{jk,t_0}$.
First, by Lemma \ref{lem:bdonradamacher1} and the Minkowski inequality, as $h=o(1)$, the right-hand-side of \eqref{eqn:newtotaldiff} is bounded by
\begin{equation*}
\begin{aligned}
& \Delta_1 + \Delta_2 + \Delta_3 + \tau_{nj} J(H_{jk})\\
&  \leq  \;  O_p \bigg\{ \left\| \widehat{\alpha}_{jk,t}- F_{jk}(x(t))\right\|_{L_2}^2+h^{2\beta_2}+ \left(\frac{nh}{\log n}\right)^{-\frac{2\beta_2}{2\beta_2+1}} + \frac{\log p}{n}  \bigg\}.
\end{aligned}
\end{equation*}
Second, by Lemma \ref{lem:bdonradamacher}, with probability at least $1-2p^{-c_1}$, the left-hand-side of \eqref{eqn:newtotaldiff} is bounded by, for some constant $C>0$,
\begin{equation*}
\begin{aligned}
& \widetilde\Delta_1 + \widetilde\Delta_2 + \widetilde\Delta_3 +\Rcal_1 + \tau_{nj} J(\widehat{H}_{jk})\\
&  \geq  \; \left\| \widehat{\alpha}_{jk,t}- F_{jk}(x(t)) \right\|_{L_2}^2 -  \bigg\{ h^{2\beta_2}+ \left(\frac{nh}{\log n}\right)^{-\frac{2\beta_2}{2\beta_2+1}} + \frac{\log p}{n}  + n^{-1/2}e^{-p}  \bigg\}\\
&\quad + o_p\left( n^{\frac{-\beta_1}{2\beta_1+1}} \left\|  \widehat{\alpha}_{jk,t}- F_{jk}(x(t)) \right\|_{L_2} \right) + o_p\left( n^{\frac{-2\beta_1}{2\beta_1+1}} \right).
\end{aligned}
\end{equation*}
Then, putting the above two bounds together, we have that 
\begin{equation} \label{eqn:bdonl3}
\begin{aligned}
\left\|  \widehat{\alpha}_{jk,t}- F_{jk}(x(t)) \right\|_{L_2}^2 = O_p\left\{ \left( \frac{nh}{\log n} \right)^{-\frac{2\beta_2}{2\beta2+1}} + h^{2\beta_2} + \frac{\log p}{n}  + n^{-\frac{2\beta_1}{2\beta_1+1}} \right\}.
\end{aligned}
\end{equation}

\medskip
\noindent
\textbf{Step 5: Estimation of $F_j(x(t))$}. 
Combining the bounds \eqref{eqn:bdonl2} and \eqref{eqn:bdonl3}, we obtain that
\begin{equation*} \label{eqn:bdonl4}
\begin{aligned}
\left\| F_j(x(t))-\widehat{F}_j(x(t)) \right\|_{L_2}^2 = O_p\left\{ \left( \frac{nh}{\log n} \right)^{-\frac{2\beta_2}{2\beta2+1}} + h^{2\beta_2} + \frac{\log p}{n}  + n^{-\frac{2\beta_1}{2\beta_1+1}} \right\}.
\end{aligned}
\end{equation*}
This leads to the desired upper bound. Letting $h=O\left((n/\log n)^{-1/(2\beta_2+2)}\right)$, then the localized kernel ODE estimator in \eqref{eqn:estnonadditivemodel} satisfies that,
\begin{equation*}
\left\| F_j(x(t))-\widehat{F}_j(x(t)) \right\|_{L_2}^2 = O_p\left\{ \left(\frac{n}{\log n}\right)^{-\frac{\beta_2}{\beta_2+1}}+\frac{\log p}{n} + n^{-\frac{2\beta_1}{2\beta_1+1}} \right\}.
\end{equation*}
This completes the proof of Theorem \ref{thm:optimalestoffunctional}.
\end{proof}

%%%%%%%%%%%%%%%%%%%%%%%%%%%%%%%%%%%%%%%%%%%%%%%%%%%
\subsubsection{Auxiliary lemmas for Theorem \ref{thm:optimalestoffunctional}}
\label{sec:auxlem}

\noindent 
For any $g\in\HH$, define the norm, $\|g(x(t))\|_n =\sqrt{(1/n) \sum_{i=1}^ng^2(x(t_i))}$.

\begin{lemma}
\label{lem:bdonradamacher1}
Suppose that $H_{jk}\in\Hcal$, and the errors $\{\epsilonbf_{ij}\}_{i=1}^n$ are i.i.d.\ Gaussian. Then there exists some constant $C>0$, such that, for any $c_1>0$ and $c_2>1$, with probability at least $1-2p^{-c_1}$,
\begin{equation*}
\begin{aligned}
& \frac{1}{n}\sum_{i=1}^nR_h\left(t_i-t_0\right)\epsilon_{ij}H_{jk}(x(t_{i}))\\
\leq \; & C h^{-1}\left\{ \|H_{jk}\|_{L_2}^2 \log^{-2}\|H_{jk}\|_{L_2} + h^{2\beta_2}+ \left( \frac{nh}{\log n} \right)^{-\frac{2\beta_2}{2\beta_2+1}}  \right.\\
& \quad\quad\quad\quad \quad\quad\quad\quad \quad\quad\left.+  \frac{\log p}{n}+ \sqrt{\frac{\log p}{n} } \|H_{jk}\|_{L_2} + n^{-1/2}e^{-p} \right\}.
\end{aligned}
\end{equation*}
\end{lemma}

\begin{proof}
Recall the RKHS $\HH$ as defined in \eqref{eqn:spaceH}. For notational simplicity, we denote $F_{jl} \equiv F_{jll}$, for $l=1,\ldots,p$. Recall that $F_{jlr}\in\HH_l\otimes\HH_r$, $l,r\neq k$ and the corresponding reproducing kernel is $K_{lr}$. Let $\lambda_\nu(K)$ denote the $\nu$th largest eigenvalue of a positive definite operator $K$.  Note that $\lambda_\nu(K_{lr}) = O\left\{(\nu\log^{-1}\nu)^{-2\beta_2}\right\}$, for $\nu\geq 1$ \citep{Bach2017}. By the Sobolev's embedding theorem, $(K_{lr}R_h)(L_2)$ can be embedded to a Sobolev space $\mathcal W$ that corresponds to a reproducing kernel $K^*$ \citep{Cucker2002}. Moreover, $\lambda_\nu(K^*) = O\left\{\lambda_\nu(K_{lr})\right\}$.
Let $\{e_\nu:\nu\geq 1\}$ denote the eigenfunctions of $K^*$; that is, $K^*e_\nu=\lambda_\nu(K^*)e_\nu$ for $\nu\geq1$. Denote by $\mathcal E_\nu$ and $\mathcal E_\nu^\perp$ the linear space spanned by $\{e_s:1\leq s\leq \nu\}$ and $\{e_s:s\geq \nu+1\}$, respectively. By the Courant-Fischer-Weyl min-max principle,
\begin{equation*}
\begin{aligned}
\lambda_\nu(K_{lr}R_h) &\geq \min_{e\in\mathcal E_\nu}\|K_{lr}^{1/2}R_h^{1/2}e\|_{L_2}^2/\|e\|_{L_2}^2 
\geq C\min_{e\in\mathcal E_\nu}\|(K^*)^{1/2}e\|_{L_2}^2/\|e\|_{L_2}^2 
\geq C\lambda_\nu(K^*)
\end{aligned}
\end{equation*}
for some constant $C>0$. On the other hand,
\begin{equation*}
\begin{aligned}
\lambda_\nu(K_{lr}R_h)  & \leq \min_{e\in\mathcal E_{\nu-1}^\perp}\|K_{lr}^{1/2}R_h^{1/2}e\|_{L_2}^2/\|e\|_{L_2}^2 
\leq C'\min_{e\in\mathcal E_{\nu-1}^\perp}\|(K^*)^{1/2}e\|_{L_2}^2/\|e\|_{L_2}^2
\leq C'\lambda_\nu(K^*)
\end{aligned}
\end{equation*}
for some constant $C'>0$. Henceforth, together with Assumption \ref{assump:kerneldensity}, we have, 
\begin{equation}
\label{eqn:eigenkr}
\begin{aligned}
\lambda_\nu(K_{lr}R_h)   =O\left\{\lambda_\nu(K^*)\right\} = O\left\{(\nu\log^{-1}\nu)^{-2\beta_2}\right\}.
\end{aligned}
\end{equation}

Since $\{\epsilon_{ij}\}_{i=1}^{n}$ are i.i.d.\ Gaussian, by Lemma 2.2 of \citet{Yuan2016} and Corollary 8.3 of \citet{Vandegeer2000}, we have that, for any $c_1>0$, with probability at least $1-p^{-c_1}$,
\begin{equation}
\label{eqn:empmean}
\begin{aligned}
& \frac{1}{n}\sum_{i=1}^nR_h\left(t_i-t_0\right)\epsilon_{ij}H_{jk}(x(t_{i})) \\
\leq \; & 2C_1n^{-1/2}\sum_{l,r=1;l,r\neq k}^p \left(\|F_{jlr}R_h\|_{n}\log^{-1}\|F_{jlr}R_h\|_{n}\right)^{1-\frac{1}{2\beta_2}}  \left( \|F_{jlr}R_h\|_{\HH}\log^{-1}\|F_{jlr}R_h\|_{\HH} \right)^{\frac{1}{2\beta_2}} \\
& + 2C_1n^{-1/2}\sqrt{(c_1+1)\log p}\sum_{l,r=1;l,r\neq k}^p  \|F_{jlr}R_h\|_{n}  + 2C_1n^{-1/2}e^{-p}\sum_{l,r=1;l,r\neq k}^p\|F_{jlr}R_h\|_{\HH} \\ 
\equiv \; & 2C_1 (\Delta_4 + \Delta_5 + \Delta_6), 
\end{aligned}
\end{equation}
for some constant $C_1$. Next, we bound the three terms $\Delta_4, \Delta_5, \Delta_6$ on the right-hand-side of \eqref{eqn:empmean}, respectively. 

For $\Delta_4$, by the Young's inequality, there exists $c_2>1$ such that, 
\begin{equation*}
\begin{aligned}
\Delta_4 \leq \; & c_2^{-\frac{4\beta_2}{2\beta_2-1}}\sum_{l,r=1;l,r\neq k}^p \left(\|F_{jlr}R_h\|_{n}\log^{-1}\|F_{jlr}R_h\|_{n}\right)^{2}\\
& + c_2^{\frac{4\beta_2}{2\beta_2+1}}n^{-\frac{2\beta_2}{2\beta2+1}}h^{-\frac{2}{2\beta_2+1}}\sum_{l,r=1;l,r\neq k}^p \left( \|F_{jlr}\|_{\HH}\log^{-1}\|F_{jlr}\|_{\HH} \right)^{\frac{2}{2\beta_2+1}}.
\end{aligned}
\end{equation*}
Note that 
\begin{equation*}
\label{eqn:bdfjlogfj}
\sum_{l,r=1;l,r\neq k}^p \left( \|F_{jlr}\|_{\HH}\log^{-1}\|F_{jlr}\|_{\HH} \right)^{\frac{2}{2\beta_2+1}} \leq C'_2 \sum_{l,r=1;l,r\neq k}^p \left( \|F_{jlr}\|_{\HH}\log^{-1}\|F_{jlr}\|_{\HH} \right)^{0}  \leq C_2, 
\end{equation*}
for some constants $C_2',C_2$, where the last step is due to Assumption \ref{assump:complexity} that the number of nonzero functional components of $H_{jk}$ is bounded. Henceforth,
\begin{equation}
\label{eqn:firststepbd}
\begin{aligned}
\Delta_4\leq c_2^{-\frac{4\beta_2}{2\beta_2-1}}\sum_{l,r=1;l,r\neq k}^p \left(\|F_{jlr}R_h\|_{n}\log^{-1}\|F_{jlr}R_h\|_{n}\right)^{2} + c_2^{\frac{4\beta_2}{4\beta_2+1}}n^{-\frac{2\beta_2}{2\beta2+1}}h^{-\frac{2}{2\beta_2+1}}C_2.
\end{aligned}
\end{equation}
By \eqref{eqn:eigenkr}, Theorem 4 of \citet{Koltchinskii2010} and Theorem 3 of \citet{fan1993local}, there exists some constant $C_3>0$, such that, with probability at least $1-p^{-c_1}$,
\begin{equation*}
\begin{aligned}
\sum_{l,r=1;l,r\neq k}^p \left(\|F_{jlr}R_h\|_{n}\log^{-1}\|F_{jlr}R_h\|_{n}\right)^{2}
\leq 2C_3^2\sum_{l,r=1;l,r\neq k}^p \left(\|F_{jlr}\|_{L_2}\log^{-1}\|F_{jlr}\|_{L_2}\right)^{2}  \\
+2C_3^2h^{2\beta_2}+ 2C_3^2\left\{ \left( \frac{nh}{\log n} \right)^{-\frac{2\beta_2}{2\beta_2+1}}+\frac{(c_1+1)\log p}{n} \right\} \sum_{l,r=1;l,r\neq k}^p \left( \|F_{jlr}\|_{\HH}\log^{-1} \|F_{jlr}\|_{\HH} \right)^2.
\end{aligned}
\end{equation*}
Note that there exists some constant $c_3>1$, such that 
\begin{equation*}
\sum_{l,r=1;l,r\neq k}^p \left( \|F_{jlr}\|_{L_2}\log^{-1}\|F_{jlr}\|_{L_2} \right)^2 \leq c_3 \left( \|H_{jk}\|_{L_2}\log^{-1}\|H_{jk}\|_{L_2} \right)^2,
\end{equation*} 
where we recall that 
$H_{jk}(x(t)) = \sum_{l,r=1;l,r\neq k}^p F_{jlr}(x_l(t), x_r(t)) $. 
Moreover,
\begin{equation*}\sum_{l,r=1;l,r\neq k}^p \left( \|F_{jkl}\|_{\HH}\log^{-1} \|F_{jkl}\|_{\HH} \right)^2 \leq \sum_{l,r=1;l,r\neq k}^p ( \|F_{jlr}\|_{\HH} \log^{-1} \|F_{jlr}\|_{\HH} )^0\leq C_2.
\end{equation*} Then, we have
\begin{equation*}
\begin{aligned}
\sum_{l,r=1;l,r\neq k}^p  \left(\|F_{jlr}R_h\|_{n}\log^{-1}\|F_{jlr}R_h\|_{n}\right)^{2} \leq 2C_3^2c_3 \left(\|H_{jk}\|_{L_2}\log^{-1}\|H_{jk}\|_{L_2} \right)^2 \\
+2C_3^2h^{2\beta_2}+ 2C_2C_3^2\left\{ \left(\frac{nh}{\log n} \right)^{-\frac{2\beta_2}{2\beta_2+1}} + \frac{(c_1+1)\log p}{n} \right\}.
\end{aligned}
\end{equation*}
Inserting into (\ref{eqn:firststepbd}) yields that
\begin{equation*}
\label{eqn:firststepbdfinal}
\begin{aligned}
\Delta_4 & \leq \;  2C_3^2c_3c_2^{-\frac{4\beta_2}{2\beta_2-1}} \left(\|H_{jk}\|_{L_2}\log^{-1}\|H_{jk}\|_{L_2} \right)^2 +2C_3^2c_2^{-\frac{4\beta_2}{2\beta_2-1}}h^{2\beta_2}\\
& + 2C_2C_3^2c_2^{-\frac{4\beta_2}{2\beta_2-1}} \left\{ \left( \frac{nh}{\log n} \right)^{-\frac{2\beta_2}{2\beta_2+1}}+\frac{(c_1+1)\log p}{n} \right\} +  C_2c_2^{\frac{4\beta_2}{4\beta_2+1}} n^{-\frac{2\beta_2}{2\beta2+1}}h^{-\frac{2}{2\beta_2+1}}.
\end{aligned}
\end{equation*}
Since $\beta_2>1$ and $h=o(1)$, we have $n^{-\frac{2\beta_2}{2\beta2+1}}h^{-\frac{2}{2\beta_2+1}} < \left( \frac{nh}{\log n} \right)^{-\frac{2\beta_2}{2\beta_2+1}}$.

For $\Delta_5$, by Theorem 4 of \citet{Koltchinskii2010} again, there exists a constant $C_4>0$, such that
\begin{equation*}
\begin{aligned}
& \sum_{l,r=1;l,r\neq k}^p  \|F_{jlr}R_h\|_{n} \\
\leq \; & C_4\sum_{l,r=1;l,r\neq k}^p\|F_{jlr}\|_{L_2}+C_4h^{\beta_2}+C_4\left\{ \left(\frac{nh}{\log n}\right)^{-\frac{\beta_2}{2\beta_2+1}}+\sqrt{\frac{(c_1+1)\log p}{n}} \right\}\sum_{l,r=1;l,r\neq k}^p\|F_{jlr}\|_{\HH}\\
\leq \; & C_4\sum_{l,r=1;l,r\neq k}^p\|F_{jlr}\|_{L_2}+C_4h^{\beta_2}+C_2C_4\left\{ \left(\frac{nh}{\log n}\right)^{-\frac{\beta_2}{2\beta_2+1}}+\sqrt{\frac{(c_1+1)\log p}{n}}\right\}.
\end{aligned}
\end{equation*}
Define the set $\mathcal Q_1 \equiv \left\{ l,r=1,\ldots,p; l,r\neq k: \|F_{jlr}\|_{L_2}>\sqrt{n^{-1}\log p} \right\}$. By the Cauchy-Schwartz inequality, we have, 
\begin{equation*}
\begin{aligned}
& \sum_{l,r\in\mathcal Q_1}\|F_{jlr}\|_{L_2} \leq \text{card}^{1/2}(\mathcal Q_1)\cdot\left(\sum_{l,r\in\mathcal Q_1}\|F_{jlr}\|^2_{L_2}\right)^{1/2}\\
\leq \; & \sum_{l,r=1;l,r\neq k}^p\|F_{jlr}\|_\HH^0\cdot\left(\sum_{l,r=1;l,r\neq k}^p\|F_{jlr}\|^2_{L_2}\right)^{1/2}\leq C_2c_4\|H_{jk}\|_{L_2}, 
\end{aligned}
\end{equation*}
The constant $c_4>1$ satisfies that $\sum_{l,r=1;l,r\neq k}^p\|F_{jlr}\|^2_{L_2}\leq c^2_4\|H_{jk}\|^2_{L_2}$, where we recall that $H_{jk}(x(t)) = \sum_{l,r=1;l,r\neq k}^p F_{jlr}(x_l(t), x_r(t)) $. Next, define the set $\mathcal Q_2 \equiv \{l,r=1,\ldots,p; l,r\neq k: \|F_{jkl}\|_{L_2}\leq \sqrt{n^{-1}\log p}\}$. By definition,
\begin{equation*}
\begin{aligned}
\sum_{l,r\in\mathcal Q_2}\|F_{jlr}\|_{L_2} &  \leq \sum_{l,r\in\mathcal Q_2}\|F_{jlr}\|^0_{L_2}\sqrt{\frac{\log p}{n}}\leq\sqrt{\frac{\log p}{n}}\sum_{l,r=1;l,r\neq k}^p\|F_{jlr}R_h\|^0_{L_2}\leq C_2 \sqrt{\frac{\log p}{n}}.
\end{aligned}
\end{equation*}
Combining $\mathcal Q_1$ and $\mathcal Q_2$ gives that,
\begin{equation*}
\begin{aligned}
\sum_{l,r=1;l,r\neq k}^p\|F_{jlr}\|_{L_2} &\leq \sum_{l,r\in\mathcal Q_1}\|F_{jlr}\|_{L_2} + \sum_{l,r\in\mathcal Q_2}\|F_{jlr}\|_{L_2}
\leq C_2c_4\|H_{jk}\|_{L_2}+ C_2 \sqrt{\frac{\log p}{n}}.
\end{aligned}
\end{equation*}
Henceforth, we can bound $\Delta_5$ as,
\begin{equation*}
\begin{aligned}
\Delta_5 \leq \sqrt{(c_1+1)}C_4\left[C_2c_4\sqrt{\frac{\log p}{n}}\|H_{jk}\|_{L_2} + C_2h^{\beta_2}+ C_2\left(\frac{nh}{\log n}\right)^{-\frac{\beta_2}{2\beta_2+1}}\sqrt{\frac{\log p}{n}} + C_2\frac{\log p}{n}\right].
\end{aligned}
\end{equation*}

For $\Delta_6$, it can be bounded as,
\begin{equation*}
\Delta_6 \leq n^{-1/2}e^{-p}\sum_{l,r=1;l,r\neq k}^p  \|F_{jlr}\|_{\HH}^0   \leq C_2n^{-1/2}e^{-p}.
\end{equation*}
Combining the bounds for $\Delta_4, \Delta_5, \Delta_6$, and applying the Cauchy-Schwarz inequality completes the proof of Lemma \ref{lem:bdonradamacher1}. 
\end{proof}

\medskip
\medskip
\begin{lemma}
\label{lem:bdonradamacher}
Suppose that $H_{jk}\in\HH$. Then there exists some constant $C>0$, such that, for any $c_1>0$ and $c_2>1$, with probability at least $1-2p^{-c_1}$,
\begin{equation*}
\begin{aligned}
\|H_{jk}\|^2_{L_2}&\leq \|H_{jk}R_h^{1/2}\|_{n}^2 + C\Big\{c_2^{-\frac{4\beta_2}{2\beta_2-1}} \|H_{jk}\|^2_{L_2}\log^{-2}\|H_{jk}\|_{L_2} +h^{2\beta_2} + c_2^{\frac{4\beta_2}{2\beta_2-1}}\left( \frac{nh}{\log n} \right)^{-\frac{2\beta_2}{2\beta2+1}} \\
& + \sqrt{c_1+1}\frac{\log p}{n} +\sqrt{(c_1+1)\frac{\log p}{n}}(\|H_{jk}\|_{L_2}+h^{\beta_2})+n^{-1/2}e^{-p} \Big\}.
\end{aligned}
\end{equation*}
\end{lemma}
\begin{proof}
Note that
\begin{equation}
\label{eqn:bdonhjrh}
\|H_{jk}\|_{L_2}^2 - \|H_{jk}R_h^{1/2}\|_n^2\leq \underset{\begin{subarray}{c} 
g\in\HH,\|g\|_\HH^0\leq \|H_{jk}\|_\HH^0 \\ 
\|g\|_{L_2}\leq \|H_{jk}\|_{L_2}
\end{subarray}}{\sup}\left(\|g\|_{L_2}^2-\|gR_h^{1/2}\|_n^2\right).
\end{equation}
By the Talagrand's concentration inequality \citep{Talagrand1996}, with probability at least $1-e^{-c_1}$, we have that, 
\begin{equation}
\label{eqn:decomp1}
\begin{aligned}
&\underset{\begin{subarray}{c}
g\in\HH,\|g\|_\HH^0\leq \|H_{jk}\|_\HH^0 \\
\|g\|_{L_2}\leq \|H_{jk}\|_{L_2}
\end{subarray}}{\sup}\left(\|g\|_{L_2}^2-\|gR_h^{1/2}\|_n^2\right) \\
&\leq 2 \E \underset{\begin{subarray}{c}
g\in\HH,\|g\|_\HH^0\leq \|H_{jk}\|_\HH^0 \\
\|g\|_{L_2}\leq \|H_{jk}\|_{L_2}
\end{subarray}}{\sup}\left(\|g\|_{L_2}^2-\|gR_h^{1/2}\|_n^2\right)+4\|H_{jk}\|_{L_2}\sqrt{\frac{c_1}{n}}+\frac{16c_1}{n}.
\end{aligned}
\end{equation}
By the symmetrization inequality for the Rademacher process \citep{Vandevaart1996} and Theorem 3 of \citet{fan1993local}, there exists a constant $C_1>0$, such that,
\begin{equation}
\label{eqn:decomp2}
\begin{aligned}
&\E \underset{\begin{subarray}{c}
  g\in\HH,\|g\|_\HH^0\leq \|H_{jk}\|_\HH^0 \\
  \|g\|_{L_2}\leq \|H_{jk}\|_{L_2}
  \end{subarray}}{\sup}\left(\|g\|_{L_2}^2-\|gR^{1/2}_h\|_n^2\right) \\
  &\leq \E \underset{\begin{subarray}{c}
  g\in\HH,\|g\|_\HH^0\leq \|H_{jk}\|_\HH^0 \\
  \|g\|_{L_2}\leq \|H_{jk}\|_{L_2}
  \end{subarray}}{\sup}\left\{ \frac{1}{n}\sum_{i=1}^n\omega_iR_h(t_i)g^2(x(t_i)) \right\}+C_1h^{2\beta_2} \\
  & \leq C_1\E \underset{\begin{subarray}{c}
  g\in\HH,\|g\|_\HH^0\leq \|H_{jk}\|_\HH^0 \\
  \|g\|_{L_2}\leq \|H_{jk}\|_{L_2}
  \end{subarray}}{\sup}\left\{ \frac{1}{n}\sum_{i=1}^n\omega_iR_h(t_i)g(x(t_i)) \right\} +Ch^{2\beta_2},
\end{aligned}
\end{equation}
where $\omega,\ldots,\omega_n$ are independent random variables drawn from the Rademacher distribution; i.e., $\P(\omega_i = 1)=\P(\omega_i=-1)=1/2$, for $i=1,\ldots,n$. The last inequality in \eqref{eqn:decomp2} is due to the contraction inequality, and the fact that $g^2$ is a Lipschitz function. Henceforth, with Talagrand's concentration inequality, there exists a constant $C_2>0$, such that,  with probability at least $1-e^{-c_1}$,
\begin{equation}
\label{eqn:decomp3}
\begin{aligned}
&\E \underset{\begin{subarray}{c}
  g\in\HH,\|g\|_\HH^0\leq \|H_{jk}\|_\HH^0 \\
  \|g\|_{L_2}\leq \|H_{jk}\|_{L_2}
  \end{subarray}}{\sup}\left\{ \frac{1}{n}\sum_{i=1}^n\omega_iR_h(t_i)g(x(t_i))\right\} \\
\leq \; & C_2\left[ \underset{\begin{subarray}{c}
  g\in\HH,\|g\|_\HH^0\leq \|H_{jk}\|_\HH^0 \\
  \|g\|_{L_2}\leq \|H_{jk}\|_{L_2}
  \end{subarray}}{\sup}\sum_{l,r=1}^p\left\{ \frac{1}{n}\sum_{i=1}^n\omega_iR_h(t_i)g_{lr}(x(t_i)) \right\} +c_1h^{\beta_2}+ \|H_{jk}\|_{L_2}\sqrt{\frac{c_1}{n}}+\frac{c_1}{n}\right].
\end{aligned}
\end{equation}
By Lemma 2.2 of \citet{Yuan2016}, and the result that the $\nu$th eigenvalue of the RKHS $\HH$ is of order $(\nu\log^{-1}\nu)^{-2\beta_2}$, for $\nu\geq 1$ \citep{Bach2017}, there exists a constant $C_3>0$, such that, with probability at least $1-d^{-c_1}$,
\begin{equation*}
\begin{aligned}
&\sum_{l,r=1}^p\left\{ \frac{1}{n}\sum_{i=1}^n\omega_iR_h(t_i)g_{lr}(x(t_i)) \right\} \\
\leq \; & C_3 n^{-1/2}\sum_{l,r=1}^p\left\{ \left( \|g_{lr}R_h\|_{\HH}\log^{-1}\|g_{lr}R_h\|_{\HH} \right)^{\frac{1}{2\beta_2}} \left(\|g_{lr}R_h\|_{L_2}\log^{-1}\|g_{lr}R_h\|_{L_2} \right)^{1-\frac{1}{2\beta_2}} \right. \\
& \left. + \|g_{lr}R_h\|_{L_2}\sqrt{(c_1+1)\log p}+e^{-p}\|g_{lr}R_h\|_\HH \right\}.
\end{aligned}
\end{equation*}
Following the arguments for bounding $\Delta_4$ in \eqref{eqn:empmean}, there exists a constant $C_4>0$ and for any $c_2>1$, such that,
\vspace{-0.01in}
\begin{equation*}
\begin{aligned}
& n^{-1/2}\underset{\begin{subarray}{c}
  g\in\HH,\|g\|_\HH^0\leq \|H_{jk}\|_\HH^0 \\
  \|g\|_{L_2}\leq \|H_{jk}\|_{L_2}
  \end{subarray}}{\sup} \sum_{l,r=1}^p\left( \|g_{lr}R_h\|_{\HH}\log^{-1}\|g_{lr}R_h\|_{\HH} \right)^{\frac{1}{2\beta_2}} \left( \|g_{lr}R_h\|_{L_2}\log^{-1}\|g_{lr}R_h\|_{L_2} \right)^{1-\frac{1}{2\beta_2}}\\
\leq \; & C_4\underset{\begin{subarray}{c}
  g\in\HH,\|g\|_\HH^0\leq \|H_{jk}\|_\HH^0 \\
  \|g\|_{L_2}\leq \|F_j\|_{L_2}
  \end{subarray}}{\sup}\sum_{l,r=1}^p\left(\|g_{lr}\|_{L_2}\log^{-1}\|g_{lr}\|_{L_2}\right)^{2}+ C_4h^{2m} + C_4\left(\frac{nh}{\log n}\right)^{-\frac{2\beta_2}{2\beta2+1}} +C_4\frac{\log p}{n}\\
\leq \; & C_4C_5\|H_{jk}\|^2_{L_2}\log^{-2}\|H_{jk}\|_{L_2} + C_4h^{2m} + C_4\left(\frac{nh}{\log n}\right)^{-\frac{2\beta_2}{2\beta2+1}} +C_4\frac{\log p}{n}.
\end{aligned}
\end{equation*}
Here the last step is due to  $\sum_{l,r=1}^p \left( \|g_{lr}\|_{L_2}\log^{-1}\|g_{lr}\|_{L_2} \right)^2 \leq C_5\left( \|H_{jk}\|_{L_2}\log^{-1}\|H_{jk}\|_{L_2} \right)^2$ for some constant $C_5>1$. Following the arguments for bounding $\Delta_5$ in  \eqref{eqn:empmean}, and by Theorem 3 of \citet{fan1993local}, there exists a constant $C_6>0$, such that, 
\begin{equation*}
\begin{aligned}
\sum_{l,r=1}^p\|g_{lr}R_h\|_{L_2} &\leq C_6\left\{ \sqrt{\frac{\log p}{n}}+\|H_{jk}\|_{L_2} + h^{\beta_2} \right\}.
\end{aligned}
\end{equation*}
Henceforth, for some constant $C_7>0$,
\begin{equation*}
\begin{aligned}
 &\underset{\begin{subarray}{c}
  g\in\HH,\|g\|_\HH^0\leq \|H_{jk}\|_\HH^0 \\
  \|g\|_{L_2}\leq \|H_{jk}\|_{L_2}
  \end{subarray}}{\sup}\sum_{l,r=1}^p\left\{\frac{1}{n}\sum_{i=1}^n\omega_iR_h(t_i)g_{lr}(x(t_i))\right\} \\
\leq \; & C_7\left\{ c_2^{-\frac{4\beta_2}{2\beta_2-1}}\|H_{jk}\|^2_{L_2}\log^{-2}\|H_{jk}\|_{L_2}+h^{2\beta_2} +c_2^{\frac{4\beta_2}{4\beta_2+1}}\left(\frac{nh}{\log n}\right)^{-\frac{2\beta_2}{2\beta2+1}} \right\}\\
& + C_7\sqrt{\frac{(c_1+1)\log p}{n}}\left\{ \sqrt{\frac{\log p}{n}}+\|H_{jk}\|_{L_2}+h^{\beta_2} \right\} + C_7n^{-1/2}e^{-p}.
\end{aligned}
\end{equation*}
Together with (\ref{eqn:decomp1}), (\ref{eqn:decomp2}), and (\ref{eqn:decomp3}), we have, with probability at least $1-2e^{-c_1}$,
\begin{equation}
\label{eqn:bdongl2grh}
\begin{aligned}
&\underset{\begin{subarray}{c}
  g\in\HH,\|g\|_\HH^0\leq \|H_{jk}\|_\HH^0 \\
  \|g\|_{L_2}\leq \|H_{jk}\|_{L_2}
  \end{subarray}}{\sup}\left(\|g\|_{L_2}^2-\|gR_h^{1/2}\|_n^2\right)\\
\leq \; & C_8\left\{ c_2^{-\frac{4\beta_2}{2\beta_2-1}}\|H_{jk}\|^2_{L_2}\log^{-2}\|H_{jk}\|_{L_2} +h^{2\beta_2}+c_2^{\frac{4\beta_2}{4\beta_2+1}}\left(\frac{nh}{\log n}\right)^{-\frac{2\beta_2}{2\beta2+1}} \right\}\\
& + C_8\sqrt{\frac{(c_1+1)\log p}{n}}\left\{ \sqrt{\frac{\log p}{n}}+\|H_{jk}\|_{L_2} +h^{\beta_2} \right\} + C_8n^{-1/2}e^{-p},\\
\end{aligned}
\end{equation}
for some constant $C_8>0$. 
Combining \eqref{eqn:bdonhjrh} and \eqref{eqn:bdongl2grh} completes the proof of Lemma \ref{lem:bdonradamacher}. 
\end{proof}

%%%%%%%%%%%%%%%%%%%%%%%%%%%%%%%%%%%%%%%%%%%%%%%%%%%
\subsection{Proof of Theorem \ref{thm:postselection}}

\begin{proof}
Define the empirical process,
\vspace{-0.01in}
\begin{equation*}
\widetilde{Z}_n(t_0) \equiv \sqrt{nh}\cdot\widehat{\sigma}^{-1}_n(t_0)\left[  \widehat{F}_{jk}(x_k(t_0))- F_{jk}(x_k(t_0))  \right],\quad\forall t_0\in\Tcal.
\end{equation*}
Define $\widetilde{V}^Z \equiv \sup_{t_0\in\Tcal}\widetilde{Z}_n(t_0)$. We divide the proof of this theorem into four steps.

\medskip
\noindent
\textbf{Step 1}. We aim to prove the following statement: There exists a Gaussian process $\widetilde{\mathbb H}_n(t_0)$, such that  $\E[\sup_{t_0\in\Tcal}\widetilde{\mathbb H}_n(t_0) ]\leq C\sqrt{\log n}$, for some constant $C>0$, and  a sequence of random variables $W_n^0$, such that $W_n^0 =\sup_{t_0\in\Tcal}\widetilde{\mathbb H}_n(t_0)$ and $\P\left( |W_n^0-\widetilde{V}^Z|>\epsilon_{1n} \right) < \delta_{1n}$, for some $(\epsilon_{1n},\delta_{1n})\to 0$, as $n\to\infty$.

We construct the Gaussian process $\widetilde{\mathbb H}_n(t_0) $ as
\begin{equation*}
\begin{aligned}
\widetilde{\mathbb H}_n(t_0) & = \frac{1}{\sqrt{nh^{-1}}}\sum_{i=1}^n \epsilon_i\frac{R_h(t_i-t_0)R_{t_0,i\cdot}^\top \Sigma_{k\cdot}}{\widehat{\sigma}_n(t_0)},
\end{aligned}
\end{equation*}
where $\epsilon_i$ is the error term in (\ref{eqn:obserdata}). Then $\widetilde{\mathbb H}_n(t_0)$ is a Gaussian variable conditional on $\{t_1,\ldots,t_n\}$. By the Jensen's inequality, there exists some constant $C>0$, such that, 
\begin{equation*}
\exp\left[ s\E(W_n^0) \right] \leq \E\exp\left( sW_n^0 \right) = \E\left\{ \sup_{t_0\in\Tcal}\exp\left[s\widetilde{\mathbb H}_n(t_0)\right] \right\} \leq n\exp\left( Cs^2 \right), 
\end{equation*}
for $s>0$, where the last inequality follows from the definition of the Gaussian moment generating function. Rewriting this inequality, we have $\E(W_n^0) \leq \log n/s + Cs$. Setting $s = \sqrt{\log n/C}$, we obtain that, 
\begin{equation*}
\E\left[ \sup_{t_0\in\Tcal}\widetilde{\mathbb H}_n(t_0) \right] \leq \sqrt{C\log n}.
\end{equation*}
By the Cauchy-Schwarz inequality, there exists a constant $c>0$, such that,
\begin{equation} \label{eqn:difofhnzn}
\begin{aligned}
\widetilde{Z}_n(t_0) - \widetilde{\mathbb H}_n(t_0)  \leq O\left( \sqrt{N}\{\E[\delta^2_0(X)]\}^{1/2} \right) + O_p(N^{-c}).
\end{aligned}
\end{equation}
Define $V_n^0 \equiv \sup_{t_0\in\Tcal}\widetilde{\mathbb H}_n(t_0)$. Recall that $\widetilde{V}^Z = \sup_{t_0\in\Tcal}\widetilde{Z}_n(t_0)$. Then by (\ref{eqn:difofhnzn}), there exists some constant $C>0$, such that, 
\begin{equation} \label{eqn:bdonvnz}
\P\left( \left| V_N^0-\widetilde{V}^Z \right| > CN^{-c} \right) \leq \P\left(\sup_{x\in\Xcal^p} \left| \widetilde{\mathbb H}_N(x)-\widetilde{Z}_N(x) \right| > CN^{-c}\right)\leq N^{-1}.
\end{equation}
Setting $\epsilon_{1N} = CN^{-c}$, $\delta_{1N} = N^{-1}$, and $W_N^0 \overset{d}{=} V_N^0$ completes the proof of Step 1.

\medskip
\noindent
\textbf{Step 2}. We aim to prove the following anti-concentration inequality for any $\epsilon>0$,
\begin{equation*}
\sup_{s\in\R}\P\left[ \left\vert\sup_{t_0\in\Tcal} \left| \widetilde{\mathbb H}_n(t_0) \right|-s\right\vert \leq \bar{\epsilon} \right] \leq C\bar{\epsilon}\sqrt{\log N}.
\end{equation*}
This is true due to the result of Step 1 and Corollary 2.1 of \citet{chernozhukov2014anti}.

\medskip
\noindent
\textbf{Step 3}.We aim to prove the following statement: Letting $c_n(\alpha)$ and $\widehat{c}_n(\alpha)$ be the $(1-\alpha)$-quantiles of $\widetilde{V}^Z$ and  $V_n^0$, respectively, there exist $\tau_n,\epsilon_{n},\delta_{n}>0$, such that,
\begin{equation*}
\P\big[ \widehat{c}_n(\alpha)<c_n(\alpha+\tau_n)-\epsilon_{n} \big] \leq \delta_{n}, \quad \P\big[ \widehat{c}_n(\alpha)>c_n(\alpha-\tau_n)+\epsilon_{n} \big] \leq \delta_{n},
\end{equation*}
and $(\tau_n, \epsilon_{n},\delta_{n})\to 0$ as $n\to\infty$. 

Recall the Gaussian multiplier process $\widehat{\mathbb H}_n(t_0)$ in Section \ref{sec:band}, which is defined as, 
\begin{equation*}
\widehat{\Hbb}_n(t_0) \equiv \frac{1}{\sqrt{nh^{-1}}}\sum_{i=1}^n\xi_i\cdot\frac{\widehat{\sigma}_jR_h(t_{i}-t_0) R_{t_0,i\cdot}^\top \Sigma_{k\cdot}}{\widehat{\sigma}_n(t_0)}, \;\; \textrm{ for any } t_0\in\Tcal,
\end{equation*}
where $\xi_1,\ldots,\xi_n$ are independent standard normal variables. We consider the process,
\begin{equation*}
\widehat{\mathbb H}^{(1)}_n(t_0) =  \frac{1}{\sqrt{nh^{-1}}}\sum_{i=1}^n\xi_i\cdot\frac{\sigma_jR_h(t_{i}-t_0) R_{t_0,i\cdot}^\top \Sigma_{k\cdot}}{\widehat{\sigma}_n(t_0)}.
\end{equation*}
Let $\widehat{V}_n = \sup_{t_0\in\Tcal}\widehat{\mathbb H}_n(t_0)$, and $\widehat{V}^{(1)}_n = \sup_{t_0\in\Tcal}\widehat{\mathbb H}_n^{(1)}(t_0)$. Denote $\Delta \mathbb H^{(1)}(t_0) = \widehat{\mathbb H}_n^{(1)}(t_0) - \widehat{\mathbb H}_n(t_0)$. By the triangle inequality,
\begin{equation}
\label{eqn:bdonsupdeltah}
\begin{aligned}
\sup_{t_0\in\Tcal} \big| \Delta \mathbb H^{(1)}(t_0) \big| & \leq   \big| \sigma_j-\widehat{\sigma}_j \big|
\sup_{t_0\in\Tcal}\sqrt{h}\cdot\widehat{\sigma}_n^{-1}(t_0)
\left[\sup_{t_0\in\Tcal}\mathcal I^{ \mathbb H}_1(t_0)+\sup_{t_0\in\Tcal}\mathcal I^{ \mathbb H}_2(t_0)\right],
\end{aligned}
\end{equation}
where 
\begin{equation*}
\begin{aligned}
\mathcal I^{ \mathbb H}_1(t_0) & =  \frac{1}{n}\sum_{i=1}^nR_h(t_{i}-t_0) \Big\vert R_{t_0,i\cdot}^\top \Sigma_{k\cdot} - F_j(x(t_0))\Big\vert\\
\mathcal I^{ \mathbb H}_2(t_0) & =  \frac{1}{n}\sum_{i=1}^nR_h(t_{i}-t_0) F_j(x(t_0)).
\end{aligned}
\end{equation*}

For $\mathcal I^{ \mathbb H}_1(t_0)$, by the Cauchy-Schwarz inequality, we have that, 
\begin{equation*}
\begin{aligned}
\sup_{t_0 \in\Tcal}|\mathcal I^{ \mathbb H}_1(t_0)| & \leq \sup_{t_0\in\Tcal}\left(\frac{1}{n}\sum_{i=1}^nR_h(t_{i}-t_0)(\Psi_{i\cdot}^\top(\widehat{\theta}_z-\theta_z))^2\right)^{1/2}\left(\frac{1}{n}\sum_{i=1}^nK_h(X_{i1}-z)\right)^{1/2} \\
&\leq C\inf_{t_0\in\Tcal}\left(\frac{\log(D_n/p_1(z))}{p^2_1(z)}\right)^{1/2}\cdot\sqrt{m}\left(\sqrt{\frac{m\log(dm)}{nh}}+\frac{m}{nh}+\sqrt{\frac{\log(1/h)}{nh}}\right)^{1/2}\\
& = O(r_{nj}^{1/2}).
\end{aligned}
\end{equation*}

For $\mathcal I^{ \mathbb H}_2(t_0)$, we have that,
\begin{equation*}
\begin{aligned}
\sup_{t_0\in\Tcal}|\mathcal I^{ \mathbb H}_2(t_0)| & \leq\sup_{t_0\in\Tcal}\frac{1}{n}\|\Psi^\top W_z\mathbf{1}\|_{2,\infty}\|\theta_z\|_1\\
&\leq \sup_{t_0\in\Tcal}\frac{1}{n}\|W_z^{1/2}\Psi_{\cdot j}\Psi_{\cdot j}^\top W_z^{1/2}\|_2\|W_z^{1/2}\mathbf{1}\|_2\|\theta_z\|_1\\
&\leq\frac{C}{m}\cdot\sup_{t_0\in\Tcal}\sqrt{p(z)}\cdot\sqrt{m}=O(1/\sqrt{m}).
\end{aligned}
\end{equation*}
Therefore,  we have that, 
\begin{equation*}
\P\left(\left|\widehat{V}_n-\widehat{V}_n^{(1)}\right|>C\sqrt{r_n}m^{1/4}\right)\leq n^{-1}.
\end{equation*}

We next bound $|\widehat{\sigma}_j-\sigma_j|$, for $j=1,\ldots,p$. We start with an upper bound on $|\widehat{\sigma}_j^2-\sigma_j^2|$. Let $\widehat{\epsilon}_{ij} = y_{ij} -\int_0^{t_i} \widehat{F}_j(\widehat{x}(t))dt$, $i=1,\ldots,n$. Using the triangle inequality, we have that, 
\begin{equation}
\label{eqn:bdonsigmahatj}
\begin{aligned}
\big| \widehat{\sigma}_j^2 -\sigma_j^2 \big| & \leq \left\vert \frac{1}{n}\sum_{i=1}^n\widehat{\epsilon}_{ij}^2-\frac{1}{n}\sum_{i=1}^n\epsilon_{ij}^2 \right\vert + \left\vert \frac{1}{n}\sum_{i=1}^n\epsilon_{ij}^2 - \sigma_j^2 \right\vert\equiv I_1 + I_2.
\end{aligned}
\end{equation}

To bound $I_{1}$, we have that, 
\begin{equation*}
\begin{aligned}
I_{1} &\leq \frac{1}{n}\sum_{i=1}^n\Big|\widehat{\epsilon}_{ij}^2 - \epsilon_{ij}^2\Big|\leq \frac{1}{n}\sum_{i=1}^n|\widehat{\epsilon}_{ij} - \epsilon_{ij}|\left(|\widehat{\epsilon}_{ij}-\epsilon_{ij}| + 2|\epsilon_{ij}|\right)\\
&\leq \left(\frac{1}{n}\sum_{i=1}^n[\widehat{\epsilon}_{ij}-\epsilon_{ij}]^2\right)^{1/2}\times\left[2\left(\frac{1}{n}\sum_{i=1}^n\epsilon_{ij}^2\right)^{1/2}+\left\{\frac{1}{n}\sum_{i=1}^n[\widehat{\epsilon}_{ij}-\epsilon_{ij}]^2\right\}^{1/2}\right].
\end{aligned}
\end{equation*}
Since $\E(\epsilon_{ij}^2)=\sigma_j^2<\infty$, we have that, 
\begin{equation*}
\left( \frac{1}{n}\sum_{i=1}^n\epsilon_{ij}^2 \right)^{1/2}= O_p(1).
\end{equation*}
Let $r_{nj}\equiv  \tau_{nj}^{-\frac{1}{2\beta_2}} \frac{\log n}{n h} + \tau_{nj} + h^{2\beta_2}+ \frac{\log p}{n} + n^{-\frac{2\beta_1}{2\beta_1+1}}$. By Theorem \ref{thm:optimalestoffunctional}, we have $n^{-1} \sum_{i=1}^n[\widehat{\epsilon}_{ij}-\epsilon_{ij}]^2 = O_p(r_{nj})$. Therefore, 
\begin{equation}
\label{eqn:bdoni1}
I_1 \leq O_p(r_{nj}^{1/2}).
\end{equation}

To bound $I_{2}$, we have that, 
\begin{equation*}
\E(I^2_{2}) \leq n^{-1} \E(\epsilon_{ij}^4) = O(n^{-1}),
\end{equation*}
where the last step is due to that $\epsilon_{ij}$ is a normal random variable and hence $\epsilon_{ij}$ has a bounded fourth moment. Therefore, 
\begin{equation}
\label{eqn:bdoni2}
I_{2}\leq O_p(n^{-1/2}).
\end{equation}
Combining (\ref{eqn:bdonsigmahatj}), (\ref{eqn:bdoni1}), and (\ref{eqn:bdoni2}), we have that, 
\begin{equation*}
\begin{aligned}
\widehat{\sigma}_j -\sigma_j& = O\left( |\widehat{\sigma}_j^2 -\sigma_j^2| \right) \leq O_p(r_{nj}^{1/2}).
\end{aligned}
\end{equation*}

By (\ref{eqn:bdonsupdeltah}), we have that, 
\begin{equation*}
\sup_{t_0\in\Tcal}|\Delta \mathbb H^{(1)}(t_0)| \leq O_p(r_{nj}^{1/2})
\end{equation*}
Then there exists a constant $C>0$, such that,
\begin{equation}
\label{eqn:bdonvnhatvn1}
\P\left( \big| \widehat{V}_n-\widehat{V}_n^{(1)} \big|>Cr_{nj}^{1/2} \right) \leq \P\left( \sup_{t_0\in\Tcal}|\Delta \mathbb H^{(1)}(t_0)|>Cr_{nj}^{1/2} \right)\leq n^{-1}.
\end{equation}
Since $\sigma\xi\overset{d}{=}(\epsilon_{1j},\ldots,\epsilon_{nj})^\top$, we have $\sup_{t_0\in\Tcal}\widehat{\mathbb H}_n^{(1)}(t_0)\overset{d}{=}\sup_{t_0\in\Tcal}\widetilde{\mathbb H}_n(t_0)$. That is, $\widehat{V}^{(1)}_n\overset{d}{=} V_n^0$. Combining (\ref{eqn:bdonvnz}) with (\ref{eqn:bdonvnhatvn1}), we have that,
\begin{equation*}
\P\left(\big| \widehat{V}_n-\widetilde{V}_n^Z \big|>Cn^{-c}\right)\leq n^{-1}.
\end{equation*}
Therefore, by the definition of $\widehat{c}_N(\alpha)$,
\begin{equation*}
\begin{aligned}
\P\left(\widetilde{V}_n^Z\leq \widehat{c}_n(\alpha)+Cn^{-c}\right) \geq \P\left(\widehat{V}_n\leq \widehat{c}_n(\alpha)\right) - \P\left(\big| \widehat{V}_n-\widetilde{V}_n^Z \big| > Cn^{-c}\right)
\geq 1-\alpha-n^{-1},
\end{aligned}
\end{equation*}
which implies that the estimated quantile is lower bounded as, 
\begin{equation*}
\widehat{c}_n(\alpha)\geq c_n(\alpha+n^{-1})-Cn^{-c},\quad\text{for some }c\in(0,c_{\min}].
\end{equation*}
Similarly, we also have $\widehat{c}_n(\alpha)\leq c_n(\alpha-n^{-1})+Cn^{-c}$. Setting $\tau_n=n^{-1}$, $\epsilon_{n} = Cn^{-c}$, and $\delta_{n}=n^{-1}$ completes the proof of Step 3.

\medskip
\noindent
\textbf{Step 4}. By verifying the statements in Steps 1 to 3, we now apply Corollary 3.1 of \citet{chernozhukov2014anti} and obtain that, 
\begin{equation*}
\mathbb P\left(F_{jk}(x_k(t_0)) \in \Ccal_{n,\alpha}, \ \forall t_0\in\Tcal\right)\geq 1-\alpha-Cn^{-c}.
\end{equation*}
Therefore, the confidence interval $\text{CI}(f_0(x))$ is asymptotic honest. This completes the proof of Theorem \ref{thm:postselection}.
\end{proof}

%%%%%%%%%%%%%%%%%%%%%%%%%%%%%%%%%%%%%%%%%%%%%%%%%%%
\subsection{Proof of Theorem \ref{thm:optimalrecovery}}
\label{sec:pfoptimalrecovery}

\noindent
We divide the proof of this theorem into two parts. We first present the main proof in Section \ref{eqn:mainproof}, then give an auxiliary lemma useful for the proof of this theorem in Section \ref{sec:auxlemnew}.

%%%%%%%%%%%%%%%%%%%%%%%%%%%%%%%%%%%%%%%%%%%%%%%%%%%
\subsubsection{Main proof}
\label{eqn:mainproof}

\begin{proof}
We use the primal-dual witness method to prove that the localized kernel ODE approach selects all significant variables, and includes no insignificant ones.  Recall that, by the representer theorem \citep{wahba1990}, the selection problem becomes \eqref{eqn:regfortheta}; i.e., 
\begin{equation}
\label{eqn:zjlasso}
\begin{aligned}
\min_{\thetabf_j}\left\{\frac{1}{n}(\zbf_j - G\thetabf_j)^\top R_{t_0}(\zbf_j - G\thetabf_j) +  \kappa_{nj} \left( \sum_{k'=1, k'\neq k}^p\theta_{jk'}+\sum_{k'=1, k'\neq k}^p\sum_{l=1,  l\neq k', k}^{p}\theta_{jk'l}  \right)\right\},
\end{aligned}
\end{equation}
subject to $\theta_{jk'}\geq 0,\theta_{jk'l} \geq 0$, where the ``response" is $z_j = (y_j - \bar{y}_j) - \widehat{\alpha}_{jk,t_0}\bar{t} - (1/2)n\eta_{nj} c_j$, and the ``predictor" is $G \in \Rbb^{n \times (p-1)^2}$.  The vector $\theta_j$ solves (\ref{eqn:zjlasso}), if it satisfies the Karush-Kuhn-Tucker (KKT) condition,
\begin{equation}
\label{eqn:kkttheta}
\frac{2}{n}G^\top R_{t_0}(G\theta_j-z_j) + \kappa_{nj}g_j = 0, \quad j=1,\ldots,p,
\end{equation}
where  $G$ contains errors in the variables due to the estimated $\widehat{x}(t)$, and
\begin{equation}
\label{eqn:kkttg}
g_j  = \text{sign}(\theta_j), \;\; \text{if }\theta_j\neq 0, \quad\textrm{ and }\quad |g_j| \leq 1, \;\; \text{otherwise}.
\end{equation}

To apply the primal-dual witness method, we next construct an oracle primal-dual pair $(\widehat{\theta}_j,\widehat{g}_j)$ satisfying the KKT conditions (\ref{eqn:kkttheta}) and (\ref{eqn:kkttg}). Specifically,
\begin{itemize}
\item[(a)] We set $\widehat{\theta}_{jk'l}=0$ for $(k',l)\not\in S^*_j$, where $S^*_j$ is as defined in Section \ref{sec:theoryconfidence}, and $s_j = \text{card}(S^*_j)$.

\item[(b)] Let $\widehat{\theta}_{S^*_j}$ be the minimizer of the partial penalized likelihood, 
\begin{equation}
\label{eqn:consttheta}
(\zbf_j - G_{S^*_j}\thetabf_{S^*_j})^\top R_{t_0} (\zbf_j - G_{S^*_j}\thetabf_{S^*_j}) + n \kappa_{nj} \left( \sum_{k'=1, k'\neq k}^p\theta_{jk'}+\sum_{k'=1,k'\neq k, l}^p\sum_{l=1,l\neq k}^{p}\theta_{jk'l} \right).
\end{equation}

\item[(c)] Let $S_j^c$ be the complement of $S^*_j$ in $\{(k',l):k',l=1,\ldots,k-1,k+1,\ldots,p\}$.
We obtain $\widehat{g}_{S_j^c}$ from (\ref{eqn:kkttheta}) by substituting in the values of $\widehat{\theta}_j$ and $\widehat{g}_{S^*_j}$.
\end{itemize}

Next, we verify the support recovery consistency; i.e., 
\begin{equation*}
\label{eqn:supportconsis}
\max_{(k,l)\in S^*_j}\|\widehat{\theta}_{jkl}-\theta_{jkl}\|_{\ell_2}\leq \frac{2}{3}\theta_{\min},
\end{equation*}
which in turn implies that the oracle estimator $\widehat{\theta}_j$ recovers the support of $\theta_j$ exactly. 

Note that the subgradient condition for the partial penalized likelihood (\ref{eqn:consttheta}) is 
\begin{equation*}
\begin{aligned}
2G_{S^*_j}^\top R_{t_0}(G_{S^*_j}\widehat{\theta}_{S^*_j}-z_j)+n\kappa_{nj}\widehat{g}_{S^*_j} = 0,
\end{aligned}
\end{equation*}  
which implies that
\begin{equation*}
\begin{aligned}
2G_{S^*_j}^\top R_{t_0}(G_{S^*_j}\widehat{\theta}_{S^*_j}-G_{S^*_j}\theta_{S^*_j})+2G_{S^*_j}^\top R_{t_0}(G_{S^*_j}\theta_{S^*_j}-z_j)+n\kappa_{nj}\widehat{g}_{S^*_j} = 0.
\end{aligned}
\end{equation*}
Define $\mathcal R_{S^*_j} \equiv 2G_{S^*_j}^\top R_{t_0}G_{S^*_j}\theta_{S^*_j} - 2G_{S^*_j}^\top R_{t_0}z_j$.  Then,
\begin{equation}
\label{eqn:diftheta}
\widehat{\theta}_{S^*_j}-\theta_{S^*_j} = -\left(2G_{S^*_j}^\top R_{t_0}G_{S^*_j}\right)^{-1}(\mathcal R_{S^*_j}+n\kappa_{nj}\widehat{g}_{S^*_j}).
\end{equation}
For each $(k,l)$, denote the corresponding column of $G$ by $G_{kl}$. Then for $(k,l) \in S^*_j$,
\begin{equation}
\label{eqn:defofrk}
\mathcal R_{kl} = 2G_{kl}^\top R_{t_0}G_{S^*_j}\theta_{S^*_j} - 2G_{kl}^\top R_{t_0}z_j.
\end{equation}
By Lemma \ref{lem:conds3}, we have $\|\mathcal R_{kl}\|_{\ell_2}\leq\eta_{\mathcal R}$ for any $(k,l)\in S^*_j$. Then,
\begin{equation}
\label{eqn:rs0}
\|\mathcal R_{S^*_j}\|_{\ell_2}\leq \eta_{\mathcal R}\sqrt{s_j}.
\end{equation}
By Assumption \ref{assump:dependency-incoherence}, we have that $\Lambda_{\min}\left(G_{S^*_j}^\top R_{t_0}G_{S^*_j}\right)\ge C_{\min}/2$, for some constant $C_{\min}>0$. Henceforth,
\begin{equation*}
 \Lambda_{\max}\left\{\left(2G_{S^*_j}^\top R_{t_0}G_{S^*_j}\right)^{-1}\right\}\leq \frac{1}{C_{\min}}.
\end{equation*}
Note that for any $(k,l)\in S^*_j$, $\|\widehat{g}_{jkl}\|_{\ell_2}\leq 1$, which implies that, 
\begin{equation}
\label{eqn:gs0hat}
\|\widehat{g}_{S^*_j}\|_{\ell_2} \leq \sqrt{s_j}.
\end{equation}
Therefore, we have that, 
\begin{equation*}
\max_{(k,l)\in S^*_j}\|\widehat{\theta}_{jkl}-\theta_{jkl}\|_{\ell_2}\leq \|\widehat{\theta}_{S^*_j}-\theta_{S^*_j}\|_{\ell_2}\leq \frac{\eta_{\mathcal R}\sqrt{s_j}}{C_{\min}}+n\kappa_{nj}\frac{\sqrt{s_j}}{C_{\min}}\leq \frac{2}{3}\theta_{\min}.
\end{equation*}
where the last inequality is due to Assumption \ref{assump:minregeffect}.

Next, we verify the strict dual feasibility; i.e., 
\begin{equation*}
\max_{(k,l)\not\in S^*_j}|\widehat{g}_{jkl}|<1,
\end{equation*}
which in turn implies that the oracle estimator $\widehat{\theta}_j$ satisfies the KKT condition of the localized kernel ODE optimization problem.  

For any $(k,l)\not\in S^*_j$, by (\ref{eqn:kkttheta}), we have, 
\begin{equation*}
2G_{kl}^\top R_{t_0}(G_{S^*_j}\widehat{\theta}_{S^*_j}-z_j)+n\kappa_{nj}\widehat{g}_{jkl} = 0,
\end{equation*}
which implies that
\begin{equation*}
\begin{aligned}
&2G_{kl}^\top R_{t_0}(G_{S^*_j}\widehat{\theta}_{S^*_j}-G_{S^*_j}\theta_{S^*_j})+2G_{kl}^\top R_{t_0}(G_{S^*_j}\theta_{S^*_j}-z_j)+n\kappa_{nj}\widehat{g}_{jkl} = 0.
\end{aligned}
\end{equation*}
By (\ref{eqn:diftheta}) and (\ref{eqn:defofrk}), we have, 
\begin{equation*}
n\kappa_{nj}\widehat{g}_{jkl} = G_{kl}^\top R_{t_0}G_{S^*_j}(G_{S_j}^\top G_{S^*_j})^{-1}(\mathcal R_{S^*_{j}}+n\kappa_{nj}\widehat{g}_{S^*_j})-\mathcal R_{kl}.
\end{equation*}
By Assumption \ref{assump:dependency-incoherence} again, and by (\ref{eqn:rs0}) and (\ref{eqn:gs0hat}), we have that,
\begin{equation*}
|\widehat{g}_{jkl}|\leq \frac{(\xi_G+1)\sqrt{s_j}}{n\kappa_{nj}}\eta_{\mathcal R}+\xi_G\sqrt{s_j},\quad (k,l)\not\in S^*_j.
\end{equation*}
By Assumption \ref{assump:minregeffect}, we obtain that, $|\widehat{g}_{jkl}|<1$, for any $(k,l)\not\in S^*_j$.

Finally, the selection consistency for $S^*_j$ implies the selection consistency for $\widehat{S}_j$. This completes the proof of Theorem \ref{thm:optimalrecovery}.
\end{proof}

%%%%%%%%%%%%%%%%%%%%%%%%%%%%%%%%%%%%%%%%%%%%%%%%%%%
\subsubsection{Auxiliary lemma for Theorem \ref{thm:optimalrecovery}}
\label{sec:auxlemnew}

\noindent 
The next lemma gives a bound similar to the deviation condition in \citet{loh2012} and \citet{dai2021kernel}. The difference is that, the noise in the variable $\widehat{x}(t)$ in our setting involves a nonlinear transformation through the kernel $K(\widehat{x}(t),\widehat{x}(s))$. Besides, we have adopted the localized learning.

\begin{lemma}
\label{lem:conds3}
For $j=1,\ldots,p$, we have, 
\begin{equation*}
\|G^\top_{kl} R_{t_0}G_{S^*_j}\theta_{S^*_j} - G_{kl}^\top R_{t_0}z_j\|_{\ell_2}\leq \eta_{\mathcal R}, 
\end{equation*}
where 
\begin{equation*}
\eta_{\mathcal R} = O_p\left( \left(\frac{n}{\log n}\right)^{-\frac{\beta_2}{2(\beta_2+1)}} + \left(\frac{\log p}{n}\right)^{1/2} + n^{-\frac{\beta_1}{2\beta_1+1}} \right).
\end{equation*}
\end{lemma}
\begin{proof}
Similar to the ``predictor" $G$ defined in \eqref{eqn:regfortheta} in Section \ref{sec:compalgo}, we first construct a noiseless version of the ``predictor", $\widetilde{G} \in \R^{n \times p^2}$, whose first $p$ columns are $\widetilde{\Sigmabf}^{k'} \cbf_j$, the last $p(p-1)$ columns are $\widetilde{\Sigma}^{k'l} \cbf_j$, and $\widetilde{\Sigma}^{k'} = (\widetilde{\Sigma}^{k'}_{ii'}), \widetilde{\Sigma}^{k'l} = (\widetilde{\Sigma}^{k'l}_{ii'})$ are both $n\times n$ matrices whose $(i,i')$th entries are,  
\begin{equation*}
\begin{aligned}
\widetilde{\Sigma}^{k'}_{ii'} & = \int_\TT\int_\TT\{ T_i(s)-\bar{T}(s) \} K_{k'}(\xbf(t),\xbf(s)) \{ T_{i'}(t)-\bar{T}(t) \} dsdt,   \;\; 1 \leq k' \leq p, 1 \leq i, i' \leq n, \\
\widetilde{\Sigma}^{k'l}_{ii'} & = \int_\TT\int_\TT\{ T_i(s)-\bar{T}(s) \} K_{k'l}(\xbf(t),\xbf(s)) \{ T_{i'}(t)-\bar{T}(t) \} dsdt, \;\; 1 \leq k' < l \leq p, 1 \leq i, i' \leq n.
\end{aligned}
\end{equation*}

Next, we consider the term $\left\|G^\top_{kl}R_{t_0}z_j - G_{kl}^\top R_{t_0}G_{S^*_j}\theta_{S^*_j} \right\|_{\ell_2}$, which can be bounded as, 
\begin{align}
\label{eqn:Gdecomp}
\begin{split}
\left\|G^\top_{kl}R_{t_0}\left(z_j - G_{S^*_j}\theta_{S^*_j} \right)\right\|_{\ell_2} 
\leq \; & \left\|G^\top_{kl}R_{t_0}\left(\E[z_j] - \widetilde{G}_{S^*_j}\theta_{S^*_j}\right)\right\|_{\ell_2} + \left\|G^\top_{kl}R_{t_0}\left(\widetilde{G}_{S^*_j}-G_{S^*_j}\right) \theta_{S^*_j} \right\|_{\ell_2} \\ 
& + \left\|G^\top_{kl}R_{t_0}\left(z_j-\E[z_j]\right)\right\|_{\ell_2} \\
\equiv \; & \Delta_7 + \Delta_8 + \Delta_9.
\end{split}
\end{align}
We next bound the three terms $\Delta_7, \Delta_8, \Delta_9$ on the right-hand-side of (\ref{eqn:Gdecomp}), respectively.

For $\Delta_7$, by the Cauchy-Schwarz inequality and Theorem \ref{thm:optimalestoffunctional}, we have, 
\begin{equation*}
\begin{aligned}
\Delta^2_7 
&\leq \left\| G^\top_{kl} \right\|_{\ell_2}^2 \left\| R_{t_0}\left(\E[z_j]-\widetilde{G}_{S^*_j}\theta_{S^*_j}\right) \right\|_{\ell_2}^2 \\
&\leq C_1\left\| R_{t_0} \left(\E[z_j]-\widetilde{G}_{S^*_j}\theta_{S^*_j}\right) \right\|_{\ell_2}^2 
= O_p\left( \left(\frac{n}{\log n}\right)^{-\frac{2\beta_2}{2(\beta_2+1)}} + \frac{\log p}{n} \right),
\end{aligned}
\end{equation*}
for some constant $C_1>0$, where the last step is by (\ref{eqn:step2term1}).

For $\Delta_8$, again by the Cauchy-Schwarz inequality, we have,
\begin{equation*}
\begin{aligned}
\Delta^2_8 \leq \left\| G^\top_{kl} \right\|_{\ell_2}^2 \left\|R_{t_0} \left( \widetilde{G}_{S^*_j}-G_{S^*_j} \right)\theta_{S^*_j} \right\|_{\ell_2}^2 \leq C_2 \left\|R_{t_0}\left( \widetilde{G}_{S^*_j}-G_{S^*_j} \right) \right\|^2_\infty \left\| \theta_{S^*_j} \right\|^2_{\ell_1} = O_p\left( n^{-\frac{2\beta_1}{2\beta_1+1}} \right),
\end{aligned}
\end{equation*}
for some constants $C_2>0$, where the last step is by (\ref{eqn:bdonerrorinx}), and the fact that $\|\theta_{S^*_j}\|_{\ell_1}$ is bounded. 

For $\Delta_9$, by Lemma \ref{lem:bdonradamacher}, we have, 
\begin{equation*}
\begin{aligned}
\Delta^2_9 = O_p\left( \left( \frac{n}{\log n} \right)^{-\frac{2\beta_2}{2(\beta_2+1)}} + \frac{\log p}{n} \right).
\end{aligned}
\end{equation*}

Combining the above three bounds, we obtain that, 
\begin{equation*}
\left\| G^\top_{kl}R_{t_0}(z_j -G_{S^*_j}\theta_{S^*_j}) \right\|_{\ell_2} = O_p\left( \left(\frac{n}{\log n}\right)^{-\frac{\beta_2}{2(\beta_2+1)}} + \left(\frac{\log p}{n}\right)^{1/2} + n^{-\frac{\beta_1}{2\beta_1+1}} \right),
\end{equation*}
which completes the proof of Lemma \ref{lem:conds3}.
\end{proof}

%%%%%%%%%%%%%%%%%%%%%%%%%%%%%%%%%%%%%%%%%%%%%%%%%%%
\section{Additional Theoretical Discussions}
\label{sec:add-theoretical}

In this section, we discuss more on the averaging operator introduced in Section \ref{sec:model}. We also analyze Assumption \ref{assump:dependency-incoherence} in Section \ref{sec:theoryconfidence} in more detail.

%%%%%%%%%%%%%%%%%%%%%%%%%%%%%%%%%%%%%%%%%%%%%%%%%%%
\subsection{Averaging Operator}
\label{sec:ave-operator}

Consider a standard one-way ANOVA model, $Y_{ti} = f_t + \epsilon_{ti}$, where $f_t$ denotes the treatment mean at treatment level $t=1,\ldots,T$, $i=1,\ldots,n$ indexes the sample observations, and $\epsilon_{ti}$ is an independent normal error.  The ANOVA decomposition is written as, 
\begin{equation*}
f_t = \theta_0 +\alpha_t,
\end{equation*} 
where $\theta_0$ is the global mean, and $\alpha_t$ is the treatment effect at level $t$. The parameters $\theta_0$ and $\alpha_t$ are identifiable through a side condition, where a common choice is that $\sum_{t=1}^T\alpha_t=0$.

Similarly, the one-way ANOVA model on a continuous domain $\mathcal T$ can be cast as $Y_i = f(t_i) + \epsilon_i$, where $t\in\mathcal T$. The ANOVA decomposition is written as,  
\begin{equation*}
f(t) = \mathcal Af + (I-\mathcal A)f,
\end{equation*}
where $\mathcal A$ is an averaging operator that ``averages out" the covariate $t$ to return a constant function, and $I$ is the identity operator.  For instance, with $\mathcal Af = \int_{\mathcal T}f(t)dt$, one has $f(t) = \int_{\mathcal T}f(t)dt+ \left\{f(x)- \int_{\mathcal T}f(t)dt\right\}$, corresponding to $\sum_{t=1}^T\alpha_t=0$ in the standard one-way ANOVA model. Note that applying $\mathcal A$ to a constant function returns that constant, hence the name ``averaging."  It follows that $\mathcal A(\mathcal A f) = \mathcal A f$, or simply, $\mathcal A^2=\mathcal A$. Write the constant term $\theta_0=\mathcal Af$, which denotes the global mean.  Write the term $\tilde{f}=(I-\mathcal A)f$, which denotes the treatment effect that satisfies the side condition $\mathcal A\tilde{f}=\int_{\mathcal T}\tilde{f}(t)dt=0$.  Following this reasoning, we define the averaging operator for our model \eqref{eqn:nonadditivemodel} as 
\begin{equation*}
\mathcal A F_j(x(t)) = \int_{\mathcal T}F_j(x(t))dt. 
\end{equation*}
Then in the construction of the RKHS $\mathcal H$ in model \eqref{eqn:spaceH}, a sufficient side condition is the zero marginal integral for each $k=1,\ldots,p$, such that
\begin{equation*}
\int_{\mathcal T} F_{jk}(x_k(t))dt = 0,\;\; \textrm{ for any } \; k=1,\ldots,p.
\end{equation*}
Such an averaging operator has been commonly used in the RKHS literature; see, e.g., \citet{Wahba1995, gu2013, LinZhang2006, dai2021kernel}.

%%%%%%%%%%%%%%%%%%%%%%%%%%%%%%%%%%%%%%%%%%%%%%%%%%%
\subsection{Discussion on Assumption \ref{assump:dependency-incoherence}}
\label{sec:assumption5}

We further study Assumption \ref{assump:dependency-incoherence}, and show that the bandwidth $h$ in the localization and $R_{t_0}$ does not affect the validity of this assumption,  as long as $h\to 0$ when $n\to\infty$. 

We start with the first part of Assumption 5. Note that,
\begin{equation*}
\begin{aligned}
\Lambda_{\min}\left(G_{S^*_j}^\top R_{t_0}G_{S^*_j}\right) & = \left[\sigma_{\min}\left(R^{1/2}_{t_0}G_{S^*_j}\right)\right]^2\\
& \geq \left[\sigma_{\min}\left(R^{1/2}_{t_0}\right)\sigma_{\min}(G_{S^*_j})\right]^2 = \Lambda_{\min}(R_{t_0})\Lambda_{\min}(G_{S^*_j}^\top G_{S^*_j}).
\end{aligned}
\end{equation*}
Here $\sigma_{\min}$ denotes the minimum singular value and $\Lambda_{\min}$ denotes the minimum eigenvalue. Since the bandwidth $h\to 0$ as $n\to\infty$, we have that $R_h(t_i-t_0)\to \frac{1}{t_i-t_0}c_R$, where $c_R=\lim_{h\to0}h^{-1}R(h^{-1})>0$ is a constant. Therefore, as $n\to\infty$, $\Lambda_{\min}(R_{t_0})=\min_i\frac{1}{t_i-t_0}c_R>0$, which does not depend on the bandwidth $h$. On the other hand, Assumption 3 in \citet{dai2021kernel} states that $\Lambda_{\min}(G_{S^*_j}^\top G_{S^*_j})$ is lower bounded by a constant. Together, it implies that the bandwidth $h$ does not affect the validity of the first part of this assumption.

Similarly, for the second part of Assumption 5, we have that, 
\begin{equation*}
\begin{aligned}
\left\|G_{kl}^\top R_{t_0}G_{S^*_j}(G_{S^*_j}^\top R_{t_0}G_{S^*_j})^{-1}\right\|_{\ell_2} 
& \leq \left[\Lambda_{\min}\left(G_{S^*_j}^\top R_{t_0}G_{S^*_j}\right) \right]^{-1}\left\|G_{kl}^\top R_{t_0}G_{S^*_j}\right\|_{\ell_2}\\
& \leq \Lambda_{\max}(R_{t_0})\left[ \Lambda_{\min}(R_{t_0})\right]^{-1}\left[\Lambda_{\min}(G_{S^*_j}^\top G_{S^*_j})\right]^{-1}\left\|G_{kl}^\top G_{S^*_j}\right\|_{\ell_2}.
\end{aligned}
\end{equation*}
Since $R_h(t_i-t_0)\to \frac{1}{t_i-t_0}c_R$,  $\Lambda_{\max}(R_{t_0})\left[ \Lambda_{\min}(R_{t_0})\right]^{-1}\leq \max_{i,j}\frac{t_i-t_0}{t_{j}-t_0}$ as $n\to\infty$, which does not depend on the bandwidth $h$.  Again, Assumption 4 in \citet{dai2021kernel} states that $\max_{(k,l)\not\in S^*_j}\left\|G_{kl}^\top G_{S^*_j}(G_{S^*_j}^\top G_{S^*_j})^{-1}\right\|_{\ell_2}\leq \left[\Lambda_{\min}(G_{S^*_j}^\top G_{S^*_j})\right]^{-1}\left\|G_{kl}^\top G_{S^*_j}\right\|_{\ell_2}$ is upper bounded by $\xi_G$. Together, it implies that the bandwidth $h$ does not affect the validity of the second part of this assumption 5.

%%%%%%%%%%%%%%%%%%%%%%%%%%%%%%%%%%%%%%%%%%%%%%%%%%%
\section{Additional Numerical Analyses}
\label{sec:add-numerical}

In this section, we report some additional numerical results for the simulation study in Section \ref{sec:simulation}. We also carry out a sensitivity analysis to investigate the effect of the choice of the local weight function and bandwidth.

%%%%%%%%%%%%%%%%%%%%%%%%%%%%%%%%%%%%%%%%%%%%%%%%%%%
\subsection{Additional Results about Enzymatic Regulation Equations}
\label{sec:add-ex1}

\begin{figure}[t!]
\centering
\includegraphics[width=\textwidth]{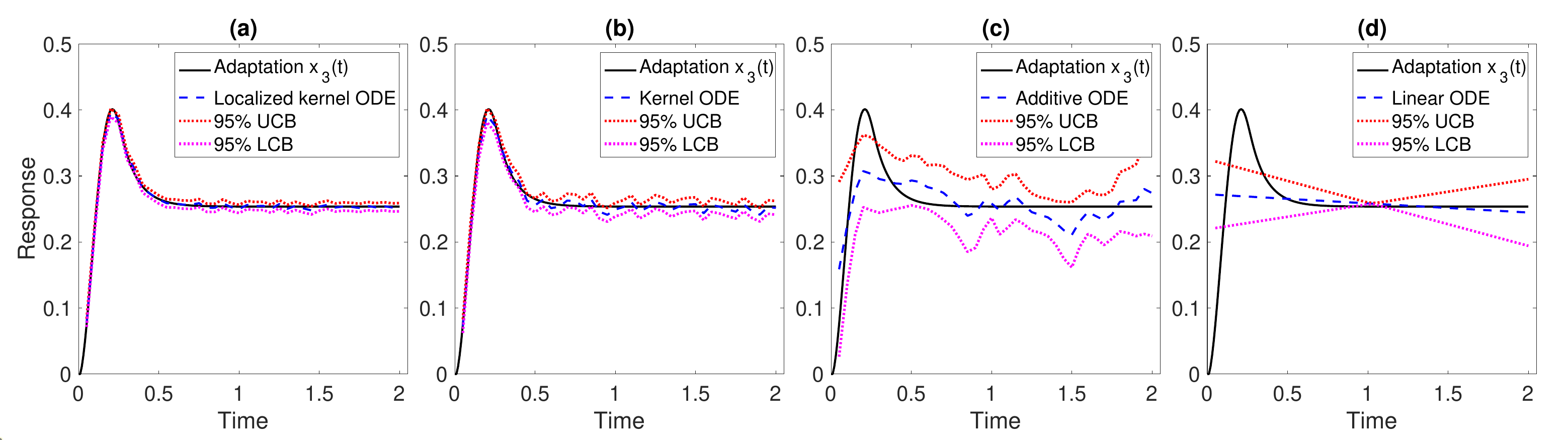}
\caption{The true (black solid line) and the estimated (blue dashed line) trajectory of $x_3(t)$, with the $95\%$ upper and lower confidence bounds (red dotted lines). The results are averaged over $500$ data replications. (a) Localized kernel ODE; (b)  Kernel ODE; (c) Additive ODE; (d) Linear ODE.}
\label{fig:visual}
\end{figure}

Figure \ref{fig:visual} reports the true and estimated trajectory of $x_3(t)$, with $95\%$ upper and lower confidence bounds, of the four ODE methods. The noise level is set as $\sigma_j = 0.1, j=1, 2, 3$, and the results are averaged over $500$ data replications. It is seen that the localized kernel ODE estimate has a smaller variance than its counterparts, including the kernel ODE method. Additionally, the confidence intervals of localized kernel ODE and kernel ODE achieve the desired coverage for the true trajectory. In contrast, the confidence intervals of additive and linear ODE models mostly fail to include the truth.

\begin{figure}[t!]
\centering
\includegraphics[width=\textwidth]{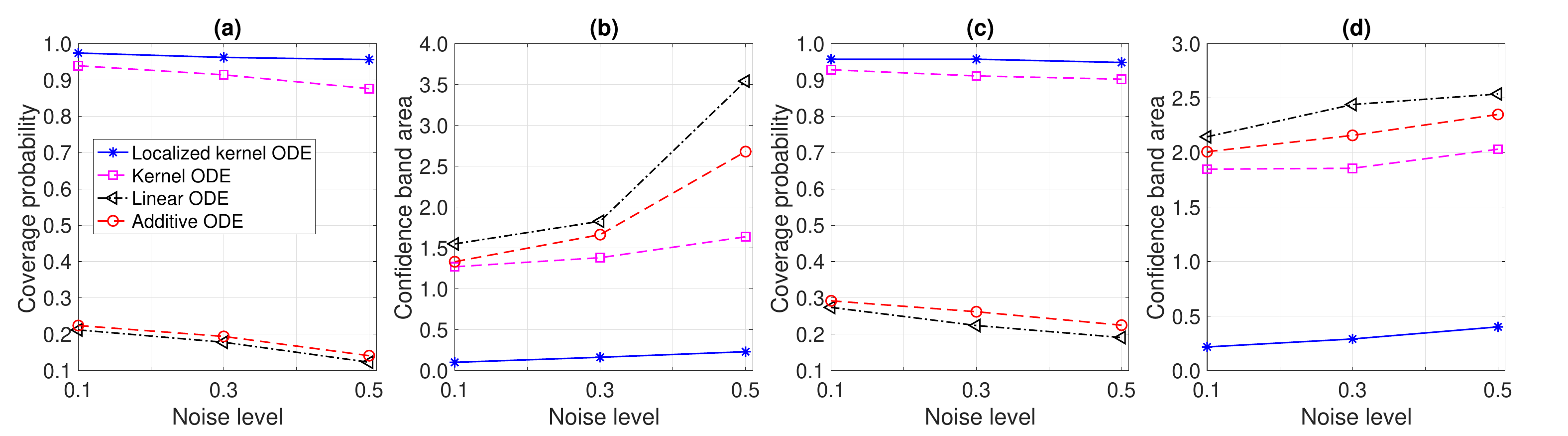}
\caption{The NFBLB example: the empirical coverage probability and confidence band area for the varying noise level $\sigma_j$. The results are averaged over 500 data replications. (a)-(b): Nonzero functional $F_{23}(x_3(t))$; (c)-(d): Zero functional $F_{12}(x_2(t))$.}
\label{fig:eg1-coverage}
\end{figure}

Figure \ref{fig:eg1-coverage} reports the coverage probability and confidence band area for the varying noise level of different ODE methods, and is a visualization of Table \ref{table:enzymatic} in Section \ref{sec:enzymatic}. We see that the inference method based on the localized kernel ODE clearly outperforms the alternative solutions with a larger coverage probability and a more tight confidence band.

\begin{figure}[t!]
\centering
\includegraphics[width=\textwidth]{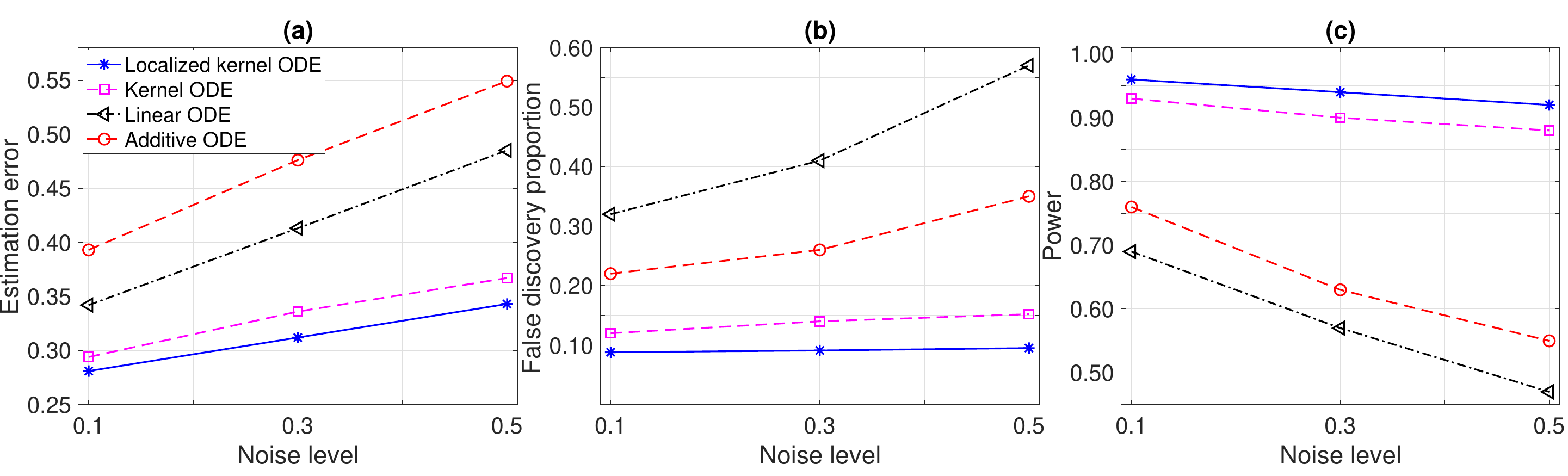}
\caption{The NFBLB example: the estimation and sparse selection performance for the varying noise level $\sigma_j$. The results are averaged over 500 data replications.}
\label{fig:eg1-prediction}
\end{figure}

Figure \ref{fig:eg1-prediction} reports the empirical FDR, power, and trajectory estimation error for the varying noise level when we aim to recover the entire regulatory system through the proposed confidence band coupled with the BH procedure for multiple testing correction at the FDR level of $10\%$. Here, the empirical FDR and power are the same as defined in Section \ref{sec:lotkavolterra}. The estimation accuracy of $\widehat{F}_{j}(\widehat{x}(t))$ is defined as the squared root of the sum of mean squared errors for $F_j(x_j(t)),j=1,2,3$, at $t\in[0,2]$, i.e., $\left\{\sum_{j=1}^{3}\int_{0}^{2}[\widehat{F}_j(\widehat{x}_j(t))-F_j(x_j(t))]^2dt\right\}^{1/2}$, where the integral is evaluated at $10000$ evenly distributed time points in $[0,2]$. We see that the inference method based on the localized kernel ODE successfully controls the FDR under the nominal level, and outperforms the three alternative solutions in terms of the empirical power and the estimation error.

\begin{figure}[t!]
\centering
\includegraphics[width=\textwidth]{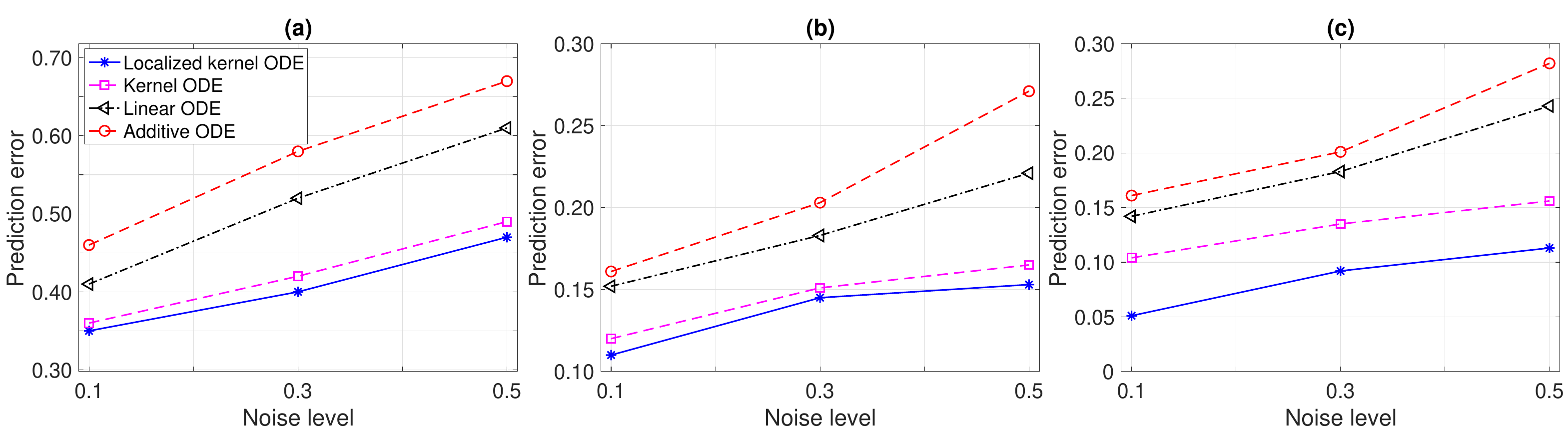}
\caption{The NFBLB example: the prediction error for the varying noise level $\sigma_j$. The results are averaged over 500 data replications. (a) Entire regulatory effect $\widehat{F}_j(x(t)), j=1,2,3$; (b) Individual regulatory effect $\widehat{F}_{23}(\widehat{x}_3(t))$; (c) Individual regulatory effect $\widehat{F}_{12}(\widehat{x}_2(t))$.}
\label{fig:eg1-individual}
\end{figure}

Figure \ref{fig:eg1-individual} reports the prediction accuracy of the entire regulatory effect $\widehat{F}_{j}(\widehat{x}(t))$, as well as the individual regulatory effects $\widehat{F}_{23}(\widehat{x}_3(t))$ and $\widehat{F}_{12}(\widehat{x}_2(t))$. For the entire regulatory effect, the predictor error is computed as the squared root of the sum of predictive mean squared errors for $F_j(x_j(t)),j=1,2,3$, at the unseen “future” time point $t\in[2,3]$, i.e., $\left\{\sum_{j=1}^{3}\int_{2}^{3}[\widehat{F}_j(\widehat{x}_j(t))-F_j(x_j(t))]^2dt\right\}^{1/2}$, where the integral is evaluated at $10000$ evenly distributed time points in $[2,3]$. Similarly, for the individual regulatory effects, the predictor error is computed as $\left\{\int_{2}^{3}[\widehat{F}_{23}(\widehat{x}_3(t))-F_{23}(x_3(t))]^2dt\right\}^{1/2}$, and $\left\{\int_{2}^{3}[\widehat{F}_{12}(\widehat{x}_2(t))-F_{12}(x_2(t))]^2dt\right\}^{1/2}$, respectively. We see that the prediction error of the localized kernel ODE estimator for the \emph{entire} regulatory effect is comparable with that of kernel ODE, which agrees with Theorem \ref{thm:optimalestoffunctional}. Moreover, Figure \ref{fig:eg1-individual}(b) and (c) show that the proposed method outperforms the kernel ODE estimator in predicting the \emph{individual} regulatory effect. This is because the kernel ODE estimator primarily targets the sum of all individual effects, whereas our proposed method directly estimates the individual functional that measures the regulatory  effect of one signal variable on another.

%%%%%%%%%%%%%%%%%%%%%%%%%%%%%%%%%%%%%%%%%%%%%%%%%%%
\subsection{Sensitivity Analysis}
\label{sec:sensitivity}

We carry out a sensitivity analysis regarding the local weight function $R_h(t)$ and the bandwidth $h$ using the enzymatic regulation equations example in Section \ref{sec:enzymatic}. We show that the inference results are not sensitive to the choice of the weight function or the bandwidth.

\begin{figure}[t!]
\centering
\includegraphics[width=\textwidth]{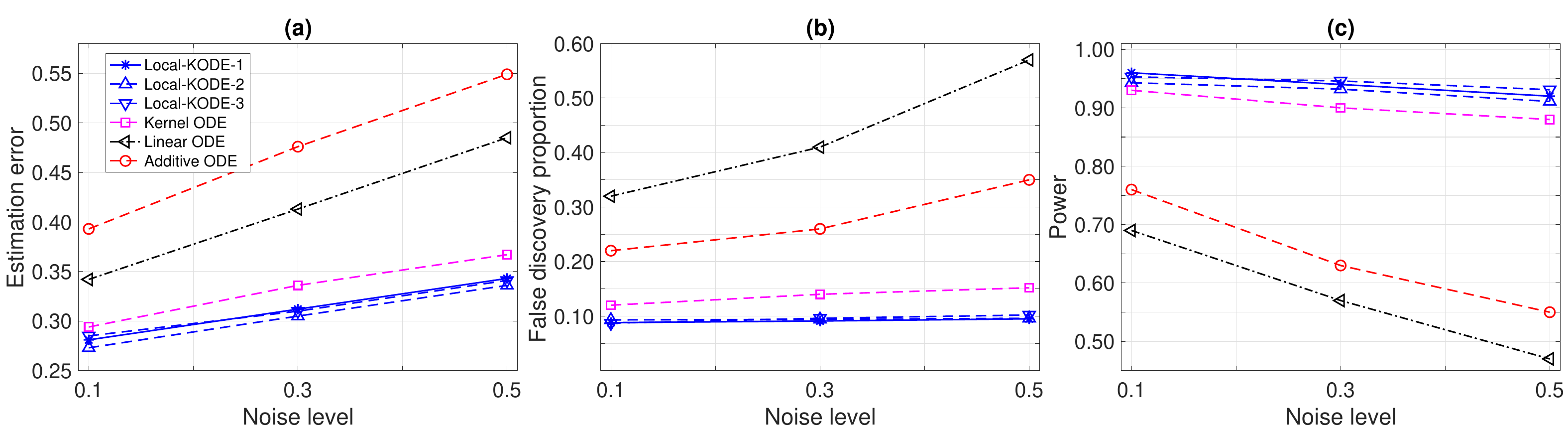}
\caption{Sensitivity analysis: the estimation and sparse selection performance for localized kernel ODE with three different local weight functions, plus kernel ODE, linear ODE, and additive ODE. The results are averaged over $500$ data replications.}
\label{fig:differentkernels}
\end{figure}

First, we consider three different local weight functions: the quadratic weight, the cubic weight, and the Gaussian weight,  
\begin{eqnarray*}
R^{(1)}_h(t) & = & (15/16)\cdot(1-t^2/h^2)^2\mathbf 1(|t|<h), \\
R^{(2)}_h(t) & = & (1-t^2/h^2)^3\mathbf 1(|t|<h), \\
R^{(3)}_h(t) & = & \exp(-t^2/2h^2).
\end{eqnarray*}
We couple them with the proposed localized kernel ODE method, while we continue to choose the bandwidth $h$ using tenfold cross-validation. We couple the proposed confidence band with the BH procedure at the FDR level of $10\%$. We consider three evaluation criteria, the false discovery proportion, the empirical power, and the estimation error, the same as those used in Figure \ref{fig:eg1-prediction}. Figure \ref{fig:differentkernels} reports the performance of the localized kernel ODE method with the three local weight functions, denoted as Local-KODE-1, 2, 3, respectively, plus the kernel ODE, linear ODE, and additive ODE methods. We see that the performances of the three localized kernel ODE methods are fairly close. The false discovery proportions differ at most $0.7\%$, the empirical powers differ at most $1.7\%$, and the estimation errors differ at most $4.2\%$, across different noise levels. Besides, they all outperform the alternative solutions. These results show that the proposed localized kernel ODE method is relatively robust to the choice of the local weight function.

\begin{figure}[t!]
\centering
\includegraphics[width=\textwidth]{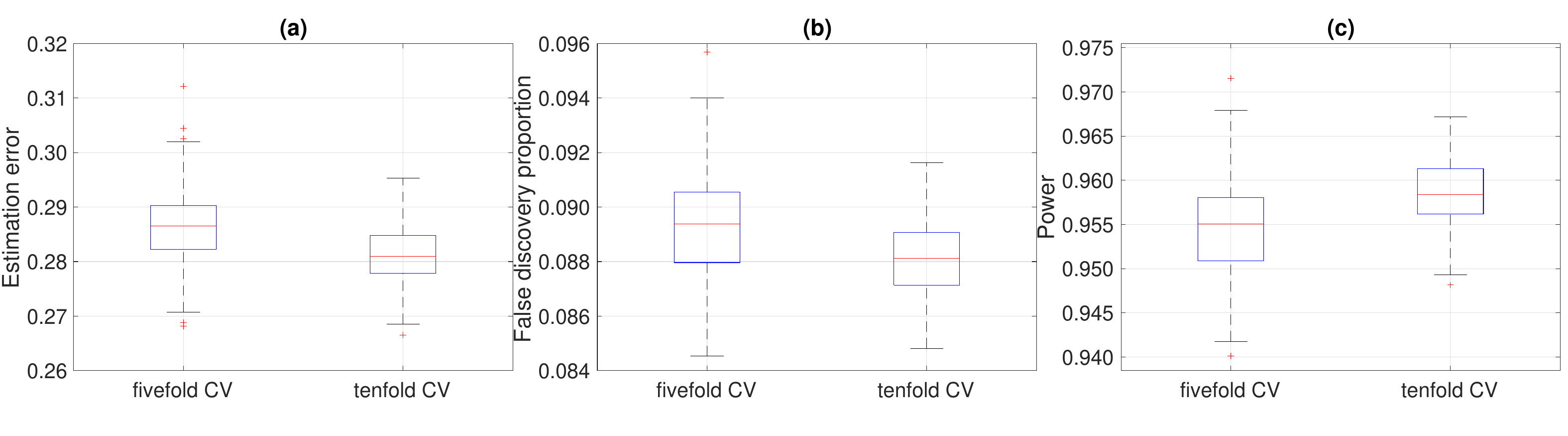}
\caption{Sensitivity analysis: the estimation and sparse selection performance for localized kernel ODE with two different choices of kernel bandwidth. The boxes range from the lower to the upper quartile, and the whiskers extend to the most extreme data point that is no more than $1.5$ times the interquartile range from the box. The results are collected over $500$ data replications.}
\label{fig:initialization}
\end{figure}

Next, we consider the selection of bandwidth $h$. We experiment with fivefold cross-validation and tenfold cross-validation of minimizing the residual sums of squares. We use the quadratic local weight function, and fix the noise level of the enzymatic regulation equations example at $\sigma_j=0.1, j=1,2,3$. Figure \ref{fig:initialization} reports the estimation and sparse selection performance of the localized kernel ODE method. We see that the performances under these two different choices of the bandwidth $h$ are close. For the medians, the false discovery proportions differ at most $0.15\%$, the empirical powers differ at most $0.3\%$, and the estimation errors differ at most $1.8\%$. These results show that the proposed localized kernel ODE method is relatively robust to the choice of bandwidth.

%%%%%%%%%%%%%%%%%%%%%%%%%%%%%%%%%%%%%%%%%%%%%%%%%%%
\vskip 0.2in
\bibliography{ref_kernel}

\end{document}